{}
{}

\newif\ifpublic\publictrue

\documentclass[12pt,a4paper]{article}

\usepackage{amsmath,amssymb}
\ifpublic\else\usepackage{showkeys}\fi
\usepackage{graphicx}
\usepackage[bookmarks=true,hyperfigures=true]{hyperref}
\usepackage[usenames, dvipsnames]{color}
\usepackage[nosort]{cite}
\usepackage[bulletsep]{collref}

\def\showkeysrefformat#1{{\normalfont\tiny\ttfamily#1}}
\makeatletter
\def\SK@@ref#1>#2\SK@{%
 {\@inlabelfalse\leavevmode\vbox to\z@{%
 \vss\SK@refcolor\rlap{\vrule\raise .75em%
  \hbox{\showkeysrefformat{#2}}}}}}
\makeatother

\usepackage[a4paper,text={160mm,247mm},centering]{geometry}

\allowdisplaybreaks[3]

\numberwithin{equation}{section}

\usepackage[font=small,labelfont=bf,width=0.85\textwidth]{caption}

\usepackage{mathfixs}
\ProvideMathFix{autobold,greekcaps,frac,root}
\ProvideMathFix{multskip}

\RequirePackage[extdef]{delimset}

\ProvideMathFix{vfrac,rfrac}
\newcommand{\half}{\rfrac{1}{2}}
\newcommand{\ihalf}{\rfrac{i}{2}}
\newcommand{\quarter}{\rfrac{1}{4}}

\newcommand{\Real}{\mathbb{R}}
\newcommand{\Complex}{\mathbb{C}}

\newcommand{\alg}[1]{\mathfrak{#1}}
\newcommand{\grp}[1]{\mathrm{#1}}
\newcommand{\gen}[1][J]{\mathrm{#1}{}}
\newcommand{\gengauge}{\gen[G]}
\newcommand{\yanghat}[1]{\widehat{#1}}
\newcommand{\genyang}[1][J]{\yanghat{\gen[#1]}{}}
\newcommand{\swdl}[1]{{(#1)}}
\newcommand{\local}{\text{loc}}
\newcommand{\bilocal}{\text{biloc}}
\newcommand{\overlap}{\text{o'lap}}
\newcommand{\copro}{\mathrm{\Delta}}
\newcommand{\yang}{\grp{Y}}
\newcommand{\envalg}{\grp{U}}
\newcommand{\appliedto}{\mathord{\cdot}}
\newcommand{\shift}{\gen[U]}

\newcommand{\superN}{\mathcal{N}}

\newcommand{\oper}[1]{\mathcal{#1}}
\newcommand{\lagrange}{\oper{L}}
\newcommand{\action}{\oper{S}}
\newcommand{\field}{Z}
\newcommand{\eomfor}[1]{\breve{#1}}

\newcommand{\der}{\mathrm{d}}
\newcommand{\del}{\partial}
\newcommand{\cdel}{\nabla}

\newcommand{\nfield}[1]{{[#1]}}

\DeclareMathOperator{\tr}{tr}

\newcommand{\nln}{\nonumber\\}

\def\[{\begin{equation}}
\def\]{\end{equation}}

\providecommand{\href}[2]{#2}
\newcommand{\arxivlink}[1]{\href{http://arxiv.org/abs/#1}{arxiv:#1}}

\makeatletter
\def\mr@ignsp#1 {\ifx\:#1\@empty\else #1\expandafter\mr@ignsp\fi}%
\newcommand{\multiref}[1]{\begingroup
\xdef\mr@no@sparg{\expandafter\mr@ignsp#1 \: }%
\def\mr@comma{}%
\@for\mr@refs:=\mr@no@sparg\do{\mr@comma\def\mr@comma{,}\ref{\mr@refs}}%
\endgroup}
\renewcommand{\eqref}[1]{(\multiref{#1})}
\makeatother

\makeatletter
\newcommand{\namedref}[2]{\hyperref[#2]{#1~\ref*{#2}}}
\newcommand{\secref}[1][Sec.]{\namedref{#1}}
\newcommand{\appref}[1][App.]{\namedref{#1}}

\makeatother

\let\oldbib=\thebibliography
\def\thebibliography{\phantomsection\addcontentsline{toc}{section}{\refname}\oldbib}

\let\oldtoc=\tableofcontents
\def\tableofcontents{\phantomsection\addcontentsline{toc}{section}{\contentsname}\oldtoc}

\providecommand{\hypersetup}[1]{}
\providecommand{\texorpdfstring}[2]{#1}

\hypersetup{plainpages=false}
\hypersetup{pdfpagemode=UseNone}
\hypersetup{bookmarksnumbered=true}
\hypersetup{pdfstartview=FitH}
\hypersetup{colorlinks=false}
\hypersetup{citebordercolor={.5 1 .5}}
\hypersetup{urlbordercolor={.5 1 1}}
\hypersetup{linkbordercolor={1 .7 .7}}

\makeatletter
\let\@keywords\@empty
\let\@subject\@empty
\providecommand{\keywords}[1]{\gdef\@keywords{#1}}
\providecommand{\subject}[1]{\gdef\@subject{#1}}
\def\thetitle{\@title}
\def\theauthor{\@author}
\def\thesubject{\@subject}
\def\thedate{\@date}
\def\thekeywords{\@keywords}
\makeatother
\AtBeginDocument{
\hypersetup{pdftitle={\thetitle}}%
\hypersetup{pdfauthor={\theauthor}}%
\hypersetup{pdfsubject={\thesubject}}%
\hypersetup{pdfkeywords={\thekeywords}}%
}

\newcommand{\remark}[2][]{}
\newcommand{\remarkref}[1][]{}

\ifpublic\else

\usepackage{ifpdf}
\ifpdf\else\RequirePackage[active]{srcltx}\fi
\RequirePackage{color}

\renewcommand{\remark}[2][]{{\normalfont\sffamily\hspace{1ex}%
  \def\emph{\textsl}\def\textbullet{$\bullet$}
  \def\tmparga{#1}%
  \def\tmpargb{MR}\ifx\tmparga\tmpargb\color{magenta}\fi%
  \def\tmpargb{NB}\ifx\tmparga\tmpargb\color{blue}\fi%
  \def\tmpargb{AG}\ifx\tmparga\tmpargb\color{Orange}\fi%
  \def\tmpargb{Refs}\ifx\tmparga\tmpargb\color{Orchid}\fi%
  \def\tmpargb{}\ifx\tmparga\tmpargb\color{red}\fi%
  \def\tmpargb{}\ifx\tmparga\tmpargb\else \textbf{#1:} \fi%
  #2\hspace{1ex}}}

\renewcommand{\remarkref}[1][]{{\def\tmparga{#1}\def\tmpargb{}%
  \ifx\tmparga\tmpargb\remark[Refs]{needed}\else\remark[Refs]{#1}\fi}}

\fi

\begin{document}

\title{Yangian Symmetry for the Action\texorpdfstring{\\}{}
of Planar \texorpdfstring{$\mathcal{N}=4$}{N=4} Super Yang--Mills and\texorpdfstring{\\}{}
\texorpdfstring{$\mathcal{N}=6$}{N=6} Super Chern--Simons Theories}

\pdfbookmark[1]{Title Page}{title}
\thispagestyle{empty}

\begingroup\raggedleft\footnotesize\ttfamily
\arxivlink{1803.06310}\\
HU-EP-18/08
\par\endgroup

\vspace*{2cm}
\begin{center}%
\begingroup\Large\bfseries\thetitle\par\endgroup
\vspace{1cm}


\begingroup\scshape
Niklas Beisert\textsuperscript{1}, Aleksander Garus\textsuperscript{1}, Matteo Rosso\textsuperscript{2}
\endgroup
\vspace{5mm}

\textit{\textsuperscript{1}Institut f\"ur Theoretische Physik,\\
Eidgen\"ossische Technische Hochschule Z\"urich,\\
Wolfgang-Pauli-Strasse 27, 8093 Z\"urich, Switzerland}
\vspace{0.1cm}

\begingroup\ttfamily\small
\verb+{+nbeisert,agarus\verb+}+@itp.phys.ethz.ch\par
\endgroup
\vspace{5mm}

\textit{\textsuperscript{2}Institut f\"ur Physik,\\
Humboldt Universit\"at zu Berlin,\\
Zum Grossen Windkanal 6, D-12489 Berlin, Germany}
\vspace{0.1cm}

\begingroup\ttfamily\small
matteorosso87@gmail.com\par
\endgroup
\vspace{5mm}
\vfill

In Honor of Joe Polchinski
\vfill

\textbf{Abstract}\vspace{5mm}

\begin{minipage}{12.7cm}
In this article we establish the notion of classical Yangian symmetry
for planar $\superN = 4$ supersymmetric Yang--Mills theory
and for related planar gauge theories.
After revisiting Yangian invariance for the equations of motion,
we describe how the bi-local generators act on the action of the model
such that the latter becomes exactly invariant.
In particular, we elaborate on the relevance of the planar limit
and how to act non-linearly with bi-local generators on the cyclic action.
\end{minipage}

\vspace*{4cm}

\end{center}

\newpage

\ifpublic
\tableofcontents
\vspace{\baselineskip}\hrule
\fi

\section{Introduction}
\label{sec:intro}

Integrability of planar $\superN=4$ supersymmetric Yang--Mills (sYM) theory,
its dual string theory on $AdS_5\times S^5$ and several other
AdS/CFT dual pairs of models
has proved to be an extremely powerful feature in computing
various quantities at weak, strong and even at intermediate coupling strength;
see \cite{Arutyunov:2009ga,Serban:2010sr,Beisert:2010jr} for reviews of the subject.
However, despite the many successes related to planar integrability,
a rigorous and unified notion of this feature is still missing.
In particular, such a notion would be highly desirable towards
finding a proof for planar integrability and thus towards firmly establishing
a solid foundation for the results alluded to above.

In this regard, the worldsheet theory of $AdS_5\times S^5$ strings
at strong 't~Hooft coupling is in rather good shape because there exist
well-established notions for integrability
in such two-dimensional field theories.
Most importantly, Joe Polchinski with Bena and Roiban showed that
the structures of integrability can be formulated in terms
of a family of flat connections \cite{Bena:2003wd}.
However, the non-ultra-locality of the algebra of worldsheet currents
poses some difficulty, see \cite{Magro:2010jx},
and the quantisation of these algebraic structures
has several unresolved issues.

The exploration of integrable structures in
planar $\superN=4$ sYM took a different path:
Various remarkable features were discovered
in the investigation of particular observables.
On a case-by-case basis,
these discoveries led to the application, refinement and development
of several integrability-based methods
to compute these quantities
much more efficiently than by ordinary field theory methods.
Some of these features and methods directly matched known integrable structures:
for example, the spectral problem of local operators
at the leading one-loop order
was related to an integrable spin chain \cite{Minahan:2002ve,Beisert:2003yb},
and the Bethe ansatz could be applied to the computation
of the spectrum in various interesting limits.

Integrability of a physics model goes hand in hand with an enhancement
of its ordinary symmetry algebra by a large amount of hidden symmetries.
A unifying structure of the integrability-related features of planar $\superN=4$ sYM
turned out to be the Yangian algebra $\yang[\alg{psu}(2,2|4)]$ \cite{Dolan:2003uh}
based on the superconformal algebra $\alg{psu}(2,2|4)$ of the model.
The Yangian $\yang[\alg{g}]$ is an infinite-dimensional quantum algebra
based on a (finite-dimensional) Lie (super)algebra $\alg{g}$
\cite{Bernard:1992ya,MacKay:2004tc,Torrielli:2010kq,Spill:2012qe,Loebbert:2016cdm}.
Yangian algebras underlie large classes of integrable models
where the Lie algebra $\alg{g}$ usually describes the ordinary symmetries of the model.
The latter is spanned by the so-called level-zero generators $\gen^A$, $A=1,\ldots,\dim \alg{g}$,
and the Yangian algebra enhances it by the level-one generators $\genyang^A$, $A=1,\ldots,\dim \alg{g}$,
which form a similar set as the $\gen^A$.
Importantly, the level-one generators act in a certain bi-local fashion
and as non-local transformations they are not easily identified as symmetries
and are therefore hidden from a standard analysis.
Beyond these, the Yangian contains infinitely many other generators
at higher levels which will not be of concern in this work.

The Yangian algebra has been observed and applied for various observables,
but in most cases some caveats exist
which prevent the action of the Yangian algebra
from being considered a proper invariance:
In the case of the spectral problem of local operators,
the spin chain Hamiltonian almost commutes
with the Yangian algebra \cite{Dolan:2003uh}.
However, the cyclic boundary conditions represent an obstacle here
because the Yangian generators typically map cyclic physical states
to non-cyclic states which have no proper meaning
in the field theory context.
Furthermore at higher loops, some issues related to
gauge symmetry have been observed in \cite{Zwiebel:2006cb}
which obscure the closure of the algebra.
The S-matrix describing scattering of magnon excitations of the spin chain
was shown to have exact Yangian symmetry \cite{Beisert:2007ds},
even at finite coupling strength,
but this symmetry is merely based on a
$\alg{psu}(2|2)\times\alg{psu}(2|2)$ subalgebra of $\alg{psu}(2,2|4)$,
or, more precisely, an extension thereof.
Also colour-ordered scattering amplitudes
of ordinary particles of the four-dimensional gauge theory
have been shown to display Yangian invariance \cite{Drummond:2009fd}
which can be understood as an interplay
between ordinary and dual conformal symmetry \cite{Drummond:2008vq}.
Unfortunately, also here an exact and complete invariance is spoilt,
this time by the presence of infrared divergences
due to the scattering of massless particles.
The divergences have been regularised in many different ways
in order to arrive at suitable results, see e.g.\
\cite{Alday:2009zm,Beisert:2010gn},
however, such regularisations obscure invariances,
and thus make it much less clear in what sense
the Yangian algebra can be regarded as a symmetry
or how to prove a corresponding statement in general.
Nevertheless, Yangian symmetry has proved useful
in constructions of scattering amplitudes, see e.g.\
\cite{Drummond:2008cr,Drummond:2010qh,ArkaniHamed:2010kv,CaronHuot:2011kk,Ferro:2016zmx}.
A similar picture arises from the study of null polygonal
Wilson loops which are T-dual to scattering amplitudes
\cite{Alday:2007hr,Drummond:2007aua,Brandhuber:2007yx}.
The main difference w.r.t.\ scattering amplitudes
is that the divergences arise from the cusps
between light-like segments of the polygon are associated
to the ultraviolet rather than the infrared regime,
and are thus potentially easier to address.
Yangian symmetry also plays a role in the construction
of many other observables such as
correlation functions and form factors of local operators.
However, here the corresponding worldsheet topology
of the observables is a clear obstruction for Yangian invariance.
Yangian invariance without any of the above restrictions
has only been observed for expectation values of smooth, non-intersecting
Maldacena--Wilson loops with disc topology \cite{Muller:2013rta,Beisert:2015uda}
(even though the analysis is limited to one loop so far).

Altogether, these results demonstrate that the Yangian algebra
appears to be some kind of symmetry of planar $\superN=4$ sYM,
but it is not evident in what sense this statement can be made precise.
They also show that the symmetry is restricted to
finite objects with the worldsheet topology of a disc
which are necessarily symmetric under ordinary superconformal symmetry.
A relevant object with this set of features is
the (unrenormalised, single-trace) action of $\superN=4$ sYM itself.
Showing invariance of the action could in fact be considered as
a proof that the theory is symmetric under
the Yangian algebra. Unfortunately, it was unclear
how to show invariance largely because it was
not understood precisely how to act with a Yangian generator
on the action and how to make sense of the planar limit
and, eventually, of quantum effects.

\bigskip

In our letter \cite{Beisert:2017pnr} we have laid the foundations
for our work to establish the Yangian algebra $\yang[\mathfrak{psu}(2,2|4)]$
as a symmetry of planar $\superN = 4$ sYM by proposing
that it is a symmetry of the equations of motion in a strong sense.
In this article, we provide a more detailed account
of our construction and present explicitly in what precise way
the action is classically invariant under the Yangian algebra.
The fact that the model possesses
a very non-trivial extended symmetry algebra
and at the same time is apparently exactly integrable
can hardly be a coincidence given the usual relationship between these two features.
It should be evident that Yangian symmetry
is a way to formally express the integrability
of planar $\superN=4$ sYM.
Hence, we may declare integrability of a planar gauge theory model
to be the presence of Yangian symmetry (or some other alike symmetry algebra).
\unskip\footnote{As there is no formal definition of integrability
applicable to this case,
it is evidently impossible to prove that Yangian invariance of the action
implies integrability (or the converse statement).
We can merely argue why it makes sense to identify the two features.}

To lend further credibility to our claim,
we will treat $\superN=6$ supersymmetric Chern--Simons theory,
also known as ABJ(M) theory \cite{Aharony:2008gk,Aharony:2008ug},
as the second main example of a gauge field theory
known to be integrable in the planar limit
following the works \cite{Minahan:2008hf,Bak:2008cp}.
Here, the Yangian $\yang[\alg{osp}(6|4)]$ has been established
as a symmetry of colour-ordered scattering amplitudes \cite{Bargheer:2010hn}.
We will show that the action of this model is indeed invariant
under a Yangian symmetry, and thus our proposed definition
of integrability matches with the expectation.
In order to convince ourselves of the non-triviality of Yangian symmetry --
after all, it might in principle be a feature of a broad range of models --
we will furthermore consider pure supersymmetric Yang--Mills theories
with $\superN<4$ supersymmetries.
These sample field theories can be addressed straight-forwardly
within our framework, and there is no indication
that they become integrable in the planar limit.
In line with this expectation, we will show that
these models do not possess Yangian symmetry in the planar limit.

\medskip

The present work is structured as follows:
In \secref{sec:symsN4} we introduce
$\superN=4$ sYM along with its superconformal symmetry
and by doing that we outline the notation used in this work.
We then review the results of our previous letter \cite{Beisert:2017pnr}
on this subject in \secref{sec:classicalYang}
with a much more detailed account of our constructions
and further explanations and discussions.
In \secref{sec:yang_action} we convert these results to a statement
on the invariance of the action
which not only serves a definition of Yangian symmetry of the model
but can also be viewed as an exact notion of integrability.
In \secref{sec:abjm} and \secref{sec:yang_counter}
we consider ABJ(M) theory and pure $\superN<4$ sYM theory
serving as two sample planar gauge theories
for which we can test our proposal.
Indeed we find that Yangian invariance of the action
coincides with our expectations on planar integrability
of these models.
Finally, we conclude in \secref{sec:conclusions}
and we present a list of open issues regarding integrability
within the AdS/CFT correspondence that can be addressed with our framework.
The appendices contain a complete account
of the superconformal algebra $\alg{psu}(2,2|4)$
and its non-linear representation on the fields
(\appref{app:superconformal})
as well as the enhancement to the Yangian algebra
(\appref{app:levelone}).

\section{\texorpdfstring{$\superN=4$}{N=4} sYM and superconformal symmetry}
\label{sec:symsN4}

We start by introducing $\superN=4$ sYM,
the corresponding superconformal algebra
and how the algebra acts as a symmetry of our model.
In particular, we shall discuss the latter point at length
by presenting several different notions of ``symmetry''.
They will later serve as the foundation for our formulation
of Yangian symmetry in this model.

\subsection{\texorpdfstring{$\superN=4$}{N=4} supersymmetric Yang--Mills theory}

The fields of $\superN=4$ sYM consist of a gauge potential $A$
with associated field strength $F$,
four Dirac fermions $\Psi$ and six real scalars $\Phi$.
All matter fields transform in the adjoint representation of the gauge group
which we assume to be $\grp{U}(N_{\text{c}})$.
We will thus express all (real) fields as $N_\text{c}\times N_\text{c}$ (hermitian) matrices,
and sequences of fields correspond to matrix products.

As we will be mainly interested in aspects of the classical field theory model
related to $\alg{psu}(2,2|4)$ superconformal symmetry,
it makes sense to use a notation
where all vector indices are expressed as bi-spinors.
In our convention, Latin indices $a,b,\ldots=1,2,3,4$ denote
$\alg{su}(4)$ internal spinors, and undotted / dotted Greek indices
$\alpha,\beta,\ldots=1,2$ / $\dot\alpha,\dot\beta,\ldots=1,2$
correspond to left / right chiral $\alg{sl}(2,\Complex)$ spacetime spinors.
The coordinates $x$ are given by a hermitian $2\times 2$ matrix,
and they carry the index structure $x^{\beta\dot\alpha}$
with the reality condition
$(x^{\alpha\dot\gamma})^\dagger = x^{\gamma\dot\alpha}$.
The corresponding partial derivatives $\del_{\dot\alpha\beta}$
are defined such that
$\del_{\dot\alpha\beta}x^{\delta\dot\gamma}=\delta^{\dot\gamma}_{\dot\alpha}\delta^\delta_\beta$.
We shall refrain from implicitly raising or lowering spinor indices.
Instead we will explicitly contract indices with the help of the
totally anti-symmetric symbols
$\varepsilon^{\alpha\gamma}$ and $\varepsilon^{\dot\alpha\dot\gamma}$
as well as $\varepsilon^{abcd}$
(all these symbols equal $+1$ when the upper or the lower indices are in proper order).

In this notation, the fields of $\superN=4$ sYM all carry two spinor indices of various kinds
\[
\Phi^{ac},
\bar\Phi_{ac},
\Psi^a{}_{\gamma},
\bar\Psi_{\dot\alpha c}{},
A_{\dot\alpha\gamma},
F_{\alpha\gamma},
\bar F_{\dot\alpha\dot\gamma}.
\]
Here, the bar denotes the hermitian conjugate of the unbarred field.
\unskip\footnote{All fields are spinor matrices,
and hermitian conjugation is meant to act on the indices
of the fields as well such that their order is reversed.
Furthermore, hermitian conjugation
raises or lowers internal $\alg{su}(4)$ indices
and exchanges undotted with dotted $\alg{sl}(2,\Complex)$ indices.
Effectively we have
$(\Phi^{ac})^\dagger = \bar\Phi_{ca} = -\bar\Phi_{ac}$,
$(\Psi^a{}_\gamma)^\dagger = \bar\Psi_{\dot\gamma a}$
and
$(F_{\alpha\gamma})^\dagger = \bar F_{\dot\gamma\dot\alpha}= \bar F_{\dot\alpha\dot\gamma}$.
In practice, we will not need these relations.}
Importantly, the two scalar fields $\Phi$ and $\bar\Phi$ are related
by the reality condition
\[
\label{eq:scalarduality}
\bar\Phi_{ab}=\half\varepsilon_{abcd}\Phi^{cd}
\qquad\iff\qquad
\Phi^{ab}=\half\varepsilon^{abcd}\bar\Phi_{cd}.
\]
The gauge-covariant derivative $\cdel$ of some matter field $\field$
and of the gauge potential $A$ itself is defined as
\unskip\footnote{It is useful to remember the commutator
of covariant derivatives
$\comm{\cdel_{\dot\alpha\beta}}{\cdel_{\dot\gamma\delta}} \field
=i\comm{\cdel_{\dot\alpha\beta}A_{\dot\gamma\delta}}{\field}$.
}
\begin{align}
\label{eq:sym_cdergauge}
\cdel_{\dot\alpha\beta} \field
&:= \partial_{\dot\alpha\beta}\field +i \comm{A_{\dot\alpha\beta}}{\field}
,\nln
\cdel_{\dot\alpha\beta} A_{\dot\gamma\delta}
&:=
\del_{\dot\alpha\beta} A_{\dot\gamma\delta}-\del_{\dot\gamma\delta} A_{\dot\alpha\beta}
+i\comm{A_{\dot\alpha\beta}}{ A_{\dot\gamma\delta}}
.\end{align}
Note that the latter definition of $\cdel A$ as the associated field strength $F$
is merely a convenient notational assignment which will later allow us to write some
expressions in a more uniform fashion.
In the spinor notation,
we can split this field strength into chiral and anti-chiral components $F$ and $\bar F$
\[
\cdel_{\dot\alpha\beta} A_{\dot\gamma\delta}
=
\varepsilon_{\dot\alpha\dot\gamma}F_{\beta\delta}
+\varepsilon_{\beta\delta}F_{\dot\alpha\dot\gamma}.
\]

The Lagrangian is tightly constrained by supersymmetry and possesses two marginal couplings,
the Yang--Mills coupling constant $g_{\text{YM}}$ as well as the topological angle $\theta$;
we shall not be interested in the latter, and the former can be expressed
as an overall factor of the Lagrangian which we will also drop.
The Lagrangian reads
\begin{align}
\label{eq:sym_action}
\lagrange
&=
-\rfrac{1}{2}\varepsilon^{\alpha\epsilon}\varepsilon^{\gamma\kappa}
 \tr\brk!{ F_{\alpha\gamma} F_{\epsilon\kappa}}
-\rfrac{1}{2}\varepsilon^{\dot\alpha\dot\epsilon}\varepsilon^{\dot\gamma\dot\kappa}
 \tr\brk!{\bar F_{\dot\alpha\dot\gamma}\bar F_{\dot\epsilon\dot\kappa}}
\nln & \qquad
+i\varepsilon^{\dot\kappa\dot\alpha}\varepsilon^{\beta\gamma}
 \tr\brk!{ \bar\Psi_{\dot\kappa d} \cdel_{\dot\alpha\beta}\Psi^d{}_{\gamma} }
-\rfrac{1}{4} \varepsilon^{\dot\alpha\dot\gamma}\varepsilon^{\beta\delta}
   \tr\brk!{\cdel_{\dot\alpha\beta}\bar\Phi_{ef}\.\cdel_{\dot\gamma\delta}\Phi^{ef}}
\nln & \qquad
+\ihalf\varepsilon^{\alpha\gamma}
  \tr\brk!{\bar\Phi_{ef}\acomm{\Psi^e{}_\alpha}{\Psi^f{}_\gamma} }
+ \ihalf\varepsilon^{\dot\alpha\dot\gamma}
  \tr\brk!{\Phi^{ef}\acomm{\bar\Psi_{\dot\alpha e}}{\bar\Psi_{\dot\gamma f}} }
\nln & \qquad
+ \rfrac{1}{16}\tr\brk!{\comm{\bar\Phi_{ab}}{\bar\Phi_{ef}}\comm{\Phi^{ab}}{\Phi^{ef}}}
.\end{align}
The equations of motion are obtained by varying the action with respect to the
fields of the theory. To this end, we introduce the notation $\eomfor{\field}$
for the variational derivative of the action w.r.t.\ a generic field $\field$
\footnote{Variational derivatives are defined by
$\delta \field(y)/\delta \field(x) = \delta^4 (x-y)$.
In this case the delta-function eliminates the integral
over spacetime within the action.}
\[
\label{eq:sym_vardef}
\eomfor{\field}:=\frac{\delta \action}{\delta \field}.
\]
More concretely, the set of all variations $\eomfor{\field}$ reads
\begin{align}
\label{eq:sym_eom}
\eomfor{\Phi}_{lk}&=\varepsilon^{\dot\alpha\dot\gamma}\varepsilon^{\beta\delta}
\cdel_{\dot\alpha\beta}\cdel_{\dot\gamma\delta}\bar\Phi_{kl}
- \half\comm!{\bar\Phi_{ab}}{\comm{\Phi^{ab}}{\bar\Phi_{kl}}}
\nln & \qquad
+\ihalf\varepsilon_{klgh}\varepsilon^{\alpha\gamma}\acomm{\Psi^g{}_\alpha}{\Psi^h{}_\gamma}
+ i\varepsilon^{\dot\alpha\dot\gamma}
  \acomm{\bar\Psi_{\dot\alpha k}}{\bar\Psi_{\dot\gamma l}}
,\nln
\eomfor{\bar\Psi}^{l\dot\kappa}&=
  i\varepsilon^{\dot\kappa\dot\alpha}\varepsilon^{\beta\gamma}\cdel_{\dot\alpha\beta}\Psi^l{}_{\gamma}
  -i\varepsilon^{\dot\kappa\dot\alpha}\comm{\Phi^{le}}{\bar\Psi_{\dot\alpha e}}
,\nln
\eomfor{\Psi}^{\lambda}{}_k&=
i\varepsilon^{\lambda\beta}\varepsilon^{\dot\alpha\dot\gamma}\cdel_{\dot\alpha\beta}\bar\Psi_{\dot\gamma k}
-i\varepsilon^{\lambda\beta}\comm{\bar\Phi_{ke}}{\Psi^e{}_{\beta}}
,\nln
\eomfor{A}^{\lambda\dot\kappa}&=
\varepsilon^{\lambda\beta}
\varepsilon^{\dot\kappa\dot\epsilon}
\brk!{
-\varepsilon^{\alpha\gamma} \cdel_{\dot\epsilon\alpha}F_{\gamma\beta}
-\varepsilon^{\dot\alpha\dot\gamma} \cdel_{\dot\alpha\beta}\bar F_{\dot\gamma\dot\epsilon}
- \ihalf\comm!{\bar\Phi_{ac}}{\cdel_{\dot\epsilon\beta}\Phi^{ac}}
-\acomm{\Psi^a{}_{\beta}}{\bar\Psi_{\dot\epsilon a}}}
.\end{align}
With this notation, the equations of motion can be concisely written as $\set{\eomfor{\field}=0}$.
Moreover, the Bianchi identity for the field strength
relates the derivatives of the chiral components
(with the opposite sign w.r.t.\ $\eomfor{A}$)
\[
\varepsilon^{\alpha\gamma} \cdel_{\dot\epsilon\alpha}F_{\gamma\beta}
=
\varepsilon^{\dot\alpha\dot\gamma} \cdel_{\dot\alpha\beta}\bar F_{\dot\gamma\dot\epsilon}.
\]

\subsection{Superconformal algebra}
\label{sec:superconformal}

The superconformal algebra $\alg{psu}(2,2|4)$ is spanned by the following generators:
the $\alg{sl}(2,\Complex)$ Lorentz generators $\gen[L]^\alpha{}_\gamma$, $\gen[\bar L]{}^{\dot\alpha}{}_{\dot\gamma}$,
the translations $\gen[P]_{\dot\alpha\beta}$,
the conformal boosts $\gen[K]^{\beta\dot\alpha}$,
the dilatation $\gen[D]$,
the internal $\alg{su}(4)$ symmetry generators $\gen[R]^a{}_b$,
the supersymmetries $\gen[Q]_{a\beta}$, $\gen[\bar{Q}]{}_{\dot\alpha}{}^b$
and the superconformal boosts $\gen[S]^{\beta a}$, $\gen[\bar S]{}_b{}^{\dot\alpha}$.

In this article we mainly consider a \emph{gauge-covariant} representation
of the superconformal symmetries $\gen\in \alg{psu}(2,2|4)$
on the fundamental fields $\field\in\set{A,\Phi,\Psi,\bar\Psi}$ of our model.
\unskip\footnote{Note that the representation on \emph{fields}
acts by replacing each fundamental field with a transformed field.
It does not act on the \emph{coordinates},
i.e.\ $\gen \appliedto x=0$ for all generators $\gen$,
and correspondingly it does not act on \emph{partial derivatives} $\del$ either.
Consequently, it does not act on functions of the coordinates
which are independent of the fundamental fields, such as,
e.g., the parameter fields for gauge transformations
(unless the latter are defined in terms of the fundamental fields).}
The representation of the translation generators $\gen[P]$ on a generic field $\field$ reads
\[
\label{eq:sym_poincare1}
\gen[P]_{\dot\alpha\beta}\appliedto\field
= i \cdel_{\dot\alpha\beta}\field .
\]
Note that in our representation,
shifts are generated by the covariant derivative $\cdel$
rather than by a plain partial derivative $\del$.
\unskip\footnote{The difference between a plain and a covariant
representation amounts to a gauge transformation
generated by the gauge field contracted to the Killing vector of the symmetry.
However, this requires a full superspace formulation where all components
(in particular the ones along the fermionic directions)
of the gauge field and Killing vectors exist.
Therefore a formulation in components must remain covariant to some extent.}
As usual, the gauge potential $\field=A$ is a somewhat special case
because it is not gauge-covariant in contradistinction to the matter fields.
Notwithstanding, the above rule also explicitly applies to $\field=A$
which is mapped to the (gauge-covariant) field strength $\cdel A$ according to
the definition \eqref{eq:sym_cdergauge}.
Lorentz rotations of the fields are generated by the rule
\begin{align}
\gen[L]^\beta{}_\delta\appliedto\field &=
- ix^{\beta\dot\alpha} \cdel_{\dot\alpha\delta}\field
+\ihalf \delta^{\beta}_{\delta} x^{\kappa\dot\alpha} \cdel_{\dot\alpha\kappa}\field
+(\gen[L]_\text{spin})^\beta{}_\delta\appliedto\field
,\nln
\gen[\bar L]^{\dot\alpha}{}_{\dot\gamma}\appliedto\field &=
-i x^{\beta\dot\alpha} \cdel_{\dot\gamma\beta}\field
+\ihalf\delta^{\dot\alpha}_{\dot\gamma} x^{\beta\dot\kappa} \cdel_{\dot\kappa\beta}\field
+(\gen[\bar L]_\text{spin})^{\dot\alpha}{}_{\dot\gamma}\appliedto\field
,
\end{align}
where $\gen[L]_\text{spin}$ and $\gen[\bar L]_\text{spin}$
denote the spin contribution of the operator
which acts non-trivially only on the spacetime indices of spinor fields
\begin{align}
(\gen[L]_\text{spin})^\beta{}_\delta\appliedto\Psi^c{}_\epsilon &=
-i \delta^\beta_\epsilon \Psi^c{}_\delta
+\ihalf \delta^\beta_\delta \Psi^c{}_\epsilon
,\nln
(\gen[\bar L]_\text{spin})^{\dot\alpha}{}_{\dot\gamma}\appliedto\bar\Psi_{\dot\epsilon d} &=
-i \delta^{\dot\alpha}_{\dot\epsilon} \bar\Psi_{\dot\gamma d}
+\ihalf  \delta^{\dot\alpha}_{\dot\gamma} \bar\Psi_{\dot\epsilon d}
.
\end{align}
Note that in particular there is no spin action
on the gauge field even though it carries spacetime indices.
Likewise, the dilatation generator $\gen[D]$
is represented on a generic field $\field$ by a universal rule
\[
\label{eq:sym_poincare2}
\gen[D]\appliedto \field
= -i x^{\beta\dot\alpha} \cdel_{\dot\alpha\beta} \field-i\Delta_{\field}\field
\]
with the coefficients
$\Delta_\Phi = 1$, $\Delta_\Psi =\Delta_{\bar\Psi} = \vfrac{3}{2}$ and $\Delta_A=0$.
Here, the case $\field=A$ is special because $\Delta_A=0$ does not
match the mass dimension $1$ of the gauge potential.
This curiosity is related to the fact that our representation
also involves a gauge transformation, as we shall discuss below.
\unskip\footnote{The assignment is perfectly consistent:
For example, one can convince oneself that the commutation relation
$\comm{\gen[D]}{\gen[P]}=-i\gen[P]$ holds for all fields $\field$
including $\field=A$. To that end, it is important to realise
that the second generator will also act on the gauge potentials $A$
hidden within covariant derivatives $\cdel$.
Furthermore, for the gauge-covariant field strength one finds $\Delta_F=2$ as expected.}
The representation of the supercharges reads
\begin{align}
\label{eq:sym_qqbar}
\gen[Q]_{a\beta}\appliedto\Phi^{cd} &=
\delta^c_a \Psi^d{}_\beta-\delta^d_a \Psi^c{}_\beta
,\nln
\gen[Q]_{a\beta}\appliedto\bar\Phi_{cd} &=
\varepsilon_{acde}\Psi^e{}_\beta
,\nln
\gen[Q]_{a\beta}\appliedto A_{\dot\gamma\delta} &=
-i \varepsilon_{\beta\delta} \bar\Psi_{\dot\gamma a}
,\nln
\gen[Q]_{a\beta}\appliedto\Psi^c{}_{\delta} &=
-2\delta^c_a F_{\beta\delta}
+i\varepsilon_{\beta\delta} \comm{\Phi^{ce}}{\bar\Phi_{ae}}
,\nln
\gen[Q]_{a\beta}\appliedto\bar\Psi_{\dot\gamma d} &=
2i \cdel_{\dot\gamma\beta}\bar\Phi_{ad}
.\end{align}
We present the representation of the remaining superconformal
generators explicitly in \appref{app:superconformal}.

This representation of the superconformal algebra closes on-shell up to
field-dependent gauge transformations. For example, it is easy to confirm that
\begin{align}
\label{eq:sym_algcomm}
\comm{\gen[P]_{\dot\alpha\beta}}{\gen[P]_{\dot\gamma\delta}} \appliedto \field
&= -i\gengauge[\cdel_{\dot\alpha\beta} A_{\dot\gamma\delta}]\appliedto\field
,\nln
\comm{\gen[Q]_{a\beta}}{\gen[P]_{\dot\gamma\delta}} \appliedto \field
&= i\varepsilon_{\beta\delta}\gengauge[\bar\Psi_{\dot\gamma a}]\appliedto\field
,\nln
\acomm{\gen[Q]_{a\beta}}{\gen[Q]_{c\delta}}\appliedto\field
&=
2i \varepsilon_{\beta\delta}\gengauge[\bar\Phi_{ac}]\appliedto\field
,\end{align}
where $\gengauge[X]$ generates a gauge transformation with
the gauge parameter field $X$ as follows
\[
\gengauge[X]\appliedto\field = \comm{X}{\field},
\qquad
\gengauge[X]\appliedto A_{\dot\alpha\beta} = i\cdel_{\dot\alpha\beta}X.
\]
The appearance of a gauge transformation in the algebra is not surprising
given the fact the we use covariant rather than partial derivatives
to generate shifts: namely, one can view the difference between
$\cdel$ and $\del$ as a (field-dependent) gauge transformation $i\gengauge[A]$.
\unskip\footnote{Incidentally, the relationship
$\cdel\field=\del \field+i\gengauge[A]\appliedto\field$
not only provides the standard covariant derivative on covariant fields $\field$,
but for $\field=A$ it also yields precisely the non-abelian field strength
as defined in \eqref{eq:sym_cdergauge}.}
\footnote{In a full superspace formulation of a supersymmetric gauge theory,
one could choose to remove the gauge transformations by
subtracting from each generator $\gen$
a gauge transformation $\gengauge[\xi\cdot A]$
sourced by the superspace gauge field
contracted to the Killing vector $\xi$ associated to $\gen$.
In a component formulation this is not possible due to the lack of
superspace coordinates and components.}

The advantage of this representation is that covariant quantities remain manifestly covariant.
The price to pay, however, is that the relations of the original algebra $\alg{psu}(2,2|4)$ are not manifest
-- to recover them, the ideal generated by gauge transformations must be quotiented out.
The related complications are minor for the superconformal algebra,
but they introduce considerable difficulty
in considering the algebraic relations for the Yangian algebra,
as will be discussed in \cite{Beisert:2018ijg}.

\subsection{Notions of symmetry}
\label{sec:sc_onshell}

In the following, we shall discuss three notions of symmetry of a model:
invariance of the action as well as
a strong and a weak version of invariance of the equations of motion.
They will serve as a starting point for our discussion
of Yangian symmetry of planar $\superN=4$ sYM
which has several complications that can be avoided
by some of the notions.
We will consider some generator $\gen\in\alg{psu}(2,2|4)$
in the context of $\superN=4$ sYM,
but the following arguments will be rather general and apply to
generic local symmetries of a field theory.

\paragraph{Invariance of the action.}

Typically, a symmetry $\gen$ of a model is understood
as an invariance of the action $\action$
\[
\label{eq:symmetryonaction}
\gen\appliedto\action = 0.
\]
This statement is strong because it implies important structures and relationships
for a field theory. For instance, one can derive
conserved currents and charges by means of Noether's theorem.
Moreover, in a quantum field theory, the Ward--Takahashi identities
imply a large set of relationships between various correlation functions.
For the case of $\superN=4$ sYM, invariance of the action \eqref{eq:sym_action}
can be checked directly using
the representations of $\gen$ such as \eqref{eq:sym_poincare1,eq:sym_poincare2,eq:sym_qqbar}.

One can also perform one step of evaluating the symmetry representation on the action
and write the above symmetry statement \eqref{eq:symmetryonaction}
in terms of the combination $\eomfor{\field}=\delta \action/\delta \field$
expressing the equation of motion as
\[
\label{eq:sym_sinvariance}
\gen\appliedto\action =
(\gen\appliedto\field^I) \frac{\delta\action}{\delta \field^I}=
(\gen\appliedto \field^I) \eomfor{\field}_I = 0.
\]
Here, the field index $I$ refers to all dependencies of all fields
including the gauge degrees of freedom as well as the full coordinate dependence.
\unskip\footnote{Due to the implicit integration over all space,
one can expect that partial integration is necessary to confirm the statement.
Analogously, one should take cyclicity of the trace over the gauge degrees of freedom
into account.}
As such it is evident that the statement must hold off shell
for it becomes trivial when the equations of motion $\eomfor{\field}=0$ are imposed.
In other words, a symmetry implies that \eqref{eq:sym_sinvariance} holds
without making use of the equations of motion.

\paragraph{Strong invariance of the equations of motion.}

An analogous statement can be obtained by considering the variation of
\eqref{eq:sym_sinvariance} with respect to a generic field $\field^K$
\[
  \gen\appliedto \field^I \frac{\delta^2\action}{\delta \field^I\.\delta \field^K}
+ \frac{\delta(\gen\appliedto \field^I)}{\delta \field^K} \frac{\delta\action}{\delta \field^I}=0.
\]
This statement can readily be expressed in a more concise form as
\[
\label{eq:sym_eomoffshell}
\gen\appliedto \eomfor{\field}_K
=
- \frac{\delta(\gen\appliedto\field^I)}{\delta \field^K} \eomfor{\field}_I.
\]
It now has an open field index $K$ which means
that it is localised at some point $x$ of spacetime and
that there is one statement for each component of each fundamental field.
In the case of a gauge theory, the statement is no longer gauge-invariant
but rather gauge-covariant.
Here the open gauge indices imply an $N_\text{c}\times N_\text{c}$ matrix
of relationships.
This statement dictates how the equation of motion on the l.h.s.\ transforms
under the symmetry $\gen$.
Importantly, the statement is still valid off shell.
Arguably, it is as powerful
as the invariance of the action \eqref{eq:symmetryonaction}
because the variation towards \eqref{eq:sym_eomoffshell}
merely discards a constant term from the statement
which plays an insignificant role for almost all purposes.
We shall therefore denote \eqref{eq:sym_eomoffshell} as
\emph{strong invariance of the equations of motion}.
Again, verification of this statement for $\superN=4$ sYM
is straight-forward using the concrete equations of motion \eqref{eq:sym_eom}
and the representations of $\gen$.

\paragraph{Weak invariance of the equations of motion.}

We can now apply the equations of motion $\eomfor{\field}=0$
to the statement \eqref{eq:sym_eomoffshell} in order to remove
the r.h.s.
\[
\label{eq:sym_Jeom}
\gen\appliedto \eomfor{\field}_K \approx 0.
\]
Note that even though the l.h.s.\ contains $\eomfor{\field}$,
it does not vanish automatically because of the
variation within $\gen$ which hits
$\eomfor{\field}$ before the e.o.m.\ are imposed.
By construction, this statement is only valid on shell,
which is expressed by the symbol `$\approx$' here and in the following.
Both relationships \eqref{eq:sym_eomoffshell} and \eqref{eq:sym_Jeom}
state that the variation of the equations of motion is proportional
to the equations of motion.
In other words, they imply that symmetries map solutions to solutions.
Nevertheless, the second version of the statement
is clearly weaker because it does not predict the
specific linear combinations on the r.h.s..
We therefore call \eqref{eq:sym_Jeom}
\emph{weak invariance of the equations of motion}.

\medskip

Let us now reason in the opposite direction and ask ourselves the following question:
Suppose we have a transformation $\gen$ such that \eqref{eq:sym_Jeom} holds.
To what extent can we consider $\gen$ a symmetry of the theory?
When can we promote this transformation to a symmetry of the action of the theory?
The following example shows that \eqref{eq:sym_Jeom} can hardly be considered
a sufficient condition for a symmetry.
Weak invariance of the equations of motion is only a necessary condition.
We propose that the strong version \eqref{eq:sym_eomoffshell}
is a sufficient condition for $\gen$ to be a symmetry of the action.

\paragraph{Example.}

Let us now demonstrate that the strong invariance condition \eqref{eq:sym_eomoffshell}
indeed allows to differentiate which invariance of the equations of motion stems
from a true symmetry of the action.
To this end consider a simple example,
the free complex scalar field $\phi$ defined by the following action
\[
\label{eq:ex_action}
\action=\int \der x^d \. \bar\phi \.\partial^2 \phi.
\]
The equations of motion following from \eqref{eq:ex_action} are the wave equations
\[
\label{eq:ex_eom_phi}
\eomfor{\bar\phi} = \partial^2 \phi = 0,
\qquad
\eomfor{\phi} = \partial^2 \bar\phi =0 .
\]
The above equations of motion
are weakly invariant as in \eqref{eq:sym_Jeom}
under a global complex rescaling of the fields
\begin{align}
\phi  \mapsto e^{\rho + i \theta} \phi,
\qquad
\bar\phi \mapsto  e^{\rho -i \theta} \bar\phi,
\end{align}
with $\rho$, $\theta$ real parameters.
However, the above transformation leaves the action invariant
\eqref{eq:symmetryonaction} only for pure rotations $\rho=0$.

Can we observe the distinction of $\rho$ and $\theta$
on the level of equations of motion,
using only the strong invariance formula \eqref{eq:sym_eomoffshell}?
Let us introduce the generators of the infinitesimal form of the above transformations,
separating the complex \emph{rotation} $\gen[R]$ and \emph{scaling} $\gen[S]$
\begin{align}
\gen[R] \appliedto \phi &= i \theta \. \phi
,&
\gen[R] \appliedto \bar\phi &= - i \theta\. \bar\phi
,\\
\gen[S] \appliedto \phi & = \rho \. \phi
,&
\gen[S] \appliedto \bar\phi & = \rho \. \bar\phi
.
\end{align}
It is then a simple exercise to see that $\gen[R]$
indeed satisfies \eqref{eq:sym_eomoffshell}, whereas for $\gen[S]$
one finds \eqref{eq:sym_eomoffshell} with the opposite sign on the r.h.s..
Hence as claimed, \eqref{eq:sym_eomoffshell}
allows us to verify whether a given symmetry of the equations of motion
is also a symmetry of the action.

\section{Yangian symmetry of the equations of motion}
\label{sec:classicalYang}

In this section we will review and elaborate on
our results of \cite{Beisert:2017pnr}
on Yangian symmetry of the equations of motion of $\superN=4$ sYM.
The goal of that letter was to establish classical Yangian symmetry of
planar $\superN=4$ sYM theory and thus to establish a clear notion
of integrability for this model which can later be carried over to
quantum field theory.

The Yangian algebra is an extension of $\alg{psu}(2,2|4)$ superconformal symmetry,
and the discussion of Yangian symmetry can follow along same lines as in
the previous section.
However, there are also several features that distinguish the Yangian algebra
and its representation on fields from the more basic superconformal symmetry:
\begin{itemize}
\item
Yangian symmetry alias integrability
ought to hold only in the planar limit;
non-planar terms are expected to violate the symmetry.
However, there is no evident notion for the planar part of the action;
the planar limit is largely related to the composition of Feynman diagrams.
How should one realise the planar limit in the context of Yangian symmetry?
\item
The Yangian is a quantum algebra rather than a Lie algebra.
As such there is a large number of presentations which are equivalent
for physical purposes. Furthermore, most of the algebraic relations
comprise many terms and thus require a lot of patience and attention to verify.
\item
Yangian representations are typically non-local.
Non-local symmetries are harder to detect,
and they may not be accessible with conventional tools of field theory.
In particular, Noether's theorem does not apply, at least not directly,
and also the notion of quantum anomalies is obscure.
In fact, non-locality does not refer to spacetime in our case,
but rather to colour space.
This again calls out for unconventional methods and ideas.
\item
A related issue is cyclicity violation. The action of $\superN=4$ sYM
is cyclic due to the trace in colour space.
Yangian representations respect cyclicity only under very special conditions.
We have to formulate these conditions and make sure that they are satisfied.
\item
Symmetry representations in interacting field theories
are often non-linear in the fields.
\unskip\footnote{In the field theory context,
non-linearity of a representation refers to non-linearity in fields.
Importantly, such a representation acts linearly on polynomials of fields
thus displaying linearity as one of its elementary properties.}
It is largely unknown
whether Yangian algebras possess such non-linear representations.
\item
A related issue is gauge symmetry. Gauge symmetry inevitably leads
to interacting field theories with non-linear representations.
The Yangian representation should be compatible with gauge symmetry such that
the unphysical degrees of freedom do not interfere with the symmetry.
Moreover, gauge symmetries have to be fixed in the process of quantisation
such that the gauge field propagator can be defined.
The process of gauge fixing can potentially violate Yangian symmetry.
\end{itemize}
Many of these features and in particular their interplay
complicate the investigation of Yangian symmetry.
We will therefore start with a thorough
discussion of the algebra and more elementary considerations of symmetry
before we move on to the two kinds of invariances of the equations of motion.
The treatment of the equations of motion will be manifestly gauge covariant
which provides some structural constraints on the novel terms that we shall encounter.
Eventually, we will address Yangian invariance of the action
in the next section where all of the complications come into play.

\subsection{Algebra and representations}
\label{sec:yang_algebra}

Given a semi-simple Lie algebra
\unskip\footnote{The discussion and notation will assume an ordinary Lie algebra
generated by bosonic operators $\gen$.
The generalisation to superalgebras is straight-forward by inserting appropriate signs;
we will write out these signs only when we discuss specific generators
of the superalgebra $\alg{g}=\alg{psu}(2,2|4)$ and their representations.}
$\alg{g}$, the associated Yangian algebra $\yang[\alg{g}]$ is an infinite-dimensional
quantum algebra generated by two sets of generators
$\gen^A$ and $\genyang^A$ with $A=1,\ldots,\dim(\alg{g})$
\cite{Drinfeld:1985rx,Drinfeld:1986in}.
The so-called \emph{level-zero} generators $\gen^A$
generate the universal enveloping algebra $\envalg[\alg{g}]$ of
the underlying Lie algebra $\alg{g}$,
and they obey the commutation relations
\[
\comm!{\gen^A}{\gen^B} = if^{A B}{}_C\. \gen^C
\]
with $f^{A B}{}_C$ the structure constants of $\alg{g}$.
The \emph{level-one} generators $\genyang^A$
transform in the adjoint representation of $\alg{g}$,
\[
\label{eq:yang_lv1rel1}
\comm!{\gen^A}{\genyang^B} = if^{A B}{}_C\. \genyang^C,
\]
and they obey the so-called Serre relation at level two
\[
\label{eq:yang_lv1rel2}
\comm!{\genyang^A}{\comm{\genyang^B}{\gen^C}} + \operatorname{cyclic}=
f^{A D}{}_E \. f^{B F}{}_H \.f^{C G}{}_L\. f_{D F G}\.
\gen^{(E}\.\gen^H\. \gen^{L)}.
\]
Here, adjoint indices are raised and lowered with the Killing form $k_{A B}$.
\unskip\footnote{%
  The Killing form in the simple Lie superalgebra $\alg{psu}(2,2|4)$ vanishes,
  but the algebra nevertheless possesses an invariant bilinear form $k'_{AB}$
  that is used to raise and lower adjoint indices.}
All the higher-level generators are given as polynomials in
the level-zero and level-one generators subject to the above commutation relations.

The Yangian $\yang[\alg{g}]$ is actually a Hopf algebra,
and as such it possesses a coalgebra structure
$\copro:\yang\to\yang\otimes\yang$ defined by
\[
\copro\genyang^A =
\genyang^A\otimes 1
+1\otimes\genyang^A
+f^A{}_{BC}\.\gen^B\otimes\gen^C.
\]
The coproduct is coassociative and compatible with the algebra product
in a certain sense. These features allow one to use the
coproduct for the definition of tensor product representations.
The $(n-1)$-fold iterated coproduct of $\gen$ and $\genyang$
acting on $n$ tensor factors reads
\[
\label{eq:yang_copro}
\copro^{n-1}\gen^A
= \sum_{k=1}^{n} \gen^A_k
,\qquad
\copro^{n-1} \genyang^A
=
\sum_{k=1}^{n}  \genyang_k^A
+ f^{A}{}_{BC} \sum_{1\leq k < l \leq n}
\gen_k^B\. \gen_l^C .
\]
Here, $\gen_k^A$ indicates the action of the generator $\gen^A$ on the
$k$-th factor or site of the tensor product
(with a trivial action on the other sites).
The above coproduct of $\gen$ is how one would conventionally
define a tensor product representation for the Lie algebra $\alg{g}$.
It acts on all sites in the same fashion, and therefore the ordering
of the sites plays no role.
The coproduct of a level-one generator $\genyang$
consists of a local part $\sum_k \genyang_k$
analogous to the level-zero coproduct
and of a bi-local contribution
$\sum_{j<k} \gen_j\otimes \gen_k$.
This latter bi-local term does depend on the
ordering of the sites of the tensor product;
in other words, the coproduct is non-cocommutative.

\medskip

As we have seen, the coproduct makes us deal with several generators
of $\alg{g}$ at the same time requiring the use of adjoint indices $A,B,\ldots$
and structure constants $f$. A convenient abbreviation to avoid indices
is Sweedler's notation: here one denotes by $\gen^\swdl{1}$
and $\gen^\swdl{2}$ the level-zero generators
in the first and second factor, respectively, of the coproduct
of some level-one generator $\genyang=\genyang^A$
with an implicit sum over all combinations
\unskip\footnote{In fact, we exclude the
trivial appearance of the level-one generator,
so that $\gen^\swdl{1}\otimes\gen^\swdl{2}$
describes a sum of pairs of level-zero generators.}
\[
\label{eq:sweedler}
\gen^\swdl{1}\otimes\gen^\swdl{2}
:=
f^A{}_{BC} \gen^B\otimes \gen^C
=
-\gen^\swdl{2}\otimes\gen^\swdl{1}.
\]
The above coproduct rule can thus be expressed more concisely as
\[
\label{eq:coprosweedler}
\copro\genyang =
\genyang\otimes 1
+1\otimes \genyang
+\gen^\swdl{1}\otimes \gen^\swdl{2}.
\]

Our analysis of Yangian symmetry can furthermore be simplified by the fact
that only a single level-one generator along with the level-zero generators
is sufficient to generate the whole Yangian algebra;
all other level-one generators follow from the adjoint property \eqref{eq:yang_lv1rel1},
and the higher-level ones from the Serre-relation \eqref{eq:yang_lv1rel2}.
We are thus free to choose a particular generator
for which the resulting expressions simplify as much as possible.
Arguably, this is the level-one momentum $\genyang[P]$
which is also known as the dual conformal generator.
Its coproduct reads
(including all appropriate signs due to fermionic terms)
\unskip\footnote{The overall factor of the bi-local terms
depends on the coefficient of the invariant form $k_{AB}$,
and it thus varies with conventions.}
\begin{align}
\label{eq:yang_pahtcopro}
\copro\genyang[P]_{\dot\alpha\beta}
&=
\genyang[P]_{\dot\alpha\beta} \otimes 1
+  1\otimes \genyang[P]_{\dot\alpha\beta}
\nln &\qquad
- \gen[L]^\gamma{}_\beta \wedge\gen[P]_{\dot\alpha\gamma}
- \gen[\bar L]^{\dot\gamma}{}_{\dot\alpha} \wedge\gen[P]_{\dot\gamma\beta}
-  \gen[D] \wedge \gen[P]_{\dot\alpha\beta}
- \tfrac{i}{2} \gen[Q]_{c\beta} \wedge \gen[\bar Q]_{\dot\alpha}{}^c
,
\end{align}
where the anti-symmetric tensor product $\wedge$ of any two objects $X$ and $Y$ is defined as
\[
\label{eq:wedge_tensor}
X\wedge Y:=X\otimes Y-(-1)^{|X||Y|} Y\otimes X.
\]
Note that the action of $\genyang[P]$ conveniently only needs the dilatation $\gen[D]$
next to the super Poincar\'e generators $\gen[L]$, $\gen[\bar L]$, $\gen[Q]$, $\gen[\bar Q]$ and $\gen[P]$.
All of their representations are reasonably simple compared to the
representations of the superconformal boosts $\gen[S]$, $\gen[\bar S]$ and $\gen[K]$.
In fact we will encounter further simplifications due to the choice of
$\genyang[P]$ later on.
Expressions for the coproducts of the level-one generators $\genyang$
with $\gen\in\set{\gen[Q],\gen[\bar Q],\gen[R]}$
as well as the level-one bonus symmetry $\genyang[B]$ introduced in \cite{Beisert:2011pn}
can be found in \appref{app:levelone}.

\subsection{Issues}

The major complication in considering Yangian symmetry
within an (interacting) field theory is that symmetry representations
are often \emph{non-linear}.
By definition, the representation is still a linear map between observables;
here, non-linearity refers to the fact that a single field
can be mapped to a product of fields.
In a gauge theory, the covariant derivative is a major source
of non-linearity, see e.g.\ \eqref{eq:sym_poincare1,eq:sym_poincare2}.
Furthermore, the representation of supersymmetry
\eqref{eq:sym_qqbar} contains non-linear terms which are not
due to covariant derivatives.

The issue is that the concept of non-linear representations is in competition with
the definition of tensor product representations via the coproduct:
In a linear representation, each field corresponds to a single tensor factor.
The representation acts on fields one-to-one thus preserving the
structure of the tensor product.
For non-linear symmetries, the representation changes the
number of fields and consequently the structure of the tensor product,
see \cite{Beisert:2003ys} for a discussion and construction
of some aspects of non-linear representations.
In fact, the representation does not split into sub-representations
with a definite number of fields,
but there is only the indecomposable representation on polynomials of the fields.
For the local action of level-zero generators this complication is minor
and one can still view the full representation as the sum
of representations on component fields.
Conversely, the construction of the bi-local action for the level-one generators
is less evident, cf.\ some comments in \cite{Agarwal:2005jj}.
Here, the action of the first constituent generator $\gen^\swdl{1}$
changes the tensor product on which the second constituent generator $\gen^\swdl{2}$
is supposed to act.
This mainly refers to the precise definition of the bounds in the double sum
in \eqref{eq:yang_copro} given that $n$ is not well-defined anymore.
For example, it is conceivable that
$\gen^\swdl{2}$ acts on the output of $\gen^\swdl{1}$
corresponding to an overlapping action.
This leaves some ambiguity for the precise definition of a non-linear level-one representation.
Unfortunately, this becomes a rather serious issue when taking
gauge symmetry into account: for instance, the covariant derivative
consists of a partial derivative and a gauge potential,
and as such it relates terms of a different number of constituent fields.
In that sense, the non-linear terms of the Yangian representations
must be chosen delicately such that they will not violate gauge symmetry.

We will thus need representations which are analogous to the actions
defined in \eqref{eq:yang_copro}. In particular, they should
reduce to the ordinary coproduct rule \eqref{eq:yang_copro}
when restricting to linear terms.
To that end, it will make sense to distinguish
the local and bi-local contribution in the level-one representation,
and write
\[
\genyang=\genyang_\local+\genyang_\bilocal
\qquad
\text{where}\quad
\genyang_\bilocal:=
\gen^\swdl{1}\otimes \gen^\swdl{2}.
\]
Here, the tensor product symbol $\otimes$ should be interpreted
such that the first factor acts
on a field which is to the left of the field
on which the second factor acts.
Formally, we define the tensor product acting on a sequence of fields by the rule
\[
\label{eq:tensoractiondef}
\brk{\gen^\swdl{1}\otimes \gen^\swdl{2}}\appliedto
(\cdots\field^I\cdots\field^J\cdots)
:=\ldots+\cdots (\gen^\swdl{1}\appliedto\field^I)\cdots
(\gen^\swdl{2}\appliedto\field^J)\cdots
+\ldots,
\]
where the dots on the r.h.s.\
represent further terms due to pairwise contractions
with the omitted fields in the sequence.
The local part $\genyang_\local$ will just act on all fields homogeneously.
Note that any overlapping contributions
that could be part of some alternative definition of a bi-local action
can and should be interpreted as local contributions.

\medskip

The other main complication is that representations of the Yangian algebra
tend to be incompatible with the boundary conditions imposed by the physical system at hand;
consequently, the Yangian can merely be considered a symmetry of the bulk rather
than of the system as a whole.
Nevertheless, for planar $\superN=4$ sYM our aim is to show exact Yangian invariance
of the action.
The colour trace within the action imposes periodic boundary conditions
on the above tensor product. Moreover, the trace projects to states which
are invariant under cyclic permutations;
the physical information within the action is cyclic.
However, the representations of the Yangian generically do not respect cyclicity.

To understand the violations of cyclicity,
let us consider a cyclic state in a tensor product of $n$ sites.
The Yangian algebra acts on this state by the (linear) representation \eqref{eq:yang_copro}.
To that end, we need to define an ordering of the sites within the tensor product.
The adjacency relationship of the sites defines a local ordering,
which is however globally inconsistent due to the periodic identification.
In order to define an ordering, we need to choose a base point where to ``cut open'' the cycle.
Let $\genyang_{(k,n+k-1)}$ denote the action of $\genyang$
on the range of sites from $k$ through $n+k-1$ (modulo $n$),
i.e.\ where we cut between sites $k-1$ and $k$.
We then have that
\[
\genyang^A_{(1,n)} - \genyang^A_{(k,n+k-1)}
=
2f^A{}_{B C} \.\gen^B_{(1,k-1)}\.\gen^C_{(k,n)}.
\]
Here $\gen_{(j,k)}$ denotes
the representation of the level-zero generator on sites $j$ through $k$.
In a generic situation, the r.h.s.\ is not zero and therefore
the resulting state is not cyclic. Likewise, the action
of $\genyang$ will typically
depend on the particular cyclic representative on which it acts.
This means that there is no universal answer to the
Yangian representation on cyclic states,
and it makes no sense to ask whether or not
any such state is Yangian-invariant.

Gladly, the above difference can be rewritten as \cite{Drummond:2009fd}
\[
\label{eq:yang_diff}
\genyang^A_{(1,n)} - \genyang^A_{(k,n+k-1)}
=
-i f^A{}_{B C} f^{B C}{}_D \.\gen^D_{(1,k-1)}
+ 2f^A{}_{B C} \.\gen^B_{(1,k-1)} \gen^C.
\]
The first term on the r.h.s.\ contains the combination
$f^A{}_{B C} f^{B C}{}_D$ which is proportional to the dual Coxeter
number of the underlying Lie algebra $\alg{g}$. For the $\superN=4$ superconformal
algebra $\alg{psu}(2,2|4)$ this number is zero,
and therefore the term does not contribute.
The second term is a product of two level-zero representations.
The level-zero generator $\gen^C$ acts on the complete state,
and annihilates it if is invariant. In other words,
the level-one representation of a Yangian with vanishing dual Coxeter number
respects cyclicity for those states which are invariant under the level-zero symmetry.
In principle, this is an ideal starting point for considering Yangian invariance
of the action of planar $\superN=4$ sYM because these two
prerequisites are satisfied.

\medskip

Unfortunately, the argument about cyclicity is based on linear representations.
For non-linear representations one may suspect something equivalent to hold,
but it is unclear how to set up the representation precisely and how to
show cyclicity in this case.
The major challenge is thus to properly define non-linear representations
of level-one generators and to learn how to work with them.
In this situation, proving invariance of the action turns out to be a rather difficult task,
and we shall at first consider invariance of the equations of motion
in order to gain a better understanding of non-linear representations.

\subsection{Weak invariance of the equations of motion}
\label{sec:yang_onshell}

We start by considering the weak notion of a symmetry
of the equations of motion \eqref{eq:sym_Jeom}:
given a generator $\gen$ and its action on the fields $\field$,
it must leave the equation of motion $\eomfor{\field}=0$ invariant
\[
\gen \appliedto \eomfor{\field} \approx 0.
\]
For the Yangian level-one generators $\genyang$, the non-trivial coproduct poses two problems:
First, if we want to act with $\genyang_\bilocal$ on the equations of motion,
we have to fix an ordering prescription for each term that appears;
second, we need to specify how $\genyang_\local$ acts on a single field.

Gladly, we can impose a ``natural'' ordering for the fields within
the equations of motion for Yang--Mills theories with $\grp{U}(N_\text{c})$ gauge group.
All the fields in $\superN=4$ sYM can be treated as $N_\text{c} \times N_\text{c}$ matrices
and the (non-commutative) matrix product provides the ordering within a monomial of the fields.
The equations of motion inherit this matrix structure supposing that
all structure constants of the gauge group are written in terms
of commutators of the fields.
This is also where the large-$N_\text{c}$ limit comes into play:
the possibility to write arbitrary (adjoint) combinations
of the fields in terms of matrix polynomials requires
a $\grp{U}(N_\text{c})$ gauge group with
correspondingly large $N_\text{c}$.
\unskip\footnote{For matrix products of $N_\text{c}$ fields or more,
there are certain identities of polynomials related to anti-symmetrisation
which eventually introduce some ambiguity for the ordering of fields.
Arguably this ambiguity does not apply to the equations
of motion when $N_\text{c}>3$ (at least).
Nevertheless, we will later want to establish Yangian
symmetry for arbitrary correlators of the fields,
in which case an arbitrarily large $N_\text{c}$ will be necessary.
Of course there is the option to reverse-engineer the ordering rule
such that the level-one generators are directly represented
on fields contracted by structure constants (as long
as $N_\text{c}$ is sufficiently large),
however any such rule will be rather messy and cumbersome.}
Note that 't~Hooft's (double) line notation for adjoint fields,
which is prominently used in the discussion of the planar limit,
directly depicts our ordering prescription.

For what concerns the single-field action $\genyang_\local$,
we will fix it by first computing the bi-local action
of $\genyang_\bilocal=\gen^\swdl{1}\otimes \gen^\swdl{2}$
on some equations of motion,
which is completely determined by the conformal representation.
We shall then require that
the local terms eliminate all remaining terms
\[
\brk[s]!{\genyang_\bilocal +\genyang_\local}\appliedto
\eomfor{\field} \approx 0.
\]
The fact that suitable local terms can be found will be a first test
for Yangian symmetry.
Subsequently, we will consider the other equations of motion.
Yangian symmetry will pass a more elaborate test
if the same local terms compensate the remainders from the bi-local action
for all equations of motion.

\medskip

In the following we will focus on the level-one momentum generator
$\genyang=\genyang[P]$
whose bi-local action is determined by the coproduct in \eqref{eq:yang_pahtcopro}
We will first compute the action of this
bi-local term on the Dirac equation \eqref{eq:sym_eom}
\[
\eomfor{\bar\Psi}^{l\dot\kappa}=
  i\varepsilon^{\dot\kappa\dot\alpha}\varepsilon^{\beta\gamma}\cdel_{\dot\alpha\beta}\Psi^l{}_{\gamma}
  -i\varepsilon^{\dot\kappa\dot\alpha}\comm{\Phi^{le}}{\bar\Psi_{\dot\alpha e}}
\approx 0.
\]
We act on the latter with the coproduct \eqref{eq:yang_pahtcopro}
by means of the tensor product action \eqref{eq:tensoractiondef}
using the superconformal representation
\eqref{eq:sym_poincare1,eq:sym_poincare2,eq:sym_qqbar},
to obtain
\unskip\footnote{Note that all explicit $x$-dependence originating from the generators
$\gen[L]$, $\gen[\bar L]$ and $\gen[D]$ cancels out exactly.
This convenient feature is related to the fact that $\genyang[P]$ commutes with $\gen[P]$
in the Yangian algebra.}
\begin{align}
\label{eq:yang_pbiloc}
\genyang[P]_{\dot\alpha\beta,\bilocal}\appliedto
\eomfor{\bar\Psi}^{d\dot\gamma}&=
i\varepsilon^{\dot\gamma \dot\epsilon}
  \acomm!{\cdel_{\dot\epsilon\beta}\Phi^{df}}{\bar\Psi_{\dot\alpha f}}
+i \varepsilon^{\dot\gamma\dot\epsilon}
  \acomm!{\Phi^{df}}{\cdel_{\dot\alpha\beta} \bar\Psi_{\dot\epsilon f}}
+i\delta^{\dot\gamma}_{\dot\alpha}
  \acomm!{\comm{\Phi^{df}}{\bar\Phi_{ef}}}{\Psi^e{}_\beta}.
\end{align}
This expression does not vanish on-shell on its own.
However, we can make an ansatz for a single-field action $\genyang[P]\appliedto \field$
based on all terms with appropriate quantum numbers and symmetries,
and add their contribution to the overall action of $\genyang[P]$.
By choosing the coefficients of the ansatz as
\begin{align}
\label{eq:yang_sfa}
\genyang[P]_{\dot\alpha\beta}\appliedto \Phi^{cd} &:= 0
,\nln
\genyang[P]_{\dot\alpha\beta}\appliedto \Psi^c{}_\delta
&:=-\varepsilon_{\beta\delta} \acomm!{\Phi^{ce}}{\bar\Psi_{\dot\alpha e}}
,\nln
\genyang[P]_{\dot\alpha\beta}\appliedto \bar\Psi_{\dot\gamma d}
&:=-\varepsilon_{\dot\alpha \dot\gamma}\acomm!{\bar\Phi_{de}}{\Psi^e{}_\gamma}
,\nln
\genyang[P]_{\dot\alpha\beta}\appliedto A_{\dot\gamma\delta}
&:= \rfrac{i}{4}\varepsilon_{\dot\alpha \dot\gamma} \varepsilon_{\beta\delta}
\acomm{\Phi^{ef}}{\bar\Phi_{ef}},
\end{align}
we get that
\[
\label{eq:yang_phatdirac}
\genyang[P]_{\dot\alpha\beta}\appliedto\eomfor{\bar\Psi}^{d\dot\gamma}
=-\delta^{\dot\gamma}_{\dot\alpha}\varepsilon_{\beta\epsilon}
  \acomm!{\Phi^{df}}{\eomfor{\Psi}^\epsilon{}_f}
\approx 0.
\]
This shows that $\genyang[P]_{\dot\alpha\beta}$ is a weak symmetry of the Dirac equation
of $\superN=4$ sYM.

One comment on the structure \eqref{eq:yang_sfa}
of the single-field action is in order:
It may appear unconventional as it is formulated in terms of
anti-commutators in places where commutators are usually expected.
However, this configuration is actually natural considering
the action of a level-one Yangian generator $\genyang_\bilocal$
on a commutator
\[
\label{eq:yang_lv1comm}
  f^A{}_{B C} \brk!{\gen^B \otimes \gen^C} \appliedto \comm{\field^I}{\field^J}
= f^A{}_{B C} \brk!{\gen^B \appliedto \field^I\. \gen^C\appliedto \field^J
             -\gen^B \appliedto \field^J\. \gen^C\appliedto \field^I}
= f^A{}_{B C} \acomm!{\gen^B \appliedto \field^I}{\gen^C\appliedto \field^J}.
\]
This means that level-one generators typically map commutators
to anti-commutators.

With the above single-field action of \eqref{eq:yang_sfa}
it is now possible to show
that \emph{all} the equations of motion of $\superN=4$ sYM are
weakly invariant under $\genyang[P]$
\begin{align}
\label{eq:yang_OSeom}
\genyang[P]_{\dot\alpha\beta}\appliedto\eomfor{\bar\Psi}^{d\dot\gamma}
&=-\delta^{\dot\gamma}_{\dot\alpha}\varepsilon_{\beta\epsilon}
  \acomm!{\Phi^{df}}{\eomfor{\Psi}^\epsilon{}_f}
,\nln
\genyang[P]_{\dot\alpha\beta} \appliedto\eomfor{\Psi}^{\delta}{}_c
&=
-\delta^\delta_\beta\varepsilon_{\dot\alpha \dot\epsilon} \acomm!{\bar\Phi_{cf}}{\eomfor{\bar\Psi}^{f\dot\epsilon}}
,\nln
\genyang[P]_{\dot\alpha\beta} \appliedto\eomfor{\Phi}_{dc} &=
-i \varepsilon_{\dot\alpha \dot\kappa} \varepsilon_{\beta\lambda}
\acomm!{\bar\Phi_{cd}}{\eomfor{A}^{\lambda \dot\kappa}}
+\rfrac{5}{2}\varepsilon_{cdef}\varepsilon_{\dot\alpha\dot\kappa}\comm!{\Psi^e{}_\beta}{\eomfor{\bar\Psi}^{f\dot\kappa}}
\nonumber \\&\qquad
+\rfrac{5}{2}\varepsilon_{\beta\lambda}\comm!{\bar\Psi_{\dot\alpha c}}{\eomfor{\Psi}^\lambda{}_d}
-\rfrac{5}{2}\varepsilon_{\beta\lambda}\comm!{\bar\Psi_{\dot\alpha d}}{\eomfor{\Psi}^\lambda{}_c}
,\nln
\genyang[P]_{\dot\alpha\beta} \appliedto\eomfor{A}^{\delta\dot\gamma}
&=
-\rfrac{i}{2} \delta_{\dot\alpha}^{\dot\gamma}\delta_\beta^\delta \acomm!{\Phi^{ef}}{\eomfor{\Phi}_{ef}}
\nonumber\\&\qquad
-\rfrac{5i}{2} \delta_{\dot\alpha}^{\dot\gamma} \comm!{\Psi^e{}_\beta}{\eomfor{\Psi}^\delta{}_e}
+\rfrac{i}{2} \delta_{\dot\alpha}^{\dot\gamma}\delta_\beta^\delta \comm!{\Psi^e{}_\lambda}{\eomfor{\Psi}^\lambda{}_e}
\nonumber\\&\qquad
-\rfrac{5i}{2} \delta_\beta^\delta \comm!{\bar\Psi_{\dot\alpha e}}{\eomfor{\bar\Psi}^{e\dot\gamma}}
+\rfrac{i}{2} \delta_{\dot\alpha}^{\dot\gamma}\delta_\beta^\delta \comm!{\bar\Psi_{\dot\kappa e}}{\eomfor{\bar\Psi}^{e\dot\kappa}}
.\end{align}
Moreover, we have verified explicitly that they are weakly invariant
under the level-one generators $\genyang$
with $\gen\in\set{\gen[Q],\gen[\bar Q],\gen[R]}$
as well as the level-one bonus symmetry $\genyang[B]$
which extends the Yangian of $\alg{psu}(2,2|4)$.
This also fixes the single-field actions of these generators,
and we present our results in \appref{app:levelone}.

Up to some issues w.r.t.\ the closure of the Yangian algebra onto
gauge transformations,
to be discussed in \cite{Beisert:2018ijg},
we conclude that the Yangian of $\alg{psu}(2,2|4)$ is a weak
symmetry of the classical planar equations of motion of $\superN=4$ sYM.
As previously discussed, this represents a necessary condition
for Yangian symmetry in planar $\superN=4$ sYM.
It is reassuring to see that it is met,
and we can continue with the construction of a sufficient condition.

\subsection{Strong invariance of the equations of motion}
\label{sec:yang_magic}

Next we would like to promote the above results
to a strong invariance of the equations of motion.
As we argued in \secref{sec:sc_onshell},
this would amount to an honest statement of Yangian symmetry
in classical planar $\superN=4$ sYM.
To that end, we need an expression which predicts the exact form of the l.h.s.\ of
$\gen \appliedto \eomfor{\field} \approx 0$
as a linear combination of the $\eomfor{\field}$
(which are zero on shell).

Let us start from the ordinary superconformal symmetry. As we have already
shown, the invariance of the action under the generators $\gen$ of
$\alg{psu}(2,2|4)$ implies an off-shell relationship for the equations of motion of
\eqref{eq:sym_eomoffshell}
\[
\label{eq:yang_eomoffshell}
\gen\appliedto \eomfor{\field}_K =
- \eomfor{\field}_I
\frac{\delta(\gen \appliedto \field^I)}{\delta \field^K}
.\]
On the r.h.s.\ of this equality we have a sum of terms $\eomfor{\field}$
whose the coefficients are completely determined by the superconformal
representation on the fields.
Note that the sequence of fields for the term on the r.h.s.\ takes a particular form
which is not completely evident from the above expression.
It can be inferred from the colour structure
which provides the correct adjacency information
due to contraction of the field indices $I$ and $K$.
In particular, the contraction of indices $I$ implies
to take a colour trace.
Let us explain it by means of an example:
suppose $\gen \appliedto \field^I=\field^L\field^M\field^N$
yields some cubic combination of fields,
the term on the r.h.s.\ would take the form
\[
\label{eq:exampleforlocalaction}
\eomfor{\field}_I \frac{\delta (\gen \appliedto \field^I)}{\delta \field^K}
=
\delta^L_K\field^M \field^N \eomfor{\field}_I
+\delta^M_K\field^N \eomfor{\field}_I \field^L
+\delta^N_K\eomfor{\field}_I \field^L \field^M.
\]
It is straight-forward to generalise this expression to
any number of fields.

We can now ask ourselves if we can find an
analogous expression for level-one Yangian generators:
an off-shell relationship for the action of $\genyang$ on $\eomfor{\field}_K$,
whose right-hand side is a sum of $\eomfor{\field}_I$
with coefficients completely determined by the
action of the generators on the fields of the theory.
With some inspiration from the expected structures related to level-one symmetry,
we find that the answer is positive, and this novel relationship reads
\[
\label{eq:sym_Jhateom}
\genyang \appliedto \eomfor{\field}_K =
- \eomfor{\field}_I \frac{\delta(\genyang \appliedto \field^I)}{\delta \field^K}
+ \eomfor{\field}_I
\brk[s]*{ \gen^\swdl{1}
\wedge \frac{\delta}{\delta \field^K}}\appliedto\brk!{\gen^\swdl{2} \appliedto \field^I}.
\]
The anti-symmetric tensor product $\wedge$ was defined in \eqref{eq:wedge_tensor}
and it acts in analogy to \eqref{eq:tensoractiondef}.
The insertion of the new field $\eomfor{\field}_I$, however,
follows different rules based on
the colour structure associated to the indices $I$, $K$
in analogy to \eqref{eq:exampleforlocalaction}.
For example, assume that $\gen^\swdl{2} \appliedto \field^I=\field^L\field^M\field^N$
again is a cubic combination of fields.
Then the bi-local term in \eqref{eq:sym_Jhateom} expands to
\begin{align}
\eomfor{\field}_I
\smash{\brk[s]*{  \gen^\swdl{1}
\wedge \frac{\delta}{\delta \field^K}}}\appliedto (\gen^\swdl{2} \appliedto \field^I)
&=- \delta^L_K\brk[s]!{(\gen^\swdl{1}\appliedto\field^M)\field^N\eomfor{\field}_I
                     +\field^M(\gen^\swdl{1}\appliedto\field^N)\eomfor{\field}_I}
\nln&\quad
+ \delta^M_K \brk[s]!{\field^N\eomfor{\field}_I(\gen^\swdl{1}\appliedto\field^L)
                    - (\gen^\swdl{1}\appliedto\field^N)\eomfor{\field}_I\field^L}
\nln&\quad
+ \delta^N_K\brk[s]!{\eomfor{\field}_I(\gen^\swdl{1}\appliedto\field^L)\field^M
                   + \eomfor{\field}_I\field^L(\gen^\swdl{1}\appliedto\field^M)}.
\end{align}
Note that the colour trace implied by the contraction of $I$ is cut open
by the operator $\delta/\delta \field^K$, so that $\eomfor{\field}_I$
can appear in the middle of the resulting polynomial.
This expression also generalises to any number of fields,
and it can be non-zero only if $\gen^\swdl{2} \appliedto \field^I$
consists of (at least) two fields.
In other words, the bi-local term in \eqref{eq:sym_Jhateom}
only sees the non-linear contributions of $\gen^\swdl{2}$.

\medskip

First, we shall show that the above relationship holds for the Dirac equation
in $\superN=4$ sYM.
The l.h.s.\ of \eqref{eq:sym_Jhateom} has been computed in \eqref{eq:yang_phatdirac}.
We need to show that it matches with the combination
of terms on the r.h.s.\ of \eqref{eq:sym_Jhateom}.
The first term involves a variation of
$\genyang[P] \appliedto \field^I$ by $\bar\Psi$.
The only single-field action \eqref{eq:yang_sfa} containing
a field $\bar\Psi$ is the one for $Z^I=\Psi$.
Thus we can set $Z^I$ to $\Psi^f{}_\epsilon{}$
in the first term.
The second term turns out to yield no
contribution for a combination of reasons:
First of all, $\gen^\swdl{2} \appliedto \field^I$
must yield an expression non-linear in the fields.
One option is
$\gen^\swdl{2}=\gen[Q],\gen[\bar Q]$ acting
on $\field^I=\Psi,\bar\Psi$,
which however never produces any terms containing $Z^K=\bar\Psi$.
Another option is $\gen^\swdl{2}=\gen[L],\gen[\bar L],\gen[D]$,
but all non-linear terms have an explicit $x$-dependence which
will eventually cancel against other terms.
It remains to check $\gen^\swdl{2}=\gen[P]$.
It must act on $\field^I=\bar\Psi$
if the result is to contain $\field^K=\bar\Psi$.
The only other field is $\field^J=A$ (within $\cdel\bar\Psi$)
which can be acted upon by $\gen^\swdl{1}=\gen[L],\gen[\bar L],\gen[D]$.
However, all these terms are explicitly $x$-dependent and as
such they are needed to cancel against other terms.
Hence the second term does not contribute,
and we are left with
\begin{align}
\genyang[P]_{\dot\alpha\beta} \appliedto \eomfor{\bar\Psi}^{d\dot\gamma}
&=
- \tr\brk[s]*{\eomfor{\Psi}^\epsilon{}_f \frac{\delta(\genyang[P]_{\dot\alpha\beta}
\appliedto \Psi^f{}_\epsilon)}{\delta \bar\Psi_{\dot\gamma d}} }
=
\varepsilon_{\beta\epsilon} \tr\brk[s]*{\eomfor{\Psi}^\epsilon{}_f
\frac{\delta\acomm{\Phi^{fg}}{\bar\Psi_{\dot\alpha g}}}{\delta \bar\Psi_{\dot\gamma d}} }
\nonumber\\
&=
\varepsilon_{\beta\epsilon} \tr\brk[s]*{\acomm!{\eomfor{\Psi}^\epsilon{}_f}{\Phi^{fg}}
\frac{\delta\bar\Psi_{\dot\alpha g}}{\delta \bar\Psi_{\dot\gamma d}} }
=
\delta^{\dot\gamma}_{\dot\alpha}
\varepsilon_{\beta\epsilon} \acomm!{\eomfor{\Psi}^\epsilon{}_f}{\Phi^{fd}}
=
\delta^{\dot\gamma}_{\dot\alpha}
\varepsilon_{\epsilon\beta} \acomm!{\Phi^{df}}{\eomfor{\Psi}^\epsilon{}_f}.
\end{align}
This agrees precisely with \eqref{eq:yang_phatdirac}.
We have also verified that the relation \eqref{eq:sym_Jhateom}
holds exactly for all equations of motion of $\superN=4$ sYM
and for the level-one generators $\genyang$
with $\gen\in\set{\gen[P],\gen[Q],\gen[\bar Q],\gen[R],\gen[B]}$,
i.e.\ it correctly reproduces the r.h.s.\ of all terms in \eqref{eq:yang_OSeom}
and corresponding relations for the other generators.

Next, let us now explain the meaning of the two terms appearing on the r.h.s.\ in \eqref{eq:sym_Jhateom}.
The first of them is a direct counterpart of the one
from the level-zero formula \eqref{eq:yang_eomoffshell}.
The other term should be viewed a result of the non-trivial coproduct
of the level-one generators \eqref{eq:coprosweedler}.
Interestingly, the action of the two constituent operators
$\gen^\swdl{1}$ and $\gen^\swdl{2}$ is overlapping,
as $\gen^\swdl{1}$ acts purely on the output of $\gen^\swdl{2}$.
We will thus call this contribution the overlapping term.
Note that the overlapping term is a purely non-linear effect.
\unskip\footnote{This is in agreement with the fact
the overlapping contribution from two linear operators
$\gen^\swdl{1}$ and $\gen^\swdl{2}$
essentially boils down to their commutator.
The latter
is equivalent to a local term and could therefore be absorbed
into the definition of the local part of the bi-local operator.}

\medskip

To conclude this discussion,
we would like to emphasise that unlike its level-zero counterpart \eqref{eq:yang_eomoffshell},
the level-one formula \eqref{eq:sym_Jhateom}
for invariance of the equations of motion
has not been derived from first principles.
We have merely verified that it holds exactly for all equations of motion
of $\superN=4$ sYM and for several level-one generators $\genyang[J]$.
In other words, we have derived a strong form of invariance of the equations
of motion for level-one generators.
This invariance amounts to a set of non-trivial off-shell identities
which hold in classical planar $\superN=4$ sYM.
Such identities are hard to get hold of,
and independently of how we obtained them and of their precise form,
they clearly indicate a (hidden) property of classical planar $\superN=4$ sYM
which is equivalent to a global symmetry.

\section{Yangian invariance of the action}
\label{sec:yang_action}

We would now like to address Yangian invariance of the action.
Our starting point is the strong invariance of the equations of motion \eqref{eq:sym_Jhateom}
which we claimed to be a valid statement of symmetry.
In the following we will rearrange the terms in the relationship
such that they take the form of a Yangian invariance of the action
\[
\genyang\appliedto\action=0.
\]
The goal is to find a precise formulation of this statement that holds
for planar $\superN=4$ sYM.

\subsection{Notation}

In order to perform the rearrangements, we will need a concise notation
for the various terms that arise. First of all, we decompose
all objects w.r.t.\ the number of fields that they contain.
The action of $\superN=4$ sYM has quadratic, cubic and quartic terms,
and the superconformal representation has terms which preserve the
number of fields as well as terms that increase the number of fields by one unit
\unskip\footnote{We define the expansion coefficients $\oper{O}_\nfield{n}$
of traced polynomials $\oper{O}$ of the fields
with an explicit symmetry factor of $1/n$.
Conversely, there is no symmetry factor
for the expansion coefficients $\oper{X}_\nfield{n}$ and $\gen_\nfield{n}$
of open polynomials $\oper{X}$ and operators $\gen$, respectively.
Even though these different assignments may be confusing at times,
they will avoid many combinatorial factors.}
\[\label{eq:expansion}
\action=\rfrac{1}{2}\action_\nfield{2}+\rfrac{1}{3}\action_\nfield{3}+\rfrac{1}{4}\action_\nfield{4},
\qquad
\gen=\gen_\nfield{0}+\gen_\nfield{1}.
\]
We choose a particular level-one generator
which minimises the non-linear terms,
such as the level-one momentum $\genyang=\genyang[P]$,
see also \appref{app:superconformal} for further calculational simplifications.
The relevant local and bi-local terms in
$\genyang=\genyang_\local+\genyang_\bilocal$
then expand as follows
\unskip\footnote{Even though there is in principle a bi-local
term which adds a field at both insertion points,
this term will not contribute in practice.}
\[
\genyang_\local
=\genyang_{\local,\nfield{1}}
,\qquad
\genyang_\bilocal
=
\gen_\nfield{0}^\swdl{1}\otimes\gen_\nfield{0}^\swdl{2}
+\gen_\nfield{1}^\swdl{1}\otimes\gen_\nfield{0}^\swdl{2}
+\gen_\nfield{0}^\swdl{1}\otimes\gen_\nfield{1}^\swdl{2}
.\]

Next we need to write out somewhat more explicitly
how the representations act on the individual fields in the action.
The action is a polynomial in the fields whose ordering matters.
The contribution $\gen_\nfield{0}$ to the representation
maps one field to one,
and it is natural to denote the action on the field at site $k$ by
$\gen_{\nfield{0},k}$.
\unskip\footnote{To avoid excessive clutter
we may suppress the symbol `$\appliedto$'
when an operator acts on a specific field.}
For the higher contributions $\gen_\nfield{m}$, $m>0$,
which map one field to $m+1$ fields, the situation is not as evident,
see \cite{Beisert:2003ys} for discussions of this matter in a similar context.
Here we make the choice that $\gen_{\nfield{m},k}$
acts on the field at site $k$ and replaces it
by an appropriate sequence of $m+1$ fields.
Consequently, the fields at sites $1,\ldots,k-1$ are mapped to themselves,
whereas the fields at sites $k+1,k+2,\ldots$ are shifted by $m$ steps to
sites $k+m+1,k+m+2,\ldots$, e.g.
\[
\label{eq:nonlinactex}
\gen_{\nfield{1},3}
(\mathord{\stackrel{1}{\field^I}}
\mathord{\stackrel{2}{\field^J}}
\mathord{\stackrel{3}{\field^K}}
\mathord{\stackrel{4}{\field^L}}
\mathord{\stackrel{5}{\field^M}})
:=
\mathord{\stackrel{1}{\field^I}}
\mathord{\stackrel{2}{\field^J}}
\mathord{\stackrel{3,4}{(\gen_\nfield{1}\field^K)}}
\mathord{\stackrel{5}{\field^L}}
\mathord{\stackrel{6}{\field^M}}.
\]
The change of length has a relevant implication on the commutation of two operators
$\gen$ and $\gen'$.
It is evident that two one-to-one operators acting on two different sites commute.
For operators which change the length, their insertion points have
to be adjusted as follows
\[
\label{eq:non-linear-commutation}
\gen_{\nfield{l},j}
\gen'_{\nfield{m},k}
=
\gen'_{\nfield{m},k+l}
\gen_{\nfield{l},j}
\qquad\text{if }j<k.
\]

\paragraph{Cyclicity.}

All the terms in the action are gauge-invariant by means of a trace.
The trace establishes a neighbouring relationship between the last and
the first sites, i.e.\ periodic boundary conditions.
Moreover, the trace is cyclic, i.e.\ it is invariant under cyclic shifts.
In order to deal with periodic boundary conditions,
we introduce $\shift$ as the operator that performs a cyclic shift by one site
to the left; e.g., it shifts the second site to the first and the first site to the last.
Cyclicity of a traced operator $\oper{O}$ is then expressed as
\[
\shift \oper{O}=\oper{O}
\qquad\text{or expanded}\qquad
\shift \oper{O}_\nfield{n}=\oper{O}_\nfield{n}.
\]
Note that we will also work with polynomials $\oper{O}$ in the fields
which are periodic but not cyclic.
The cyclic shift operator shifts the insertion of an operator $\gen$
according to the rule
\[
\label{eq:shift-non-linear}
\gen_{\nfield{m},k} \shift \oper{O}_\nfield{n}
=\begin{cases}
\shift\gen_{\nfield{m},k+1} \oper{O}_\nfield{n}&\text{for }k\neq n,
\\
\shift^{m+1}\gen_{\nfield{m},1} \oper{O}_\nfield{n}&\text{for }k=n.
\end{cases}
\]
It it worth emphasising that a
term of the form $\shift^{k}\gen_{\nfield{m},1}$ with $k\leq m$
never appears on the r.h.s.\ of \eqref{eq:shift-non-linear}.
Such a term cannot be written with the shift $\shift$
residing to the right of $\gen$,
the shift must remain to the left of $\gen$.
This is because the term has a form
which is not covered by $\gen_{\nfield{m},k}$ alone
and which is different from \eqref{eq:nonlinactex}.
In this term, the output of $\gen_{\nfield{m}}$ extends past the last
site of the polynomial and continues periodically
to the first site or beyond.
An example analogous to \eqref{eq:nonlinactex}
could be written as
\[
\shift\gen_{\nfield{1},1}
(\mathord{\stackrel{1}{\field^I}}
\mathord{\stackrel{2}{\field^J}}
\mathord{\stackrel{3}{\field^K}}
\mathord{\stackrel{4}{\field^L}}
\mathord{\stackrel{5}{\field^M}})
:=
\mathord{\stackrel{6,1}{(\gen_\nfield{1}\field^I)}}
\mathord{\stackrel{2}{\field^J}}
\mathord{\stackrel{3}{\field^K}}
\mathord{\stackrel{4}{\field^L}}
\mathord{\stackrel{5}{\field^M}}
=
\mathord{\stackrel{1}{(\gen_\nfield{1}\field^I)_2}}
\mathord{\stackrel{2}{\field^J}}
\mathord{\stackrel{3}{\field^K}}
\mathord{\stackrel{4}{\field^L}}
\mathord{\stackrel{5}{\field^M}}
\mathord{\stackrel{6}{(\gen_\nfield{1}\field^I)_1}},
\]
where $(\gen_\nfield{m}\field)_j$ denotes
the $j$-th site of the output of $\gen_\nfield{m}$ on $\field$
(with an implicit sum over all $m$-tuples of fields
in the polynomial $\gen_\nfield{m}\field$
in similarity to Sweedler's notation).

\subsection{Level zero}

Equipped with this notation, we will first address superconformal symmetry of the action.
This will provide us with some identities that we shall need later
when we consider Yangian symmetry.
\unskip\footnote{We would like to remind the reader of a general issue w.r.t.\
homogeneous non-linear representations acting on periodic objects:
On the one hand, the periodic shift operator $\shift$ classifies periodic objects
according to the eigenvalue $e^{2\pi i n/L}$ where $n=0,\ldots,L-1$.
On the other hand, the non-linear representation changes the length $L$
of the object.
Now the spectra of $\shift$ for different lengths $L$ are largely distinct,
the only eigenvalue which is present for all lengths is $1$
corresponding to cyclic objects.
Therefore non-linear representation can be homogeneous,
i.e.\ commute with the shift operator $\shift$,
only on the subspace of cyclic objects.
In other words, the periodic object on which they act must be cyclic
and the representation must be constructed
such that the result is cyclic as well.}
First, we decompose the statement $\gen\appliedto\action=0$
by the number of fields, and we obtain the following set of 4 statements
\[
\rfrac{1}{2}\gen_\nfield{0}\appliedto\action_\nfield{2}
=\rfrac{1}{3}\gen_\nfield{0}\appliedto\action_\nfield{3}+\rfrac{1}{2}\gen_\nfield{1}\appliedto\action_\nfield{2}
=\rfrac{1}{4}\gen_\nfield{0}\appliedto\action_\nfield{4}+\rfrac{1}{3}\gen_\nfield{1}\appliedto\action_\nfield{3}
=\rfrac{1}{4}\gen_\nfield{1}\appliedto\action_\nfield{4}
=0.
\]

When making the fields explicit, the first statement reads
\[
\label{eq:sym_lev0_2}
\rfrac{1}{2}\gen_{\nfield{0},1}\action_\nfield{2}
+\rfrac{1}{2}\gen_{\nfield{0},2}\action_\nfield{2}=0.
\]
Using cyclicity of the trace, we may as well write this even more concisely as
$\gen_{\nfield{0},1}\action_\nfield{2}\simeq 0$ where the
symbol `$\simeq$' denotes equality up to cyclic permutations.
The cubic relationship following from superconformal symmetry reads
\[
\rfrac{1}{3}\gen_{\nfield{0},1}\action_\nfield{3}
+\rfrac{1}{3}\gen_{\nfield{0},2}\action_\nfield{3}
+\rfrac{1}{3}\gen_{\nfield{0},3}\action_\nfield{3}
+\rfrac{1}{2}\gen_{\nfield{1},1}\action_\nfield{2}
+\rfrac{1}{2}\gen_{\nfield{1},2}\action_\nfield{2}
\simeq 0.
\]
This relationship can be rewritten in two alternative ways using cyclicity:
Collecting terms we arrive at the simpler form
$\gen_{\nfield{0},1}\action_\nfield{3}
+\gen_{\nfield{1},1}\action_\nfield{2}\simeq 0$.
However, we can also write the relationship in a manifestly cyclic fashion as
\[
\label{eq:sym_lev0_3}
\rfrac{1}{3}\gen_{\nfield{0},1}\action_\nfield{3}
+\rfrac{1}{3}\gen_{\nfield{0},2}\action_\nfield{3}
+\rfrac{1}{3}\gen_{\nfield{0},3}\action_\nfield{3}
+\rfrac{1}{3}\gen_{\nfield{1},2}\action_\nfield{2}
+\rfrac{1}{3}\shift\gen_{\nfield{1},1}\action_\nfield{2}
+\rfrac{1}{3}\gen_{\nfield{1},1}\action_\nfield{2}=0.
\]
Here it is necessary to use the cyclic shift operator $\shift$
for one term to distribute the sequence of fields resulting from
$\gen_\nfield{1}$ over the last and first site.

\medskip

It is also useful to consider the strong invariance of the equations of motion
\eqref{eq:yang_eomoffshell}.
For any $\field^K$ specifying the equation of motion,
it yields a relationship consisting of a polynomial of fields.
In order to handle the identities for all fields $\field^K$ at the same time,
we prepend this field to the relationship polynomial and sum over all fields
\[
\label{eq:defY}
\oper{Y}:=  \field^K\brk[s]*{(\gen\appliedto\field^I) \frac{\delta^2\action}{\delta \field^I\.\delta \field^K}
+ \frac{\delta(\gen\appliedto\field^I)}{\delta \field^K} \frac{\delta\action}{\delta \field^I}}=0.
\]
In this definition we explicitly do not take the colour trace
such that the above field $\field^K$ will always reside at site $1$ of the polynomial by construction.
The expansion $\sum_n\oper{Y}_\nfield{n}$ of the open polynomial $\oper{Y}$
in the number $n$ of fields takes the form
\[
\oper{Y}=\sum_n\oper{Y}_\nfield{n}
,\qquad
\oper{Y}_\nfield{n}=
\sum_{m=0}^{n-2}
\sum_{j=2}^{n-m}
\gen_{\nfield{m},j}
\action_\nfield{n-m}
+
\sum_{m=0}^{n-2}
\sum_{j=1}^{m+1}
\shift^{j-1}
\gen_{\nfield{m},1}
\action_\nfield{n-m}.
\]
Here, the symmetry factors $1/(n-m)$ for $\action_\nfield{n-m}$ in \eqref{eq:expansion}
have been cancelled
by varying the component of the (cyclic) action $\action_\nfield{n-m}$ consisting of $n-m$ fields by $\field^K$.
\unskip\footnote{We assume that $\delta\action/\delta\field^K$
pulls one of the $n-m$ fields and moves the empty spot within the polynomial to site 1.
The subsequent variation $\delta/\delta\field^I$ pulls another field
(but not from the empty site 1) and replaces it by $\gen\appliedto\field^I$.}
Note that the statements $\half \oper{Y}_\nfield{2}=0$ and $\rfrac{1}{3}\oper{Y}_\nfield{3}=0$ are
precisely the above \eqref{eq:sym_lev0_2} and \eqref{eq:sym_lev0_3}, respectively.
In fact, the expression $\oper{Y}$ is cyclic; by rewriting the first term
using \eqref{eq:shift-non-linear}
such that $\gen_\nfield{m}$ will always act on site $1$,
we can make cyclicity manifest
\[
\label{eq:sym_lev0_n}
\oper{Y}_\nfield{n}=
\sum_{m=0}^{n-2}
\sum_{j=1}^{n}
\shift^{j-1}
\gen_{\nfield{m},1}
\action_\nfield{n-m}.
\]
Therefore, we lose no information by identifying terms by cyclic permutations,
and we find a more concise statement
\[
\oper{Y}_\nfield{n}\simeq
n\sum_{m=0}^{n-2}
\gen_{\nfield{m},1}
\action_\nfield{n-m}\simeq 0.
\]

To wrap this discussion up, we can rewrite the above definition \eqref{eq:defY}
as
\[
\oper{Y}=Z^K \frac{\delta (\gen\appliedto\action)}{\delta Z^K}.
\]
As the operation $Z^K (\delta/\delta Z^K)$ just counts the number of fields,
it is not surprising that the symmetry variation of the action
expands precisely to the coefficients $\oper{Y}_\nfield{n}$
\[
\gen\appliedto\action=\sum_n\frac{1}{n} \oper{Y}_\nfield{n}.
\]
Therefore, the above transformations are a bit of a detour in this case,
but they will help us find a corresponding expression $\genyang\appliedto\action$
for the level-one Yangian generators $\genyang$ acting on the action $\action$.

\subsection{Level one}

We would now like to construct a suitable
level-one Yangian representation on the action, $\genyang\appliedto\action$,
such that invariance amounts to $\genyang\appliedto\action=0$.
We will approach this construction by expanding in the number of fields,
\[
\genyang\appliedto\action=\sum_n\frac{1}{n} (\genyang\appliedto\action)_\nfield{n},
\]
where the prefactors $1/n$ account for cyclic symmetry.
We will start with the simplest cases at $n=2,3$
and later address arbitrary lengths.
Before doing so,
we derive some more identities which are needed
to transform the expressions.

\paragraph{Commuting constituents.}

Consider the commutator
\[
\label{eq:Hcombination}
\gen[H]:=\half\comm!{\gen^\swdl{1}}{\gen^\swdl{2}},
\]
which we can expand in fields as usual as $\gen[H]=\gen[H]_\nfield{0}+\gen[H]_\nfield{1}$ with
\[
\label{eq:sym_dual_cox}
\gen[H]_{\nfield{m},k}=
\sum_{l=0}^m\sum_{j=0}^l
\gen^\swdl{1}_{\nfield{m-l},k+j}\gen^\swdl{2}_{\nfield{l},k}.
\]
The leading term $\gen[H]_\nfield{0}$
contains the combination $f^A{}_{BC}f^{BC}{}_{D}$
which is proportional to the dual Coxeter number of
the level-zero algebra.
The dual Coxeter number for our superconformal algebra $\alg{psu}(2,2|4)$
is zero, and consequently the linear term vanishes, $\gen[H]_\nfield{0}=0$.
The non-linear term $\gen[H]_\nfield{1}$ is also zero
for $\superN=4$ sYM by explicit computation.
One can relate this finding to the G-identity for $\superN=4$ sYM
found in \cite{Beisert:2015uda}.
Altogether, the constituent operators of the bi-local
level-one generator commute
(when summed over all pairs as implied by Sweedler's notation)
\[
\label{eq:constcomm}
\gen[H]=0.
\]
%

\paragraph{Two fields.}

Next, we address level-one Yangian symmetry.
The starting point is the strong invariance statement \eqref{eq:sym_Jhateom}
of the equations of motion which we have already shown to hold.
We will treat it analogously to the strong invariance of the equation
of motion at level-zero \eqref{eq:sym_Jhateom}:
We prepend the field that selects the equation of motion
to the relationship polynomial and denote the resulting expression by
\[
\label{eq:yhatdef}
\yanghat{\oper{Y}}:=
\field^K
\brk[s]*{
\genyang \appliedto \eomfor{\field}_K
+ \eomfor{\field}_I \frac{\delta(\genyang \appliedto \field^I)}{\delta \field^K}
- \eomfor{\field}_I
\brk*{ \gen^\swdl{1}
\wedge \frac{\delta}{\delta \field^K}}\appliedto\brk!{\gen^\swdl{2} \appliedto \field^I}
}
=0.
\]
It can be expanded in fields leading
to some additional identities of a similar kind as
\eqref{eq:sym_lev0_2,eq:sym_lev0_3}
\[
\yanghat{\oper{Y}}=\sum_{n}\yanghat{\oper{Y}}_\nfield{n}.
\]
However, the relationship $\yanghat{\oper{Y}}_\nfield{2}=0$ at the level of two fields is empty,
the first non-trivial relationship is at three fields.

What are the implications of Yangian symmetry at two fields?
We use the fact that the combination $\oper{Y}_\nfield{2}$ in \eqref{eq:sym_lev0_n}
is exactly zero,
\unskip\footnote{Importantly, the relationship $\oper{Y}_\nfield{2}=0$
holds without assuming the equation of motion to hold
or without applying cyclic permutations.}
and apply another generator $\gen$ to it
to obtain a new relationship
\[
\label{eq:J0Y2}
\gen^\swdl{1}_{\nfield{0},1}\oper{Y}^\swdl{2}_\nfield{2}
=
\gen^\swdl{1}_{\nfield{0},1}
\gen^\swdl{2}_{\nfield{0},1}\action_\nfield{2}
+\gen^\swdl{1}_{\nfield{0},1}
\gen^\swdl{2}_{\nfield{0},2}\action_\nfield{2}=0.
\]
The first term has the form
$\gen[H]_{\nfield{0},1}\action_\nfield{2}$ and therefore it vanishes by \eqref{eq:constcomm}.
The second term is the bi-local part of the level-one representation
on the quadratic part of the action.
Furthermore, there is no linear contribution to the local part
of the level-one representation.
Altogether, \eqref{eq:J0Y2} boils down to the statement
\[
\genyang_\nfield{0}\appliedto\action_\nfield{2}
=\brk!{\gen_\nfield{0}^\swdl{1}\otimes\gen_\nfield{0}^\swdl{2}}\appliedto\action_\nfield{2}
=0.
\]
In other words, the level-one symmetry of the action at quadratic order in the fields
is a plain consequence of level-zero symmetry and the vanishing of the dual Coxeter number.
This agrees with the fact that the equations of motions
are trivially invariant at linear order.

\paragraph{Three fields.}

The strong invariance relationship \eqref{eq:sym_Jhateom} at three fields reads
$\yanghat{\oper{Y}}_\nfield{3}=0$ with
\begin{align}
\label{eq:yhat3}
\yanghat{\oper{Y}}_\nfield{3}&:=
\genyang_{\nfield{1},2}\action_\nfield{2}
+\shift\genyang_{\nfield{1},1}\action_\nfield{2}
+\genyang_{\nfield{1},1}\action_\nfield{2}
\nln &\qquad
+\gen^\swdl{1}_{\nfield{0},2}\gen^\swdl{2}_{\nfield{0},3}\action_\nfield{3}
-\gen^\swdl{2}_{\nfield{0},2}\gen^\swdl{1}_{\nfield{1},1}\action_\nfield{2}
+\shift\gen^\swdl{2}_{\nfield{0},1}\gen^\swdl{1}_{\nfield{1},1}\action_\nfield{2}
.\end{align}
We would now like to reformulate this combination such that it looks
more like the cubic term in the expansion of $\genyang\appliedto\action$.
To that end, let us first address cyclicity of $\yanghat{\oper{Y}}_\nfield{3}$.
The first three terms are manifestly cyclic; the latter three are not.
We therefore look at the violation of cyclicity
\begin{align}
(\shift-1)\yanghat{\oper{Y}}_\nfield{3}&=
-\gen^\swdl{1}_{\nfield{0},2}\oper{Y}^\swdl{2}_\nfield{3}
-\shift\gen^\swdl{1}_{\nfield{1},1}\oper{Y}^\swdl{2}_\nfield{2}
+\gen[H]_{\nfield{0},2}\action_\nfield{3}
+\shift \gen[H]_{\nfield{1},1}\action_\nfield{2}
.\end{align}
Here we have made use of the algebraic identities
\eqref{eq:non-linear-commutation} and \eqref{eq:shift-non-linear}
and the implicit anti-symmetry in Sweedler's notation \eqref{eq:sweedler}.
All remaining terms could then be collected
in the combinations $\oper{Y}$ \eqref{eq:sym_lev0_n} and $\gen[H]$ \eqref{eq:sym_dual_cox}
which are zero as discussed above.
Therefore the relationship $\yanghat{\oper{Y}}_\nfield{3}=0$
is effectively cyclic;
all the non-cyclic contributions to the relationship are zero
for lesser reasons than Yangian symmetry.
It also allows us to compare modulo cyclic identifications (`$\simeq$')
without losing relevant information
\begin{align}
\label{eq:yhat3cyclic}
\yanghat{\oper{Y}}_\nfield{3}&\simeq
3\genyang_{\nfield{1},1}\action_\nfield{2}
+\gen^\swdl{1}_{\nfield{0},2}\gen^\swdl{2}_{\nfield{0},3}\action_\nfield{3}
-\gen^\swdl{2}_{\nfield{0},2}\gen^\swdl{1}_{\nfield{1},1}\action_\nfield{2}
+\gen^\swdl{2}_{\nfield{0},1}\gen^\swdl{1}_{\nfield{1},1}\action_\nfield{2}
.\end{align}

Let us now turn to the reformulation.
Due to the effective cyclicity of $\yanghat{\oper{Y}}_\nfield{3}$
it is reasonable to define $(\genyang\appliedto\action)_\nfield{3}$
to be proportional to it.
To figure out the factor of proportionality,
we will compare the terms in \eqref{eq:yhat3cyclic}
to some canonical terms in $(\genyang\appliedto\action)_\nfield{3}$.
The first term in \eqref{eq:yhat3cyclic} is equivalent to
the local part of the level-one representation on $\action_\nfield{2}$
\[
\genyang_{\local,\nfield{1}}\appliedto\action_\nfield{2}
:\simeq
2\genyang_{\nfield{1},1}\action_\nfield{2},
\]
while the second term in \eqref{eq:yhat3cyclic}
is equivalent to the bi-local part on $\action_\nfield{3}$
\[
\genyang_{\bilocal,\nfield{0}}\appliedto\action_\nfield{3}
:=
 \gen^\swdl{1}_{\nfield{0},1}\gen^\swdl{2}_{\nfield{0},2}\action_\nfield{3}
+\gen^\swdl{1}_{\nfield{0},1}\gen^\swdl{2}_{\nfield{0},3}\action_\nfield{3}
+\gen^\swdl{1}_{\nfield{0},2}\gen^\swdl{2}_{\nfield{0},3}\action_\nfield{3}
\simeq
 \gen^\swdl{1}_{\nfield{0},1}\gen^\swdl{2}_{\nfield{0},2}\action_\nfield{3}.
\]
The remaining two terms in \eqref{eq:yhat3cyclic} have a special form, one may interpret
them as a non-linear bi-local contribution where both legs overlap.
Such terms are not provided by the usual coproduct rule for tensor product representations,
but our non-linear representation is not exactly a tensor product,
and hence it is conceivable to have them
\[
\label{eq:overlap3}
\genyang_{\overlap,\nfield{1}}\appliedto\action_\nfield{2}
:\simeq
\rfrac{2}{3}\gen^\swdl{1}_{\nfield{0},2}\gen^\swdl{2}_{\nfield{1},1}\action_\nfield{2}
-\rfrac{2}{3}\gen^\swdl{1}_{\nfield{0},1}\gen^\swdl{2}_{\nfield{1},1}\action_\nfield{2}.
\]
The (insignificant) prefactors in this definition were chosen to agree with a desirable
form of expression further below.
Finally, we notice that $\yanghat{\oper{Y}}_\nfield{3}$ does not contain
the standard bi-local contribution to the representation on $\action_\nfield{2}$.
However, this term is zero modulo cyclic identifications
\[
\gen^\swdl{1}_{\nfield{1},1}\gen^\swdl{2}_{\nfield{0},2}\action_\nfield{2}
+\gen^\swdl{1}_{\nfield{0},1}\gen^\swdl{2}_{\nfield{1},2}\action_\nfield{2}
\simeq 0.
\]

Altogether we derive an invariance statement for the action at cubic order in the fields
\[
\label{eq:sym_lev1_3}
\rfrac{1}{3}(\genyang\appliedto\action)_\nfield{3}:\simeq
\rfrac{1}{2}\genyang_{\local,\nfield{1}}\appliedto\action_\nfield{2}
+\rfrac{1}{3}\genyang_{\bilocal,\nfield{0}}\appliedto\action_\nfield{3}
+\rfrac{1}{2}\genyang_{\overlap,\nfield{1}}\appliedto\action_\nfield{2}
\simeq\rfrac{1}{3}\yanghat{\oper{Y}}_\nfield{3}
\simeq 0.
\]
Note that the prefactors of all terms agree with the symmetry factors $1/m$
of the $\action_\nfield{m}$ which is acted upon.
\unskip\footnote{For the novel overlapping terms
this is a choice which was used to fix the perviously chosen
prefactors in its definition \eqref{eq:overlap3}.}
We have verified explicitly that \eqref{eq:sym_lev1_3} holds
for the level-one momentum $\genyang=\genyang[P]$.

\paragraph{Non-linear level-one invariance.}

Now let us turn to a general number of fields $n$
in order to understand the precise structure of the unconventional terms.
The combination $\yanghat{\oper{Y}}_\nfield{n}$
of the strong invariance condition \eqref{eq:sym_Jhateom}
can be written in our present notation as
\begin{align}
\yanghat{\oper{Y}}_\nfield{n}&=
\sum_{m=0}^{n-2}
\sum_{j=2}^{n-m}
\genyang_{\nfield{m},j}
\action_\nfield{n-m}
+
\sum_{m=0}^{n-2}
\sum_{j=1}^{m+1}
\shift^{j-1}
\genyang_{\nfield{m},1}
\action_\nfield{n-m}
\nln&\quad
+
\sum_{m=0}^{n-2}
\sum_{l=0}^{m}
\sum_{j=2}^{n-m-1}\sum_{k=j+1}^{n-m}
\gen^\swdl{1}_{\nfield{l},j}
\gen^\swdl{2}_{\nfield{m-l},k}
\action_\nfield{n-m}
\nln&\quad
+
\sum_{m=0}^{n-2}
\sum_{l=0}^{m}
\sum_{j=1}^{l}\sum_{k=j+1}^{l+1}
\brk!{
\shift^{j-1}\gen^\swdl{1}_{\nfield{m-l},k}
-\shift^{k-1+m-l}\gen^\swdl{1}_{\nfield{m-l},j}
}
\gen^\swdl{2}_{\nfield{l},1}
\action_\nfield{n-m}.
\end{align}
Using the same identities as at the level of 3 fields,
we find that violations of cyclicity are given by
terms containing $\oper{Y}$ and $\gen[H]$
\begin{align}
(\shift-1)\yanghat{\oper{Y}}_\nfield{n}&=
\sum_{m=0}^{n-2}
\brk!{1+\shift^{m+1}}\brk[s]*{
\gen^\swdl{1}_{\nfield{m},1}
\oper{Y}^\swdl{2}_\nfield{n-m}
-
\gen[H]_{\nfield{m},1}
\action_\nfield{n-m}
}.
\end{align}
This means that $\yanghat{\oper{Y}}$ is in fact cyclic
provided that the action is invariant under level-zero
and that the combined generator $\gen[H]$ defined in \eqref{eq:Hcombination} is zero.
We may thus interpret the statement $\yanghat{\oper{Y}}=0$
as level-one invariance of the action.

To bring this statement somewhat closer to the
expected form of $\genyang\appliedto\action$
we introduce a modified combination $\yanghat{\oper{Y}}'$
with some convenient extra terms which are zero due to $\oper{Y}=\gen[H]=0$
\begin{align}
\label{eq:Yprime}
\yanghat{\oper{Y}}'_\nfield{n}&:=
\yanghat{\oper{Y}}_\nfield{n}
+\sum_{m=0}^{n-2}
\brk[s]*{
\gen^\swdl{1}_{\nfield{m},1}\oper{Y}^\swdl{2}_\nfield{n-m}
+\gen[H]_{\nfield{m},1}\action_\nfield{n-m}
}
\nln
&=
\sum_{m=0}^{n-2}
\sum_{j=1}^{n}
\shift^{j-1}\genyang_{\nfield{m},1}
\action_\nfield{n-m}
\nln&\quad
+\sum_{m=0}^{n-2}
\sum_{l=0}^{m}
\sum_{k=l+2}^{n-m+l} \sum_{j=l+2}^{k}
\shift^{j-1}
\gen^\swdl{1}_{\nfield{m-l},k}
\gen^\swdl{2}_{\nfield{l},1}
\action_\nfield{n-m}
\nln&\quad
+\sum_{m=0}^{n-2}\sum_{l=0}^{m}\sum_{k=1}^{l+1}
\brk[s]*{\sum_{j=2}^{m-l+k}-\sum_{j=k+1}^{m+1}}
\shift^{j-1}
\gen^\swdl{1}_{\nfield{m-l},k}
\gen^\swdl{2}_{\nfield{l},1}
\action_\nfield{n-m}.
\end{align}
The terms in the first line of the result represent the
local terms of the level-one representation.
The terms in the second line are clearly bi-local
with two non-overlapping insertions.
The terms in the third line take a similar form
as the bi-local terms, but here the insertion
of $\gen^\swdl{1}$ is within the range of $\gen^\swdl{2}$
and thus their actions overlap.
\unskip\footnote{Consequently, the combination of both insertions is effectively local,
nevertheless we will still associate them to the bi-local part due to
their structural similarity.}

Some comments regarding the bi-local and overlapping terms are in order:
Even through the prefactors for all terms in \eqref{eq:Yprime} are $1$,
those of the non-linear bi-local terms appear somewhat unnatural:
when compared to their linear counterparts,
they are off by a factor of $(n-m)/n$.
On the one hand, this may be worrisome in view of gauge symmetry
because such relative factors could easily upset the composition
of covariant derivatives and thus spoil this essential symmetry.
On the other hand, such factors can be compensated
by the relative length of the objects
and thus by the multiplicity of operator insertions.
Moreover, the overlapping terms
have no analog in the conventional level-one representation
based on the coproduct rule.

How to make sense of this behaviour?
Clearly, the novel non-linear representation of level-one symmetry
on cyclic polynomials can have new features,
which do not need to follow the coproduct rule strictly.
What matters is that the above form follows by elementary transformations
from the strong invariance relationship \eqref{eq:sym_Jhateom}.
The latter holds in planar $\superN=4$ sYM for
the level-one momentum $\genyang[J]=\genyang[P]$.
It therefore makes sense to view $\yanghat{\oper{Y}}'_\nfield{n}=0$
as given by \eqref{eq:Yprime}
as the definition of the non-linear level-one invariance of the action.
Whether or not it has the expected from,
it certainly does describe a non-trivial relationship of planar $\superN=4$ sYM.
In any case, the structures in \eqref{eq:Yprime}
clearly deserve further theoretical scrutiny.

\paragraph{Cyclic level-one representation.}

As the expression \eqref{eq:Yprime} is effectively cyclic,
we may furthermore identify terms
related by cyclic permutation. This yields a simpler expression
\begin{align}
\label{eq:Yprimereduced}
\yanghat{\oper{Y}}'_\nfield{n}&\simeq
\sum_{m=0}^{n-2}
n
\genyang_{\nfield{m},1}
\action_\nfield{n-m}
\nln&\quad
+\sum_{m=0}^{n-2}
\sum_{l=0}^{m}
\sum_{k=1}^{n-m+2l+1} (k-l-\half n+\half m-1)
\gen^\swdl{1}_{\nfield{m-l},k}
\gen^\swdl{2}_{\nfield{l},1}
\action_\nfield{n-m}.
\end{align}
Note that all bi-local terms could be combined in a uniform expression.
Here the first and the last $l+1$ terms labelled
by $k=1,\ldots,l+1$ and $k=n-m+l+1,\ldots, n-m+2l+1$
represent overlapping bi-local terms whereas the insertions
do not overlap in the remaining middle range $k=l+2,\ldots,n-m+l$.

Since invariance of the action can and should be expressed modulo
cyclicity of the trace, we may write the symmetry statement as
\[
\label{eq:yangianoncyclic}
\genyang\appliedto\action\simeq 0
\qquad\text{with}\qquad
\genyang\appliedto\action :\simeq \sum_n \frac{1}{n}\yanghat{\oper{Y}}'_\nfield{n} .
\]
This form is a complete expression for level-one invariance of
the action including all standard and non-standard terms.
We have verified explicitly by computer algebra
that it holds for planar $\superN=4$ sYM
for the level-one generators $\genyang$
with $\gen\in\set{\gen[P],\gen[Q], \gen[\bar Q],\gen[R],\gen[B]}$.
Unfortunately, the calculations produce hundreds of intermediate terms
which are subject to cyclic identifications and integrations by parts.
Only the invariance under the level-one bonus symmetry $\genyang[B]$
has a reasonably simple structure, and we shall show it explicitly below.

As such we have not yet fully established that the extended symmetries
generate a Yangian algebra;
we would need to show that the adjoint property \eqref{eq:yang_lv1rel1}
as well as the Serre-relation \eqref{eq:yang_lv1rel2} hold.
A complication, to be addressed in \cite{Beisert:2018ijg}, is that the algebra is
mixed with gauge transformations of a novel kind.
Independently of which algebraic relations
the symmetry generators obey,
they give rise to novel relations for planar $\superN=4$ sYM.
\unskip\footnote{By construction, these symmetries form some algebra.
If it is not of Yangian kind,
it will inevitably be much larger than that.
So the default assumption is that the algebra
is as small as possible, and thus of Yangian kind.
Nevertheless it intrinsically interesting to
understand the actual symmetry algebra and its relations in detail.}

\paragraph{Level-one bonus symmetry.}

In order to show invariance of the action
under the level-one bonus symmetry $\genyang[B]$
introduced in \cite{Beisert:2011pn}, see also \cite{deLeeuw:2012jf},
we first introduce the additional algebraic structures,
see \appref{app:superconformal} \secref[and]{app:levelone} for further details.
The coproduct depends only on the odd level-zero generators
\[
\label{eq:Bhatcopro}
\copro \genyang[B]
= \genyang[B] \otimes 1
+ 1 \otimes \genyang[B]
- \quarter\gen[S]^{\alpha b} \wedge \gen[Q]_{b \alpha}
- \quarter\gen[\bar{S}]_b {}^{\dot\alpha} \wedge \gen[\bar{Q}]_{\dot\alpha}{}^b ,
\]
and the single-field action is trivial
\[
\genyang[B] \appliedto \field =0.
\]
Moreover, the representation of the superconformal boosts $\gen[S]$ and $\gen[\bar S]$
are almost completely given in terms of the supersymmetries $\gen[\bar Q]$ and $\gen[Q]$,
respectively
\[
\label{eq:SintermsofQ}
\gen[S]^{\alpha b} \sim
ix^{\alpha\dot\epsilon}\gen[\bar Q]_{\dot\epsilon}{}^b
+\gen[S]'^{\alpha b},
\qquad
\gen[\bar S]_{a}{}^{\dot\gamma} \sim
-ix^{\beta\dot\gamma}\gen[Q]_{a\beta}
+\gen[\bar S]'_{a}{}^{\dot\gamma}.
\]
Both of the additional operators $\gen[S]'$ and $\gen[\bar S]'$
act non-trivially only on a single type of field
\[
\gen[S]'^{\alpha b}\appliedto\Psi^c{}_\delta
= -2\delta^\alpha_\delta\Phi^{bc},
\qquad
\gen[\bar S]'_{a}{}^{\dot\gamma}\appliedto \bar\Psi_{\dot\epsilon d}
= 2\delta^{\dot\gamma}_{\dot\epsilon}\bar\Phi_{ad}.
\]

Based on these relationships, we can in fact show that almost all contributions
to the invariance condition are trivially zero.
First of all, by combining \eqref{eq:Bhatcopro} and \eqref{eq:SintermsofQ}
one can observe that the $x$-dependent terms
due to $\gen[\bar S]\wedge\gen[\bar Q]$ and $\gen[S]\wedge\gen[Q]$
mutually cancel irrespectively of what they act on.
Therefore, the only non-zero contributions to the invariance condition
can originate from the $x$-independent operators
$\gen[S]'$ and $\gen[\bar S]'$ as well as when
$\gen[S]$ and $\gen[\bar S]$ act on a partial derivative
which subsequently removes the $x$-dependence
\begin{align}
\gen[S]^{\alpha b} \appliedto (\partial_{\dot\gamma\delta}\field)&=
i\partial_{\dot\gamma\delta}(x^{\alpha\dot\epsilon}\gen[\bar Q]_{\dot\epsilon}{}^b \appliedto\field)
+\partial_{\dot\gamma\delta}(\gen[S]'^{\alpha b} \appliedto \field)
\nln&=
ix^{\alpha\dot\epsilon}\partial_{\dot\gamma\delta}(\gen[\bar Q]_{\dot\epsilon}{}^b \appliedto\field)
+\partial_{\dot\gamma\delta}(\gen[S]'^{\alpha b} \appliedto \field)
+i\delta_\delta^\alpha\gen[\bar Q]_{\dot\gamma}{}^b \appliedto\field,
\nln
\qquad
\gen[\bar S]_{a}{}^{\dot\gamma}\appliedto (\partial_{\dot\epsilon\delta}\field) &=
-ix^{\beta\dot\gamma}\partial_{\dot\epsilon\delta}(\gen[Q]_{a\beta}\appliedto \field)
+\partial_{\dot\epsilon\delta}(\gen[\bar S]'_{a}{}^{\dot\gamma}\appliedto \field)
-i\delta_{\dot\epsilon}^{\dot\gamma}\gen[Q]_{a\delta}\appliedto \field.
\end{align}

First we consider overlapping terms in the action of $\genyang[B]$:
Overlapping terms require the first level-zero generator to act non-linearly.
However all non-linear contributions
from the odd generators act on fermions only
and they produce two bosons without derivatives.
When acting further on the result,
all $x$-dependent contributions must cancel by the above arguments, and,
as we have seen, bosons without derivatives cannot generate $x$-independent terms.
Moreover, extra terms cannot be generated from the action of the first generator
because $\gen[S]'$ and $\gen[\bar S]'$ are purely linear
and any partial derivatives acting on the original field
can be pulled out from the calculation.
Hence there are no overlapping terms.
Analogously, the single-field action of $\genyang[B]$ is zero.

Let us now consider the various terms in the action:
The quadratic terms $\action_\nfield{2}$ can only
contribute via overlapping terms
and therefore they are all trivially invariant on their own.
Furthermore, all the quartic terms $\action_\nfield{4}$ are purely bosonic
and they do not involve partial derivatives.
Since the operators $\gen[S]'$ and $\gen[\bar S]'$ act on fermions only,
and the absence of partial derivatives prevents the generation of
further $x$-independent terms,
also all terms in $\action_\nfield{4}$ are invariant on their own.
It remains to consider the cubic terms $\action_\nfield{3}$.

By elementary transformations we can summarise the
(non-linear) action \eqref{eq:Yprimereduced} on the cubic terms as
\begin{align}
\sum\nolimits_n\yanghat{\oper{Y}}'_\nfield{n}&\simeq
\gen^\swdl{1}_1\gen^\swdl{2}_2\action_\nfield{3}
\nln &\simeq
- \quarter (\gen[S]^{\alpha b})_1 (\gen[Q]_{b \alpha})_2\action_\nfield{3}
- \quarter (\gen[Q]_{b \alpha})_1 (\gen[S]^{\alpha b})_2\action_\nfield{3}
\nln &\qquad
- \quarter (\gen[\bar{S}]_b {}^{\dot\alpha})_1 (\gen[\bar{Q}]_{\dot\alpha}{}^b)_2\action_\nfield{3}
- \quarter (\gen[\bar{Q}]_{\dot\alpha}{}^b)_1 (\gen[\bar{S}]_b {}^{\dot\alpha})_2\action_\nfield{3}.
\end{align}
Effectively, only $x$-independent terms can potentially contribute,
and therefore we need to consider only two types of terms from this expression:
The generators $\gen[S]$ and $\gen[\bar S]$
can act on a derivative $\partial\field$
in such a way that the derivative eliminates the $x$-dependence
from the action of $\gen[S]$ and $\gen[\bar S]$.
Alternatively, $\gen[S]$ and $\gen[\bar S]$ can act on fermions
and yield terms via the extra operators $\gen[S]$ and $\gen[\bar S]$.
Curiously, the cubic terms in the action \eqref{eq:sym_action}
split into two corresponding classes,
namely terms with fermions and terms with derivatives
\begin{align}
\lagrange_\nfield{3}
&=
-i \varepsilon^{\dot\alpha\dot\gamma}\varepsilon^{\beta\delta}
\varepsilon^{\dot\epsilon\dot\kappa}\varepsilon^{\zeta\lambda}
   \tr\brk!{\comm{A_{\dot\kappa\lambda}}{\del_{\dot\alpha\beta}A_{\dot\epsilon \zeta}}A_{\dot\gamma\delta}}
-\rfrac{i}{2} \varepsilon^{\dot\alpha\dot\gamma}\varepsilon^{\beta\delta}
   \tr\brk!{\comm{\Phi^{ef}}{\del_{\dot\alpha\beta}\bar\Phi_{ef}}A_{\dot\gamma\delta}}
\nln & \qquad
+\varepsilon^{\dot\kappa\dot\alpha}\varepsilon^{\beta\gamma}
 \tr\brk!{ \acomm{\Psi^d{}_{\gamma}}{\bar\Psi_{\dot\kappa d}} A_{\dot\alpha\beta}}
+\ihalf\varepsilon^{\alpha\gamma}
  \tr\brk!{\acomm{\Psi^e{}_\alpha}{\Psi^f{}_\gamma} \bar\Phi_{ef}}
+ \ihalf\varepsilon^{\dot\alpha\dot\gamma}
  \tr\brk!{\acomm{\bar\Psi_{\dot\alpha e}}{\bar\Psi_{\dot\gamma f}} \Phi^{ef}}
.\end{align}

Let us first consider the fermionic terms on the second line
on which $\gen[S]$ and $\gen[\bar S]$
can effectively act only via the extra operators
$\gen[S]'$ and $\gen[\bar S]'$.
By considering the types which can potentially be generated
along with the parity-reversing nature of the level-one generators,
we find terms of the types
\begin{align}
&
\varepsilon^{\alpha\gamma}
  \tr\brk!{\comm{\Psi^e{}_\alpha}{\Psi^f{}_\gamma} \bar\Phi_{ef}},
&&
\varepsilon^{\alpha\gamma}
  \tr\brk!{\acomm{\bar\Phi_{ef}}{\Phi^{ef}}F_{\alpha\gamma}},
&&
  \tr\brk!{\acomm{\bar\Phi_{ef}}{\Phi^{de}}{\comm{\bar\Phi_{db}}{\Phi^{bf}}}},
\nln
&
\varepsilon^{\dot\alpha\dot\gamma}
  \tr\brk!{\comm{\bar\Psi_{\dot\alpha e}}{\bar\Psi_{\dot\gamma f}} \Phi^{ef}},
&&
\varepsilon^{\dot\alpha\dot\gamma}
  \tr\brk!{\acomm{\bar\Phi_{ef}}{\Phi^{ef}}\bar F_{\dot\alpha\dot\gamma}},
&&
\varepsilon^{\dot\alpha\dot\gamma}\varepsilon^{\beta\delta}
  \tr\brk!{\acomm{A_{\dot\alpha\beta}}{\bar\Phi_{ef}}\comm{A_{\dot\gamma\delta}}{\Phi^{ef}}},
\end{align}
as well as
\[
\varepsilon^{\dot\alpha\dot\gamma}\varepsilon^{\beta\delta}
  \tr\brk!{\del_{\dot\gamma\delta}\acomm{\bar\Phi_{ef}}{\Phi^{ef}}A_{\dot\alpha\beta}}.
\]
Now the former six terms are trivially zero because they all
involve simultaneous symmetrisation and anti-symmetrisation of indices.
Only the last term is non-zero. It can only be generated from
the term $\acomm{\Psi}{\bar\Psi} A$ in the action and the contributions read
\[
\rfrac{1}{12}\varepsilon^{\dot\kappa\dot\alpha}\varepsilon^{\beta\gamma}
 \tr\brk!{ \acomm{\gen[S]'^{\epsilon f}\Psi^d{}_{\gamma}}
{\gen[Q]_{f\epsilon}\bar\Psi_{\dot\kappa d}} A_{\dot\alpha\beta}+
  \acomm{\gen[\bar{S}]'_f{}^{\dot\epsilon}\bar\Psi_{\dot\kappa d}}
{\gen[\bar{Q}]_{\dot\epsilon}{}^f\Psi^d{}_{\gamma}} A_{\dot\alpha\beta}}.
\]
However, by explicit computation, the two terms cancel precisely.

By similar arguments as above, the purely bosonic cubic terms of the action
can yield terms of the kinds
\[
\varepsilon^{\alpha\gamma}
  \tr\brk!{\comm{\Psi^e{}_\alpha}{\Psi^f{}_\gamma} \bar\Phi_{ef}},
\quad
\varepsilon^{\dot\alpha\dot\gamma}
  \tr\brk!{\comm{\bar\Psi_{\dot\alpha e}}{\bar\Psi_{\dot\gamma f}} \Phi^{ef}},
\quad
\varepsilon^{\dot\kappa\dot\alpha}\varepsilon^{\beta\gamma}
 \tr\brk!{ \comm{\Psi^d{}_{\gamma}}{\bar\Psi_{\dot\kappa d}} A_{\dot\alpha\beta}}.
\]
As before the former two terms have incompatible symmetrisations and are zero.
Only the last term is non-zero.
By acting on the term $\comm{\Phi}{\del\Phi}A$
we find
\[
-\rfrac{i}{24} \varepsilon^{\dot\alpha\dot\gamma}\varepsilon^{\beta\delta}
   \tr\brk!{\comm{\del_{\dot\alpha\beta}(\gen[S]^{\kappa g}\bar\Phi_{ef})}
   {\gen[Q]_{g \kappa}\Phi^{ef}} A_{\dot\gamma\delta}
+
\comm{\del_{\dot\alpha\beta}\gen[\bar{S}]_g {}^{\dot\kappa}\Phi^{ef}}
  {\gen[\bar{Q}]_{\dot\kappa}{}^g\bar\Phi_{ef}}A_{\dot\gamma\delta}},
\]
which evaluates to
\[
-\rfrac{1}{2} \varepsilon^{\dot\alpha\dot\gamma}\varepsilon^{\beta\delta}
   \tr\brk!{\comm{\Psi^f{}_\beta }{\bar\Psi_{\dot\alpha f} }
    A_{\dot\gamma\delta}}.
\]
Conversely, the action on the term $\comm{A}{\del A}A$ yields
\begin{align}
&\rfrac{i}{12} \varepsilon^{\dot\alpha\dot\gamma}\varepsilon^{\beta\delta}
\varepsilon^{\dot\epsilon\dot\kappa}\varepsilon^{\zeta\lambda}
   \tr\brk!{ \comm{\del_{\dot\alpha\beta}\gen[S]^{\eta g}A_{\dot\epsilon \zeta}}
  { \gen[Q]_{g \eta}A_{\dot\gamma\delta}} A_{\dot\kappa\lambda}+
\comm{\del_{\dot\alpha\beta}\gen[\bar{S}]_g {}^{\dot\eta}A_{\dot\epsilon \zeta}}
{\gen[\bar{Q}]_{\dot\eta}{}^gA_{\dot\gamma\delta}} A_{\dot\kappa\lambda}}
\nln
&-\rfrac{i}{12} \varepsilon^{\dot\alpha\dot\gamma}\varepsilon^{\beta\delta}
\varepsilon^{\dot\epsilon\dot\kappa}\varepsilon^{\zeta\lambda}
   \tr\brk!{\comm{\del_{\dot\epsilon \zeta}\gen[S]^{\eta g}A_{\dot\alpha\beta}}
{\gen[Q]_{g \eta}A_{\dot\gamma\delta}} A_{\dot\kappa\lambda}
+\comm{\del_{\dot\epsilon \zeta}\gen[\bar{S}]_g {}^{\dot\eta}A_{\dot\alpha\beta}}
{ \gen[\bar{Q}]_{\dot\eta}{}^gA_{\dot\gamma\delta}} A_{\dot\kappa\lambda}},
\end{align}
which amounts to
\[
\rfrac{1}{2}
\varepsilon^{\dot\alpha\dot\gamma}\varepsilon^{\beta\delta}
   \tr\brk!{ \comm{\Psi^f{}_\beta}{\bar\Psi_{\dot\alpha f}} A_{\dot\gamma\delta}}.
\]
Therefore, both remaining contributions cancel,
and altogether this proves that the bonus level-one Yangian generator $\genyang[B]$
is a symmetry of planar $\superN=4$ sYM.

\section{\texorpdfstring{$\superN=6$}{N=6} supersymmetric Chern--Simons theory}
\label{sec:abjm}

Now that we have established an extended symmetry related
to planar integrability of classical $\superN=4$ sYM theory,
we should consider further gauge theories
where planar integrability has been observed.
The main example of a supposedly integrable
gauge theory in the planar limit
which is substantially different from $\superN=4$ sYM
is ABJ(M) theory.
This model is a superconformal field theory in three dimensions
with an AdS/CFT dual string theory,
it appears to be integrable in the planar limit
and its observables display Yangian symmetry.

ABJ(M) theory will serve as a crucial testing ground for our proposal:
if our definition of integrability in the form of Yangian symmetry is correct,
we should be able to show that its holds also for this theory.

\subsection{Action}
\label{sec:abjm_action}

ABJ(M) theory is a $\superN=6$ supersymmetric Chern--Simons theory
with gauge group $\grp{U}(M)\times \grp{U}(N)$.
The vector multiplet is coupled to two scalar multiplets $(\Phi,\Psi)$ and
$(\bar\Phi,\bar\Psi)$, transforming in the
$(M,\bar{N})$ and $(\bar{M},N)$
representations of the gauge group, respectively.

As for $\superN=4$ sYM, we use a spinor-matrix notation $x^{\alpha\beta}$
to denote the coordinates of three-dimensional spacetime
where $\alpha,\beta,\ldots=1,2$ denote
$\alg{sl}(2,\Real)$ spacetime spinor indices.
The coordinate matrices are symmetric
$x^{\alpha\beta}=x^{\beta\alpha}$ and real.
The corresponding partial derivatives $\partial_{\alpha\beta}=\partial_{\beta\alpha}$ are
normalised by the rule $\partial_{\alpha\beta}x^{\gamma\delta}
=\delta_\alpha^\gamma\delta_\beta^\delta+\delta_\alpha^\delta\delta_\beta^\gamma$.
Furthermore, Latin letters $a,b,\ldots = 1,2,3,4$
denote $\alg{su}(4)$ internal indices.

For simplicity, we introduce two gauge fields $A_{\alpha\beta}$ and $\tilde{A}_{\alpha\beta}$,
one for each $\grp{U}(K)$ component of the gauge group.
The fields of the theory can be represented as matrices: the field $A_{\alpha\beta}$
($\tilde{A}_{\alpha\beta}$) is a hermitian $M\times M$ ($N\times N$) matrix, the fields
$\Phi^a$ and $\Psi_{\alpha b}$ ($\bar\Phi_a$ and $\bar\Psi_\alpha{}^b$) are
complex $M\times N$ ($N\times M$) matrices; any sequence of fields is to be understood
as the appropriate matrix product.

The gauge-covariant derivatives act
on a covariant $M\times N$ field $\field$
($N\times M$ field $\bar\field$) as
\begin{align}
\label{eq:abjm_der}
\cdel_{\alpha\beta} \field
&:=
 \partial_{\alpha\beta} \field
+i A_{\alpha\beta} \field
- i \field \tilde{A}_{\alpha\beta},
\nln
\cdel_{\alpha\beta} \bar{\field}
&:=
\partial_{\alpha\beta} \bar{\field}
+i \tilde{A}_{\alpha\beta} \bar{\field}
- i \bar{\field}  A_{\alpha\beta},
\end{align}
and we define the field strengths as for $\superN=4$ sYM
\begin{align}
\cdel_{\alpha\beta} A_{\gamma\delta}
&:=
\partial_{\alpha\beta} A_{\gamma\delta}
-\partial_{\gamma\delta} A_{\alpha\beta}
+i \comm{A_{\alpha\beta}}{A_{\gamma\delta}}
,\nln
\cdel_{\alpha\beta} \tilde A_{\gamma\delta}
&:=
\partial_{\alpha\beta} \tilde A_{\gamma\delta}
-\partial_{\gamma\delta} \tilde A_{\alpha\beta}
+i \comm{\tilde A_{\alpha\beta}}{\tilde A_{\gamma\delta}}
.
\end{align}
The action of the theory is completely fixed by supersymmetry and
its Lagrangian reads
\begin{align}
\lagrange
&=
\varepsilon^{\beta\gamma} \varepsilon^{\delta\epsilon} \varepsilon^{\kappa\alpha}
\tr \brk!{
-\rfrac{1}{4} A_{\alpha\beta}\partial_{\gamma\delta} A_{\epsilon\kappa}
- \rfrac{i}{6} A_{\alpha\beta} A_{\gamma\delta} A_{\epsilon\kappa}
+ \rfrac{1}{4} \tilde{A}_{\alpha\beta} \partial_{\gamma\delta} \tilde{A}_{\epsilon\kappa}
+ \rfrac{i}{6} \tilde{A}_{\alpha\beta} \tilde{A}_{\gamma\delta} \tilde{A}_{\epsilon\kappa} }
\nln&\qquad
+i\varepsilon^{\alpha\beta}\varepsilon^{\gamma\delta} 
 \tr\brk!{\bar\Psi_\alpha{}^e \.\cdel_{\beta\gamma} \Psi_{\delta e}}
+\rfrac{1}{2} \varepsilon^{\alpha\gamma} \varepsilon^{\beta\delta}
 \tr\brk!{
 \cdel_{\alpha\beta}\bar\Phi_e \.\cdel_{\gamma\delta} \Phi^e}
\nln&\qquad
+\varepsilon^{\epsilon\kappa} \tr \brk!
{i\varepsilon^{a b c d} \.\bar\Phi_a \Psi_{\epsilon b} \bar\Phi_c \Psi_{\kappa d}
-i\varepsilon_{a b c d} \.\Phi^a \bar\Psi_\epsilon{}^b \Phi^c \bar\Psi_\kappa{}^d}
\nln&\qquad
+\varepsilon^{\epsilon\kappa} \tr \brk!{
-i\.\Phi^a \bar\Phi_a \Psi_{\epsilon b} \bar\Psi_\kappa{}^b
+i\.\bar\Phi_a \Phi^a \bar\Psi_\epsilon{}^b \Psi_{\kappa b}
+2i\. \Phi^a \bar\Phi_b \Psi_{\epsilon a} \bar\Psi_\kappa{}^b
-2i\. \bar\Phi_a \Phi^b  \bar\Psi_\epsilon{}^a \Psi_{\kappa b}
}
\nln&\qquad
+\tr\brk!{
  \rfrac{1}{3}\. \Phi^a \bar\Phi_a \Phi^b \bar\Phi_b \Phi^c \bar\Phi_c
+ \rfrac{1}{3}\. \Phi^a \bar\Phi_b \Phi^b \bar\Phi_c \Phi^c \bar\Phi_a
+ \rfrac{4}{3}\. \Phi^a \bar\Phi_b \Phi^c \bar\Phi_a \Phi^b \bar\Phi_c
- 2\. \Phi^a \bar\Phi_b \Phi^c \bar\Phi_c \Phi^b \bar\Phi_a
}.
\end{align}
%

\subsection{Symmetries}

As already stated, ABJ(M) theory is a superconformal theory; the algebra of
superconformal transformations is
\unskip\footnote{The maximal even subalgebra is $\alg{sp}(4) \times \alg{so}(6)$, where
$\alg{sp}(4)= \alg{so}(2,3)$ is the conformal algebra in three
dimensions, and $\alg{so}(6)$ is the R-symmetry algebra.}
$\alg{osp}(6|4)$.
The action of the Poincar\'e and dilatation generators reads
\begin{align}
\gen[P]_{\alpha\beta}\appliedto \field
&= i \.\cdel_{\alpha\beta}\field
,\nln
\gen[L]^\alpha{}_\beta\appliedto \field
&= -i x^{\alpha\epsilon} \cdel_{\epsilon\beta}\field
   +\rfrac{i}{2} \delta^\alpha_\beta x^{\delta\epsilon} \cdel_{\delta\epsilon}\field
+(\gen[L]_\text{spin})^\alpha{}_\beta\appliedto \field
,\nln
\gen[D]\appliedto \field
&= -\tfrac{i}{2} x^{\alpha\beta} \. \cdel_{\alpha\beta}\field - i\Delta_{\field} \field
,
\end{align}
where the conformal dimensions are $\Delta_\Phi = \vfrac{1}{2}$,
$\Delta_\Psi = 1$, $\Delta_A = 0$,
and where the spin operator $\gen[L]_\text{spin}$
acts non-trivially only on the spinor fields $\field_\alpha\in\set{\Psi_{\alpha c},\bar\Psi_\alpha{}^c}$
\[
(\gen[L]_\text{spin})^\alpha{}_\beta\appliedto \field_{\gamma}
=
-i\delta^\alpha_\gamma  \field_{\beta}
+\rfrac{i}{2} \delta^{\alpha}_\beta \field_{\gamma}.
\]
The supersymmetry charges transform in the six-dimensional real irreducible
representation of the internal algebra $\alg{su}(4)$.
They act as
\begin{align}
\gen[Q]_{\epsilon a b} \appliedto \Phi^c
&=
2\delta^c_{\bar a} \Psi_{\epsilon \bar b}
,\nln
\gen[Q]_{\epsilon a b} \appliedto \bar\Phi_c
&=
\varepsilon_{a b c d} \bar\Psi_\epsilon{}^d
,\nln
\gen[Q]_{\epsilon a b} \appliedto \Psi_{\kappa c}
&=
i\varepsilon_{a b c d} \. \cdel_{\epsilon\kappa} \Phi^d
+\varepsilon_{\epsilon\kappa} \varepsilon_{a b c d}
\brk!{-i\.\Phi^d \bar\Phi_f \Phi^f + i\.\Phi^f \bar\Phi_f \Phi^d}
-2i\varepsilon_{\epsilon\kappa} \varepsilon_{a b d f} \Phi^d \bar\Phi_c \Phi^f
,\nln
\gen[Q]_{\epsilon a b} \appliedto \bar\Psi_\delta{}^c
&=
2i\delta^c_{\bar a}\. \cdel_{\epsilon\delta} \bar\Phi_{\bar b}
+ 4i\varepsilon_{\epsilon\delta} \.\bar\Phi_{\bar a} \Phi^c \bar\Phi_{\bar b}
+ 2i \varepsilon_{\epsilon\delta} \delta^c_{\bar a}\. \bar\Phi_{\bar b} \Phi^f \bar\Phi_f
-2i\varepsilon_{\epsilon\delta}\delta^c_{\bar a}\. \bar\Phi_f \Phi^f \bar\Phi_{\bar b}
,\nln
\gen[Q]_{\epsilon a b} \appliedto A_{\gamma\delta}
&=
-4i\varepsilon_{\epsilon\bar\gamma} \.\Psi_{\bar\delta \bar a}\bar\Phi_{\bar b}
-2i \varepsilon_{\epsilon\bar\gamma} \varepsilon_{a b f g}\. \Phi^f \bar\Psi_{\bar\delta}{}^g
,\nln
\gen[Q]_{\epsilon a b} \appliedto \tilde{A}_{\gamma\delta}
&=
4i \varepsilon_{\epsilon \bar\gamma} \.\bar\Phi_{\bar a} \Psi_{\bar \delta \bar b}
+2i\varepsilon_{\epsilon \bar\gamma}  \varepsilon_{a b f g}\. \bar\Psi_{\bar\delta}{}^f \Phi^g
.
\end{align}
Here and in the following we use a shorthand notation
$X_{\bar\alpha\bar\gamma}:=\half X_{\alpha\gamma}\pm\half X_{\gamma\alpha}$
for symmetrisation of indices,
where the sign is determined by the manifest symmetries
on the defining l.h.s.\ of the equation.

\subsection{Yangian invariance}
\label{sec:abjm_yang}

In order to check the classical Yangian invariance of ABJ(M) theory, we can proceed the
same way we did for $\superN=4$ sYM: we start from the simplest equation of
motion, we act on it with the bi-local part of a level-one Yangian generator, we
fix the single-field action and then we check that the relationship of
\eqref{eq:sym_Jhateom} is satisfied.

The most convenient choice for the level-one generator of $\yang[\alg{osp}(6|4)]$
is again $\genyang[P]_{\alpha\beta}$. The bi-local part of the coproduct reads
\begin{equation}
\label{eq:abjm_phatbil}
(\genyang[P]_{\alpha\beta})_\bilocal
=
\gen[P]_{\gamma\bar\alpha}\wedge \gen[L]^\gamma{}_{\bar\beta}
+ \gen[P]_{\alpha\beta}\wedge \gen[D]
+ \rfrac{i}{8} \varepsilon^{c d e f} \gen[Q]_{\alpha cd} \wedge \gen[Q]_{\beta ef}.
\end{equation}
Furthermore, we choose the simplest of the equations of motion,
the Dirac equation $\eomfor{\bar\Psi}=0$ with
\begin{align}
\eomfor{\bar\Psi}^\alpha{}_b
&=
\varepsilon^{\alpha\epsilon}
\brk!{i\varepsilon^{\gamma\delta}\. \cdel_{\epsilon\gamma}\Psi_{\delta b}
+2i \varepsilon_{b c d f}\.\Phi^c \bar\Psi_\epsilon{}^d \Phi^f}
\nln&\quad
+\varepsilon^{\alpha\epsilon}\brk!{
i \.\Psi_{\epsilon b} \bar\Phi_c \Phi^c
-2i \.\Psi_{\epsilon c} \bar\Phi_b \Phi^c
-i \.\Phi^c \bar\Phi_c \Psi_{\epsilon b}
+2i \.\Phi^c \bar\Phi_b \Psi_{\epsilon c}
}.
\end{align}
We find that the Dirac equation is weakly invariant under
the above level-one momentum generator
with the following unique choice of single-field action
\begin{align}
\label{eq:abjm_phatphi}
\genyang[P]_{\alpha\beta} \appliedto \Phi^a &=0
,\nln
\genyang[P]_{\alpha\beta} \appliedto \bar\Phi_a &=0
,\nln
\genyang[P]_{\alpha\beta} \appliedto \Psi_{\gamma d}
&=
\varepsilon_{\bar\alpha \gamma}
\brk!{
-\Psi_{\bar\beta d} \bar\Phi_e \Phi^e
+2\. \Psi_{\bar\beta e} \bar\Phi_d \Phi^e
- \Phi^e \bar\Phi_e \Psi_{\bar \beta d}
+2\. \Phi^e \bar\Phi_d \Psi_{\bar\beta e}
}
,\nln
\genyang[P]_{\alpha\beta} \appliedto \bar\Psi_{\gamma}{}^d
&=
\varepsilon_{\bar\alpha\gamma}
\brk!{
\bar\Psi_{\bar\beta}{}^d \Phi^e \bar\Phi_e
-2\. \bar\Psi_{\bar\beta}{}^e \Phi^d \bar\Phi_e
+\bar\Phi_e \Phi^e \bar\Psi_{\bar\beta}{}^d
-2\. \bar\Phi_e \Phi^d \bar\Psi_{\bar\beta}{}^e
}
,\nln
\genyang[P]_{\alpha\beta} \appliedto A_{\gamma\delta}
&=
i\varepsilon_{\bar\alpha\bar\gamma}\. \cdel_{\bar\beta\bar\delta}\brk{ \Phi^e \bar\Phi_e }
+\varepsilon_{\bar\alpha\bar\gamma} \varepsilon_{\bar\beta\bar\delta}
\brk!{-2i \. \Phi^e \bar\Phi_f \Phi^f \bar\Phi_e + \varepsilon^{\kappa\lambda}\Psi_{\kappa e} \bar\Psi_\lambda{}^e}
,\nln
\genyang[P]_{\alpha\beta} \appliedto \tilde A_{\gamma\delta}
&=
-i\varepsilon_{\bar\alpha\bar\gamma} \cdel_{\bar\beta\bar\delta}\brk{ \bar\Phi_e \Phi^e }
+\varepsilon_{\bar\alpha\bar\gamma} \varepsilon_{\bar\beta\bar\delta}
\brk!{-2i \. \bar\Phi^e \Phi_f \bar\Phi^f \Phi_e - \varepsilon^{\kappa\lambda}\bar\Psi_\kappa{}^e \Psi_{\lambda e}}
.
\end{align}
More concretely, we find that the following equality holds
\begin{align}
  \label{eq:abjm_diracYI}
  \genyang[P]_{\alpha\beta}\appliedto \eomfor{\bar\Psi}^\gamma{}_d
&=
\varepsilon_{\bar\alpha\epsilon}\brk!{
-\rfrac{1}{2}\delta_{\bar\beta}^\gamma\.\eomfor{A}^{\epsilon\kappa} \Psi_{\kappa d}
-\rfrac{5}{2}\.\eomfor{A}^{\epsilon\gamma} \Psi_{\bar\beta d}
+\rfrac{1}{2}\delta_{\bar\beta}^\gamma\.\Psi_{\kappa d} \eomfor{\tilde A}^{\epsilon\kappa}
+\rfrac{5}{2}\. \Psi_{\bar\beta d} \eomfor{\tilde A}^{\epsilon\gamma}
}
\nln&\quad
+\varepsilon_{\bar\alpha\epsilon}\delta_{\bar\beta}^\gamma\brk!{
2 \. \eomfor{\bar\Psi}^\epsilon{}_f \bar\Phi_d \Phi^f
- \. \eomfor{\bar\Psi}^\epsilon{}_d \bar\Phi_f \Phi^f
+2 \. \Phi^f \bar\Phi_d \eomfor{\bar\Psi}^\epsilon{}_f
- \. \Phi^f \bar\Phi_f \eomfor{\bar\Psi}^\epsilon{}_d
}.
\end{align}
Here, $\eomfor{A}$ and $\eomfor{\tilde A}$
denote the variations of the action w.r.t.\ the gauge fields,
\begin{align}
\eomfor{A}^{\alpha\beta}
&=
\varepsilon^{\bar\alpha\gamma}\varepsilon^{\bar\beta\delta} \brk!{
\half \varepsilon^{\epsilon\kappa}\cdel_{\gamma\epsilon}A_{\delta\kappa}
+ i\.\Phi^e\. \cdel_{\gamma\delta}\bar\Phi_e
- i\.\cdel_{\gamma\delta}\Phi^e\.\bar\Phi_e
- 2i\. \Psi_{\gamma e} \bar\Psi_\delta{}^e
}
,\nln
\eomfor{\tilde A}^{\alpha\beta}
&=
\varepsilon^{\bar\alpha\gamma}\varepsilon^{\bar\beta\delta} \brk!{
-\half\varepsilon^{\epsilon\kappa}\cdel_{\gamma\epsilon}\tilde A_{\delta\kappa}
+ i\. \bar\Phi_e \.\cdel_{\gamma\delta}\Phi^e
- i\. \cdel_{\gamma\delta}\bar\Phi_e \Phi^e
- 2i\. \bar\Psi_{\gamma}{}^e \Psi_{\delta e}}
,
\end{align}
and they vanish on-shell as does $\eomfor{\bar\Psi}$
so that the Dirac equation is weakly invariant.
Moreover, we have verified that the r.h.s.\ of \eqref{eq:abjm_diracYI}
matches exactly the r.h.s.\ of \eqref{eq:sym_Jhateom},
and we have performed this check for all the equations
of motion of ABJ(M) theory. Therefore the equations of motion
display Yangian symmetry in a strong sense.
Finally, have shown by computer algebra that the ABJ(M) action is invariant
in the sense of \eqref{eq:yangianoncyclic}.
Therefore classical planar ABJ(M) theory has Yangian symmetry.

\section{Pure \texorpdfstring{$\superN<4$}{N<4} sYM}
\label{sec:yang_counter}

In this section we would like to address a possible concern that our results discussed above,
especially the formula \eqref{eq:sym_Jhateom},
are a mere result of level-zero (superconformal)
and gauge symmetries of the theories.
If that was the case, our work would not provide any criterion
for establishing Yangian symmetry of physical models.
As we will now demonstrate, this is fortunately not the case
and for formula \eqref{eq:sym_Jhateom} to hold increased symmetry is indeed required.

To this end let us consider a pure sYM theory, but keep $\superN$ arbitrary.
The action is again given by:

\begin{align}
 \label{eq:n2sym_action}
\lagrange
&=
-\rfrac{1}{2}\varepsilon^{\alpha\epsilon}\varepsilon^{\gamma\kappa}
 \tr\brk!{ F_{\alpha\gamma} F_{\epsilon\kappa}}
-\rfrac{1}{2}\varepsilon^{\dot\alpha\dot\epsilon}\varepsilon^{\dot\gamma\dot\kappa}
 \tr\brk!{\bar F_{\dot\alpha\dot\gamma}\bar F_{\dot\epsilon\dot\kappa}}
\nln & \qquad
+i\varepsilon^{\dot\kappa\dot\alpha}\varepsilon^{\beta\gamma}
 \tr\brk!{ \bar\Psi_{\dot\kappa d} \cdel_{\dot\alpha\beta}\Psi^d{}_{\gamma} }
-\rfrac{1}{4} \varepsilon^{\dot\alpha\dot\gamma}\varepsilon^{\beta\delta}
   \tr\brk!{\cdel_{\dot\alpha\beta}\bar\Phi_{ef}\.\cdel_{\dot\gamma\delta}\Phi^{ef}}
\nln & \qquad
+\ihalf\varepsilon^{\alpha\gamma}
  \tr\brk!{\bar\Phi_{ef}\acomm{\Psi^e{}_\alpha}{\Psi^f{}_\gamma} }
+ \ihalf\varepsilon^{\dot\alpha\dot\gamma}
  \tr\brk!{\Phi^{ef}\acomm{\bar\Psi_{\dot\alpha e}}{\bar\Psi_{\dot\gamma f}} }
\nln & \qquad
+ \rfrac{1}{16}\tr\brk!{\comm{\bar\Phi_{ab}}{\Phi^{ef}}\comm{\Phi^{ab}}{\bar\Phi_{ef}}} .
\end{align}
The action \eqref{eq:n2sym_action} looks exactly like \eqref{eq:sym_action} of $\superN=4$ sYM.
The only difference is the range of indices $a,\ b,\ldots = 1,\ldots, \superN$, where $\superN \leq 4 $.
As the reality condition \eqref{eq:scalarduality} exists only for $\superN=4$,
the scalar potential is written here in a different form, valid also for $\superN \neq 4$.
Moreover, for $\superN=1$ the scalar fields do not exist at all,
which here is implicitly taken care of by the anti-symmetry of their indices,
$\Phi^{ab}=-\Phi^{ba}=\Phi^{11}=0$.
The symmetry algebra of \eqref{eq:n2sym_action} is $\mathfrak{psu}(2,2|\superN)$
and for any $\superN \geq 0$ its Yangian exists.
What we want to demonstrate is that
$\yang[\mathfrak{psu}(2,2| \superN)]$ is a symmetry of the theory
only for the special value $\superN=4$.
To that end it is already enough to show that for other values of $\superN$
the weak invariance of the equations of motion does not hold,
as its failure will assure that none of the (aforementioned) stronger
criteria stand either.

We thus can repeat the computation from \secref{sec:yang_onshell} keeping $\superN$ arbitrary.
The action of all the generators carries over from \secref{sec:symsN4} \secref[and]{sec:classicalYang}
(see equations \eqref{eq:sym_qqbar} and \eqref{eq:yang_sfa} for supersymmetry and local Yangian action respectively),
of course up to the restricted range of indices mentioned above
(e.g.\ the local action of $\genyang[P]$
vanishes completely for $\superN=1$).
For general $\superN\geq 1$ we act with $\genyang[P]$
on the Dirac equation of motion
\unskip\footnote{For $\superN=0$ the absence of fermionic fields $\Psi$
requires a more elaborate consideration of the Yang--Mills equations.}
\[
\eomfor{\bar\Psi}^{l\dot\kappa}=
  i\varepsilon^{\dot\kappa\dot\alpha}\varepsilon^{\beta\gamma}\cdel_{\dot\alpha\beta}\Psi^l{}_{\gamma}
  -i\varepsilon^{\dot\kappa\dot\alpha}\comm{\Phi^{le}}{\bar\Psi_{\dot\alpha e}}.
\]
On shell we are left with a term of the form
\[
\genyang[P]_{\dot\alpha\beta}\appliedto
\eomfor{\bar\Psi}^{d\dot\gamma} \approx
(4-\superN)\varepsilon^{\dot\gamma\dot\kappa}\acomm{\bar F_{\dot\alpha\dot\kappa}}{\Psi^d{}_\beta}.
\]
Even though we have the principal freedom to adjust the local action of $\genyang[P]$
to the case $\superN<4$, it is easy to see that this cannot remove the residual term:
In order to cancel the term, one would need extra contributions of the form
$\genyang[P]A\sim \bar F$ or $\genyang[P]\Psi\sim \cdel \Psi$.
However, these would merely produce commutators $\comm{\bar F}{\Psi}$
rather than the desired anti-commutator $\acomm{\bar F}{\Psi}$.

Thus we see that the value $\superN=4$ is special as it warrants
the cancellation of the residual term
and the on-shell invariance of the equations of motion of $\superN=4$ sYM.
More importantly, we observe that superconformal
and gauge symmetries by themselves are not sufficient to warrant
our results presented in the preceding sections,
and indeed Yangian invariance of a theory is a non-trivial statement.

\section{Conclusions and outlook}
\label{sec:conclusions}

In this work we have elaborated upon the results of our previous letter \cite{Beisert:2017pnr}
by spelling out in detail the different kinds of planar invariance conditions
for the equations of motion of $\superN=4$ supersymmetric
Yang--Mills theory and ABJ(M) theory under a Yangian algebra.
Subsequently, we have reformulated the strong version of this invariance
as an invariance of the action
and proved explicitly that this invariance holds indeed.
This not only shows that these two classical gauge theory models
possess Yangian symmetry in the planar limit,
but it may also be viewed as a formal definition for their integrability.

\medskip

A follow-up work \cite{Beisert:2018ijg}
will address Yangian-invariance of field correlators
at tree level:
for a field theory with some set of symmetries,
one should expect correlation functions of the fields to obey
a corresponding set of Ward--Takahashi-identities reflecting these symmetries.
An important aspect in the computation of correlators
within a gauge theory is gauge fixing.
It will be shown how Yangian symmetry can be made compatible
with Faddeev--Popov gauge fixing and BRST symmetry
and how to formulate Slavnov--Taylor identities for Yangian symmetry
which properly take into account the effects of gauge fixing.
To that end, the algebraic relations of the Yangian need
to be studied in detail because they are intertwined with
gauge transformations which are henceforth deformed
by the gauge fixing procedure.
A curious side effect is that this will amount
to an extension of the gauge algebra
by bi-local and non-local gauge symmetries.

\medskip

Our works lay the foundations for many directions of further study.
Most importantly, firm contact should be made
with corresponding symmetry structures
of the world-sheet theory for strings on $AdS_5\times S^5$.
By the AdS/CFT correspondence such structures should exist,
but quantum effects on both sides, potentially including quantum anomalies,
may play a significant role for a precise matching,
see \cite{Berkovits:2004xu} for considerations
on the string theory side.
For example, the algebraic complications due to non-ultra-locality
of the string theory sigma model \cite{Delduc:2012vq}
may well have a counterpart in the Yangian algebra for gauge theory.

Another obvious question is which other models beyond $\superN=4$ sYM and
ABJ(M) theory enjoy a Yangian (or related) symmetry in the planar limit?
There are several deformations and orbifolds of $\superN=4$ sYM
which are apparently integrable \cite{Zoubos:2010kh}
and which should therefore possess an extended symmetry.
This has been shown explicitly for the
beta-deformation in \cite{Garus:2017bgl}.
The simplistic fishnet theory \cite{Gurdogan:2015csr} represents a particular
contraction limit of an integrable deformation;
however due to the non-vanishing Coxeter number of the
level-zero symmetry, some adjustments to our treatment will be inevitable,
see \cite{Chicherin:2017frs}.
Furthermore, massive deformations of supersymmetric Chern--Simons theories
\cite{Hosomichi:2008jd}
have an interesting non-conformal supersymmetry algebra \cite{Bergshoeff:2008ta}
on which a Yangian algebra could in principle be established.
One may also wonder whether and how the partial integrability
of $\superN<4$ sYM theories within certain sectors, see \cite{Belitsky:2004cz},
can be formulated in analogy to our framework.

Yangian symmetry could also prove helpful in constructing
gauge theory models which are predicted by the AdS/CFT correspondence
but remain to be formulated.
Such models include two-dimensional gauge theories as duals
of the integrable string theories
on $AdS_3\times S^3\times S^3\times S^1$
and $AdS_3\times S^3\times T^4$,
a q-deformation of $\superN=4$ sYM as a dual
of the eta/kappa-deformation of string theory on $AdS_5\times S^5$
(where the Yangian symmetry would be deformed to a quantum affine group)
as well as the infamous
six-dimensional $\superN=(2,0)$ theory
which is intrinsically non-perturbative.

\smallskip

Another direction for research is to firmly derive the applications
of integrability to various observables
from our framework of Yangian symmetry.
For example, the Bethe equations
for the spectrum of one-loop anomalous dimensions
follow directly from the well-established
Bethe ansatz framework for quantum integrable spin chains
in connection to (conventional) quantum algebra.
More interestingly, the methods developed for the spectrum
at higher loops and at finite coupling
should have a formal justification
in Yangian symmetry of the model
(potentially after implementing quantum corrections).
Along the same lines, it will be useful to understand
the origin of the extended $\alg{psu}(2|2)$ Yangian
for the magnon scattering picture
including the various master, boost and secret symmetries.
Similarly, symmetries and techniques
for other observables such as Wilson loops,
scattering amplitudes, correlation functions of local operators
and form factors might be traced back to Yangian symmetry.

\smallskip

Further open questions beyond generalisations and applications include:
How to generalise Yangian symmetry to a superspace
formulation of $\superN=4$ sYM
(either in a light-cone superspace \cite{Mandelstam:1982cb}
or in a full superspace \cite{Beisert:2015uda})?
Does the (somewhat non-local) Yangian symmetry
have some associated Noether charges?
What is the role of the Casimir operators (central elements)
of the Yangian algebra?
Can they be applied to a wider class of observables?

Finally, our framework for Yangian symmetry raises
several questions of mathematical nature:
First and foremost,
does the Yangian algebra actually close,
i.e.\ do the Serre-relations hold
and how to formulate the closure precisely?
How to construct the single-field action of the level-one generators abstractly?
Does it follow from some fundamental principles
beyond the consistency requirements employed to derive it here?
Last but not least,
in what sense is the proposed action of the Yangian generators a representation,
in particular when acting on cyclic states such as a the action functional.

\pdfbookmark[1]{Acknowledgements}{ack}
\section*{Acknowledgements}

We thank N.~Drukker, J.~Plefka and M.~Staudacher
for discussions related to the work.
The work of NB and AG is partially supported by grant no.\ 615203
from the European Research Council under the FP7
and by the Swiss National Science Foundation
through the NCCR SwissMAP.
The work of MR was supported by the grant PL 457/3-1
``Yangian Symmetry in Quantum Gauge Field Theory''
of the German Research Foundation.

\appendix

\section{\texorpdfstring{$\superN=4$}{N=4} superconformal symmetry}
\label{app:superconformal}

In this appendix we summarise the $\superN=4$ superconformal algebra
$\alg{psu}(2,2|4)$
and its representation on the fields.

\paragraph{Algebra.}

The supersymmetry algebra is spanned by the supersymmetry generators
$\gen[Q]_{a\gamma}$,
$\gen[\bar Q]_{\dot\alpha}{}^c$
and the momentum generator
$\gen[P]_{\dot\alpha\gamma}$.
The special conformal generators are given by
two fermionic generators
$\gen[S]^{\alpha c}$, $\gen[\bar S]_a{}^{\dot\gamma}$
and the bosonic generator
$\gen[K]^{\alpha\dot\gamma}$.
Furthermore, the superconformal algebra includes the
Lorentz and internal rotation generators
$\gen[L]^\alpha{}_\gamma$,
$\gen[\bar L]^{\dot\alpha}{}_{\dot\gamma}$,
$\gen[R]^a{}_c$
(whose trace over the indices vanishes)
and the dilatation generator $\gen[D]$.
Finally, we will also need gauge transformations
to discuss the gauge-covariant representations.
These are generated by $\gengauge[X]$
where the field $X$ serves as the gauge parameter matrix.

Although we do not explicitly refer to real algebras,
a suitable set of reality conditions
for $\superN=4$ sYM is given by
\[
(\gen[P]_{\dot\alpha\gamma})^\dagger =  \gen[P]_{\dot\gamma\alpha}
,\qquad
(\gen[K]^{\alpha\dot\gamma})^\dagger =  \gen[K]^{\gamma\dot\alpha}
,\qquad
(\gen[Q]_{a\gamma})^\dagger = \gen[\bar Q]_{\dot\gamma}{}^a
,\qquad
(\gen[S]^{\alpha c})^\dagger = \gen[\bar S]_c{}^{\dot\alpha}
,\]
as well as
\[
(\gen[L]^\alpha{}_\gamma)^\dagger = \gen[\bar L]^{\dot\alpha}{}_{\dot\gamma}
,\qquad
(\gen[R]^a{}_c)^\dagger = \gen[R]^{c}{}_{a}
,\qquad
\gen[D]^\dagger = \gen[D]
,\qquad
\gengauge[X]^\dagger=\gengauge[X^\dagger].
\]
The below representations will be unitary w.r.t.\ these reality conditions.

In the following we will list the most relevant algebra relations.
The Lorentz and internal algebra relations take the form
\begin{align}
\comm{\gen[L]^\alpha{}_\beta}{\gen[L]^\gamma{}_\delta}
&=
i
(\delta^\gamma_\beta \gen[L]^\alpha{}_\delta-\delta^\alpha_\delta \gen[L]^\gamma{}_\beta)
+\gengauge[\ldots]
,\nln
\comm{\gen[\bar L]^{\dot\alpha}{}_{\dot\beta}}{\gen[\bar L]^{\dot\gamma}{}_{\dot\delta}}
&=
i
(\delta^{\dot\gamma}_{\dot\beta} \gen[\bar L]^{\dot\alpha}{}_{\dot\delta}
  -\delta^{\dot\alpha}_{\dot\delta} \gen[\bar L]^{\dot\gamma}{}_{\dot\beta})
+\gengauge[\ldots]
,\nln
\comm{\gen[R]^a{}_b}{\gen[R]^c{}_d}
&=
i(\delta^c_b \gen[R]^a{}_d-\delta^a_d \gen[R]^c{}_b)
,\end{align}
Here, the Lorentz algebra relations involve
gauge transformations $\gengauge[\ldots]$,
where the omitted gauge parameter
is a term of the form $xx\cdel A$
representing the field strength contracted with
the Killing spinors of the two rotations.
We do not present the long list of algebra relations
with the remaining generators,
as these can easily be inferred as the transformations of spinor indices
compatible with the above relations.

The algebra of the scaling generator $\gen[D]$ measures
the scaling dimension $\Delta_{\gen}$ of the other generators $\gen$
\[
\comm{\gen[D]}{\gen}=-i\Delta_{\gen}\.\gen + \gengauge[\ldots]
\]
with the dimensions
\[
\Delta_{\gen[L]} = \Delta_{\gen[\bar L]} = \Delta_{\gen[R]} = \Delta_{\gen[D]} = 0,
\qquad
\Delta_{\gen[Q]}= \Delta_{\gen[\bar Q]}=-\Delta_{\gen[S]} = -\Delta_{\gen[\bar S]} = \rfrac{1}{2},
\qquad
\Delta_{\gen[P]} = - \Delta_{\gen[K]} = 1.
\]
The remaining purely bosonic conformal algebra relations read
\begin{align}
\comm{\gen[P]_{\dot\alpha\beta}}{\gen[P]_{\dot\gamma\delta}}
&=
-i\gengauge[\cdel_{\dot\alpha\beta} A_{\dot\gamma\delta}],
\nln
\comm{\gen[P]_{\dot\alpha\beta}}{\gen[K]^{\gamma\dot\epsilon}}
&=
i\delta^{\dot\epsilon}_{\dot\alpha} \gen[L]^\gamma{}_\beta
+i\delta_\beta^\gamma \gen[\bar L]^{\dot\epsilon}{}_{\dot\alpha}
+i\delta_\gamma^\beta\delta^{\dot\epsilon}_{\dot\alpha}\gen[D]
+\gengauge[\ldots],
\nln
\comm{\gen[K]^{\beta\dot\alpha}}{\gen[K]^{\gamma\dot\epsilon}}
&=
\gengauge[\ldots].
\end{align}
Again, these relations may involve some gauge transformation $\gengauge[\ldots]$
in addition to the pure conformal generators.

The non-trivial relations of the fermionic generators read
\begin{align}
\acomm{\gen[Q]_{b\alpha}}{\gen[\bar Q]_{\dot\gamma}{}^d}
&=
2 \delta^d_b  \gen[P]_{\dot\gamma\alpha},
&
\acomm{\gen[Q]_{a\beta}}{\gen[S]^{\gamma d}}
&=
2i\delta^\gamma_\beta \gen[R]^d{}_a
-2i\delta_a^d \gen[L]^\gamma{}_\beta
-i\delta_a^d\delta_\beta^\gamma \gen[D],
\nln
\acomm{\gen[S]^{\alpha b}}{\gen[\bar S]_c{}^{\dot\epsilon}}
&=
2 \delta^b_c  \gen[K]^{\alpha\dot\epsilon}.
&
\acomm{\gen[\bar Q]_{\dot\alpha}{}^b}{\gen[\bar S]_c{}^{\dot\epsilon}}
&=
2i\delta^{\dot\epsilon}_{\dot\alpha} \gen[R]^b{}_c
+2i\delta_c^b \gen[\bar L]^{\dot\epsilon}{}_{\dot\alpha}
+i\delta_c^b\delta^{\dot\epsilon}_{\dot\alpha}\gen[D] ,
\end{align}
while the non-trivial mixed relations read
\begin{align}
\comm{\gen[P]_{\dot\alpha\beta}}{\gen[S]^{\gamma e}}
&=
\delta_\beta^\gamma\gen[\bar Q]_{\dot\alpha}{}^e
-\varepsilon_{\dot\alpha\dot\kappa}\gengauge[x^{\gamma\dot\kappa}\Psi^e{}_\beta],
&
\comm{\gen[P]_{\dot\alpha\beta}}{\gen[\bar S]_c{}^{\dot\epsilon}}
&=
-\delta_{\dot\alpha}^{\dot\epsilon}\gen[Q]_{c\beta}
+\varepsilon_{\beta\delta}\gengauge[x^{\delta\dot\epsilon}\bar\Psi_{\dot\alpha c}],
\nln
\comm{\gen[K]^{\alpha\dot\gamma}}{\gen[Q]_{d\epsilon}}
&=
-\delta^\alpha_\epsilon\gen[\bar S]_d{}^{\dot\gamma}
+\gengauge[\ldots],
&
\comm{\gen[K]^{\alpha\dot\gamma}}{\gen[\bar Q]_{\dot\epsilon}{}^{d}}
&=
\delta_{\dot\epsilon}^{\dot\gamma}\gen[S]^{\alpha d}
+\gengauge[\ldots].
\end{align}
The remaining relations involving the fermionic generators are trivial
modulo gauge transformations
\begin{align}
\comm{\gen[P]_{\dot\alpha\beta}}{\gen[Q]_{d\gamma}}
&=i\varepsilon_{\beta\gamma}\gengauge[\bar\Psi_{\dot\alpha d}],
&
\comm{\gen[P]_{\dot\alpha\beta}}{\gen[\bar Q]_{\dot\gamma}{}^d}
&=i\varepsilon_{\dot\alpha\dot\gamma}\gengauge[\Psi^d{}_\beta],
\nln
\acomm{\gen[Q]_{b\alpha}}{\gen[Q]_{d\gamma}}
&=2i \varepsilon_{\alpha\gamma}\gengauge[\bar\Phi_{bd}],
&
\acomm{\gen[\bar Q]_{\dot\alpha}{}^b}{\gen[\bar Q]_{\dot\gamma}{}^d}
&=2i \varepsilon_{\dot\alpha\dot\gamma}\gengauge[\Phi^{bd}],
\nln
\acomm{\gen[Q]_{b\alpha}}{\gen[\bar S]_d{}^{\dot\gamma}}
&=\gengauge[\ldots],
&
\acomm{\gen[\bar Q]_{\dot\alpha}{}^b}{\gen[S]^{\gamma d}}
&=\gengauge[\ldots],
\nln
\acomm{\gen[\bar S]_b{}^{\dot\alpha}}{\gen[\bar S]_d{}^{\dot\gamma}}
&=\gengauge[\ldots],
&
\acomm{\gen[S]_b{}^{\alpha}}{\gen[S]^{\gamma d}}
&=\gengauge[\ldots],
\nln
\comm{\gen[\bar S]_b{}^{\dot\alpha}}{\gen[K]^{\delta\dot\gamma}}
&=\gengauge[\ldots],
&
\comm{\gen[S]_b{}^{\alpha}}{\gen[K]^{\delta\dot\gamma}}
&=\gengauge[\ldots].
\end{align}
The relations involving the fermionic generators
typically hold only on-shell, i.e.\ modulo the equations of motion.

\paragraph{Representation.}

The representation of the supersymmetries on the fields reads
\begin{align}
\gen[Q]_{a\beta}\appliedto\Phi^{cd} &=
\delta^c_a \Psi^d{}_\beta-\delta^d_a \Psi^c{}_\beta,
&
\gen[\bar Q]_{\dot\alpha}{}^b\appliedto \bar\Phi_{cd} &=
 \delta^b_c \bar\Psi_{\dot\alpha d}- \delta^b_d \bar\Psi_{\dot\alpha c},
\nln
\gen[Q]_{a\beta}\appliedto\bar\Phi_{cd} &=
\varepsilon_{acde}\Psi^e{}_\beta ,
&
\gen[\bar Q]_{\dot\alpha}{}^b\appliedto \Phi^{cd} &=
\varepsilon^{bcde} \bar\Psi_{\dot\alpha e},
\nln
\gen[Q]_{a\beta}\appliedto A_{\dot\gamma\delta} &=
-i \varepsilon_{\beta\delta} \bar\Psi_{\dot\gamma a},
&
\gen[\bar Q]_{\dot\alpha}{}^b\appliedto A_{\dot\gamma\delta} &=
-i\varepsilon_{\dot\alpha\dot\gamma} \Psi^b{}_\delta,
\nln
\gen[Q]_{a\beta}\appliedto\Psi^c{}_\delta &=
-2\delta^c_a F_{\beta\delta}
+i\varepsilon_{\beta\delta} \comm{\Phi^{ce}}{\bar\Phi_{ae}},
&
\gen[\bar Q]_{\dot\alpha}{}^b\appliedto \bar\Psi_{\dot\gamma d} &=
-2\delta^b_d \bar F_{\dot\alpha\dot\gamma}
-i\varepsilon_{\dot\alpha\dot\gamma} \comm{\Phi^{be}}{\bar\Phi_{de}},
\nln
\gen[Q]_{a\beta}\appliedto\bar\Psi_{\dot\gamma d} &=
2i \cdel_{\dot\gamma\beta}\bar\Phi_{ad},
&
\gen[\bar Q]_{\dot\alpha}{}^b\appliedto \Psi^c{}_\delta &=
2i \cdel_{\dot\alpha\delta}\Phi^{bc}.
\end{align}
The standard rules \eqref{eq:sym_poincare1} and
\eqref{eq:sym_poincare2} for the representation of the
momentum and dilatation generators read explicitly
\begin{align}
\gen[P]_{\dot\alpha\beta}\appliedto\Phi^{cd} &=
 i \cdel_{\dot\alpha\beta}\Phi^{cd},
&
\gen[D]\appliedto\Phi^{cd} &=
-i x^{\beta\dot\alpha} \cdel_{\dot\alpha\beta}\Phi^{cd}
-i\Phi^{cd}
,
\nln
\gen[P]_{\dot\alpha\beta}\appliedto\Psi^c{}_\delta &=
 i \cdel_{\dot\alpha\beta}\Psi^c{}_\delta,
&
\gen[D]\appliedto\Psi^c{}_\delta &=
-i x^{\beta\dot\alpha} \cdel_{\dot\alpha\beta}\Psi^c{}_\delta
-\rfrac{3}{2}i\Psi^c{}_\delta
,
\nln
\gen[P]_{\dot\alpha\beta}\appliedto\bar\Psi_{\dot\gamma d} &=
 i \cdel_{\dot\alpha\beta}\bar\Psi_{\dot\gamma d},
&
\gen[D]\appliedto\bar\Psi_{\dot\gamma d} &=
-i x^{\beta\dot\alpha} \cdel_{\dot\alpha\beta}\bar\Psi_{\dot\gamma d}
-\rfrac{3}{2}i\bar\Psi_{\dot\gamma d} ,
\nln
\gen[P]_{\dot\alpha\beta}\appliedto A_{\dot\gamma\delta} &=
 i \cdel_{\dot\alpha\beta}A_{\dot\gamma\delta},
&
\gen[D]\appliedto A_{\dot\gamma\delta} &=
- ix^{\beta\dot\alpha} \cdel_{\dot\alpha\beta}A_{\dot\gamma\delta}
.
\end{align}
For the Lorentz generators one finds
\unskip\footnote{In analogy to the curious assignment $\Delta_A=0$
for scaling transformations,
the spacetime indices of the gauge field $A$ are not transformed explicitly.}
\begin{align}
\gen[L]^\beta{}_\delta\appliedto\Phi^{ce} &=
- ix^{\beta\dot\alpha} \cdel_{\dot\alpha\delta}\Phi^{ce}
+\ihalf \delta^{\beta}_{\delta} x^{\kappa\dot\alpha} \cdel_{\dot\alpha\kappa}\Phi^{ce}
,
\nln
\gen[L]^\beta{}_\delta\appliedto\Psi^c{}_\epsilon &=
-i x^{\beta\dot\alpha} \cdel_{\dot\alpha\delta}\Psi^c{}_\epsilon
+\ihalf \delta^{\beta}_{\delta} x^{\kappa\dot\alpha} \cdel_{\dot\alpha\kappa}\Psi^c{}_\epsilon
-i \delta^\beta_\epsilon \Psi^c{}_\delta
+\ihalf \delta^\beta_\delta \Psi^c{}_\epsilon
,
\nln
\gen[L]^\beta{}_\delta\appliedto\bar\Psi_{\dot\gamma e} &=
-i x^{\beta\dot\alpha} \cdel_{\dot\alpha\delta}\bar\Psi_{\dot\gamma e}
+\ihalf \delta^{\beta}_{\delta} x^{\kappa\dot\alpha} \cdel_{\dot\alpha\kappa}\bar\Psi_{\dot\gamma e}
,
\nln
\gen[L]^\beta{}_\delta\appliedto A_{\dot\gamma\epsilon} &=
-i x^{\beta\dot\alpha} \cdel_{\dot\alpha\delta}A_{\dot\gamma\epsilon}
+\ihalf \delta^{\beta}_{\delta} x^{\kappa\dot\alpha} \cdel_{\dot\alpha\kappa}A_{\dot\gamma\epsilon}
,
\end{align}
analogously for the conjugate Lorentz generators
\begin{align}
\gen[\bar L]^{\dot\alpha}{}_{\dot\gamma}\appliedto\Phi^{ed} &=
-i x^{\beta\dot\alpha} \cdel_{\dot\gamma\beta}\Phi^{ed}
+\ihalf\delta^{\dot\alpha}_{\dot\gamma} x^{\beta\dot\kappa} \cdel_{\dot\kappa\beta}\Phi^{ed}
,
\nln
\gen[\bar L]^{\dot\alpha}{}_{\dot\gamma}\appliedto\Psi^e{}_\delta &=
-i x^{\beta\dot\alpha} \cdel_{\dot\gamma\beta}\Psi^e{}_\delta
+\ihalf\delta^{\dot\alpha}_{\dot\gamma} x^{\beta\dot\kappa} \cdel_{\dot\kappa\beta}\Psi^e{}_\delta
,
\nln
\gen[\bar L]^{\dot\alpha}{}_{\dot\gamma}\appliedto\bar\Psi_{\dot\epsilon d} &=
-i x^{\beta\dot\alpha} \cdel_{\dot\gamma\beta}\bar\Psi_{\dot\epsilon d}
+\ihalf\delta^{\dot\alpha}_{\dot\gamma} x^{\beta\dot\kappa} \cdel_{\dot\kappa\beta}\bar\Psi_{\dot\epsilon d}
-i \delta^{\dot\alpha}_{\dot\epsilon} \bar\Psi_{\dot\gamma d}
+\ihalf  \delta^{\dot\alpha}_{\dot\gamma} \bar\Psi_{\dot\epsilon d}
,
\nln
\gen[\bar L]^{\dot\alpha}{}_{\dot\gamma}\appliedto A_{\dot\epsilon\delta} &=
- ix^{\beta\dot\alpha} \cdel_{\dot\gamma\beta}A_{\dot\epsilon\delta}
+\ihalf\delta^{\dot\alpha}_{\dot\gamma} x^{\beta\dot\kappa} \cdel_{\dot\kappa\beta}A_{\dot\epsilon\delta}
,
\end{align}
as well as for the internal rotation generators
\begin{align}
\gen[R]^a{}_b\appliedto \Phi^{cd} &=
i\delta^c_b\Phi^{ad}+i\delta^c_b\Phi^{ad}-\rfrac{i}{2}\delta^a_b\Phi^{cd},
\nln
\gen[R]^a{}_b\appliedto \Psi^c{}_\delta &=
i\delta^c_b\Psi^a{}_\delta-\rfrac{i}{4}\delta^a_b\Psi^c{}_\delta,
\nln
\gen[R]^a{}_b\appliedto \bar\Psi_{\dot\gamma d} &=
-i\delta^a_d\bar\Psi_{\dot\gamma b}-\rfrac{i}{4}\delta^a_b\bar\Psi_{\dot\gamma d},
\nln
\gen[R]^a{}_b\appliedto A_{\dot\gamma\delta} &= 0.
\end{align}
The representation of the special superconformal generators
can be summarised as
\begin{align}
\gen[S]^{\alpha b}\appliedto\Phi^{cd} &=
ix^{\alpha\dot\epsilon}\gen[\bar Q]_{\dot\epsilon}{}^b\appliedto\Phi^{cd},
&
\gen[\bar S]_{a}{}^{\dot\gamma}\appliedto \Phi^{ed} &=
-ix^{\beta\dot\gamma}\gen[Q]_{a\beta}\appliedto\Phi^{ed},
\nln
\gen[S]^{\alpha b}\appliedto\Psi^c{}_\delta &=
ix^{\alpha\dot\epsilon}\gen[\bar Q]_{\dot\epsilon}{}^b \Psi^c{}_\delta
-2\delta^\alpha_\delta\Phi^{bc},
&
\gen[\bar S]_{a}{}^{\dot\gamma}\appliedto \Psi^e{}_\delta &=
-ix^{\beta\dot\gamma}\gen[Q]_{a\beta}\appliedto \Psi^e{}_\delta,
\nln
\gen[S]^{\alpha b}\appliedto\bar\Psi_{\dot\gamma d} &=
ix^{\alpha\dot\epsilon}\gen[\bar Q]_{\dot\epsilon}{}^b\bar\Psi_{\dot\gamma d} ,
&
\gen[\bar S]_{a}{}^{\dot\gamma}\appliedto \bar\Psi_{\dot\epsilon d} &=
-ix^{\beta\dot\gamma}\gen[Q]_{a\beta}\appliedto \bar\Psi_{\dot\epsilon d}
+2\delta^{\dot\gamma}_{\dot\epsilon}\bar\Phi_{ad},
\nln
\gen[S]^{\alpha b}\appliedto A_{\dot\gamma\delta} &=
ix^{\alpha\dot\epsilon}\gen[\bar Q]_{\dot\epsilon}{}^b A_{\dot\epsilon\delta},
&
\gen[\bar S]_{a}{}^{\dot\gamma}\appliedto A_{\dot\epsilon\delta} &=
-ix^{\beta\dot\gamma}\gen[Q]_{a\beta}\appliedto A_{\dot\epsilon\delta},
\end{align}
whereas the one of the special conformal generators
takes the explicit form
\begin{align}
\gen[K]^{\alpha\dot\gamma}\appliedto \Phi^{ed} &=
ix^{\alpha\dot\kappa}x^{\beta\dot\gamma}\cdel_{\beta\dot\kappa}\Phi^{ed}
+ix^{\alpha\dot\gamma}\Phi^{ed},
\nln
\gen[K]^{\alpha\dot\gamma}\appliedto \Psi^e{}_\delta &=
ix^{\alpha\dot\kappa}x^{\beta\dot\gamma}\cdel_{\beta\dot\kappa}\Psi^e{}_\delta
+ix^{\alpha\dot\gamma}\Psi^e{}_\delta
+i\delta^\alpha_\delta x^{\kappa\dot\gamma}\Psi^e{}_\kappa,
\nln
\gen[K]^{\alpha\dot\gamma}\appliedto \bar\Psi_{\dot\epsilon d} &=
ix^{\alpha\dot\kappa}x^{\beta\dot\gamma}\cdel_{\beta\dot\kappa}\bar\Psi_{\dot\epsilon d}
+ix^{\alpha\dot\gamma}\bar\Psi_{\dot\epsilon d}
+i\delta^{\dot\gamma}_{\dot\epsilon}x^{\alpha\dot\kappa}\bar\Psi_{\dot\kappa d},
\nln
\gen[K]^{\alpha\dot\gamma}\appliedto A_{\dot\epsilon\delta} &=
ix^{\alpha\dot\kappa}x^{\beta\dot\gamma}\cdel_{\beta\dot\kappa}A_{\dot\gamma\delta}.
\end{align}
Finally, the representation of a gauge transformation by the field $X$ is defined by
\begin{align}
\gengauge[X]\appliedto\Phi^{cd} &= \comm{X}{\Phi^{cd}}
,\nln
\gengauge[X]\appliedto\Psi^c{}_\delta &= \comm{X}{\Psi^c{}_\delta}
,\nln
\gengauge[X]\appliedto\bar\Psi_{\dot\gamma d} &= \comm{X}{\bar\Psi_{\dot\gamma d}}
,\nln
\gengauge[X]\appliedto A_{\dot\gamma\delta} &= i \cdel_{\dot\gamma\delta}X
.\end{align}

\paragraph{Extension.}

When discussing the level-one momentum generator $\genyang[P]$
it turns out useful to superficially extend
the level-zero algebra by an operator $\gen[B]$.
This operator neither represents a symmetry of the action
nor does its action describe a proper representation of some algebra.
We simply define its action as
\begin{align}
\label{eq:lvl0_genb}
\gen[B]\appliedto \Phi^{cd} &= 0,
&
\gen[B]\appliedto \Psi^c{}_\delta &= -\ihalf\Psi^c{}_\delta,
&
\gen[B]\appliedto \bar\Psi_{\dot\gamma d} &= +\ihalf\bar\Psi_{\dot\gamma d},
&
\gen[B]\appliedto A_{\dot\gamma\delta} &= 0.
\end{align}
Furthermore, we define two $2\times 2$ matrices of operators
$\gen[L']$ and $\gen[\bar L']$
and a $4\times 4$ matrix of operators $\gen[R']$
as the combinations of Lorentz and internal rotations,
scale transformations and $\gen[B]$
\begin{align}
\gen[L']^\beta{}_\delta&:=
\gen[L]^\beta{}_\delta
+\half\delta^\beta_\delta(\gen[D]+\gen[B]),
\nln
\gen[\bar L']^{\dot\alpha}{}_{\dot\gamma}&:=
\gen[\bar L]^{\dot\alpha}{}_{\dot\gamma}
+\half\delta^{\dot\alpha}_{\dot\gamma}(\gen[D]-\gen[B]),
\nln
\gen[R']^a{}_c&:=
\gen[R]^a{}_c
+\half\delta^a_c\gen[B].
\end{align}

These operators have a reasonably simpler action on the fields
of $\superN=4$ sYM than the corresponding Lorentz
generators $\gen[L]$
\begin{align}
\gen[L']^\beta{}_\delta\appliedto \Phi^{ce} &=
-i x^{\beta\dot\alpha} \cdel_{\dot\alpha\delta}\Phi^{ce}
-\ihalf\delta^\beta_\delta\Phi^{ce}
,\nln
\gen[L']^\beta{}_\delta\appliedto \Psi^c{}_\epsilon &=
-i x^{\beta\dot\alpha} \cdel_{\dot\alpha\delta}\Psi^c{}_\epsilon
-\ihalf\delta^\beta_\delta\Psi^c{}_\epsilon
-i \delta^\beta_\epsilon \Psi^c{}_\delta
,\nln
\gen[L']^\beta{}_\delta\appliedto \bar\Psi_{\dot\gamma e} &=
-i x^{\beta\dot\alpha} \cdel_{\dot\alpha\delta}\bar\Psi_{\dot\gamma e}
-\ihalf\delta^\beta_\delta\bar\Psi_{\dot\gamma e}
,\nln
\gen[L']^\beta{}_\delta\appliedto A_{\dot\gamma\epsilon} &=
- ix^{\beta\dot\alpha} \cdel_{\dot\alpha\delta}A_{\dot\gamma\epsilon}
,
\end{align}
the conjugate Lorentz generator $\gen[\bar L]$
\begin{align}
\gen[\bar L']^{\dot\alpha}{}_{\dot\gamma}\appliedto\Phi^{ed} &=
-i x^{\beta\dot\alpha} \cdel_{\dot\gamma\beta}\Phi^{ed}
-\ihalf \delta^{\dot\alpha}_{\dot\gamma} \Phi^{ed}
\nln
\gen[\bar L']^{\dot\alpha}{}_{\dot\gamma}\appliedto\Psi^e{}_\delta &=
-i x^{\beta\dot\alpha} \cdel_{\dot\gamma\beta}\Psi^e{}_\delta
-\ihalf \delta^{\dot\alpha}_{\dot\gamma} \Psi^e{}_\delta
,\nln
\gen[\bar L']^{\dot\alpha}{}_{\dot\gamma}\appliedto\bar\Psi_{\dot\epsilon d} &=
-i x^{\beta\dot\alpha} \cdel_{\dot\gamma\beta}\bar\Psi_{\dot\epsilon d}
-\ihalf  \delta^{\dot\alpha}_{\dot\gamma} \bar\Psi_{\dot\epsilon d}
-i \delta^{\dot\alpha}_{\dot\epsilon} \bar\Psi_{\dot\gamma d}
,\nln
\gen[\bar L']^{\dot\alpha}{}_{\dot\gamma}\appliedto A_{\dot\epsilon \delta} &=
- ix^{\beta\dot\alpha} \cdel_{\dot\gamma\delta}A_{\dot\epsilon\delta},
\end{align}
as well as the internal rotation generators $\gen[R]$
\begin{align}
\gen[R']^a{}_b\appliedto \Phi^{cd} &=
i\delta^c_b\Phi^{ad}+i\delta^c_b\Phi^{ad}-\rfrac{i}{2}\delta^a_b\Phi^{cd},
\nln
\gen[R']^a{}_b\appliedto \Psi^c{}_\delta &=
i\delta^c_b\Psi^a{}_\delta-\rfrac{i}{2}\delta^a_b\Psi^c{}_\delta,
\nln
\gen[R']^a{}_b\appliedto \bar\Psi_{\dot\gamma d} &=
-i\delta^a_d\bar\Psi_{\dot\gamma b}
+\rfrac{i}{2}\delta^a_b\bar\Psi_{\dot\gamma d},
\nln
\gen[R']^a{}_b\appliedto A_{\dot\gamma\delta} &= 0.
\end{align}
Moreover several terms of the coproduct of level-one generators
naturally combine using $\gen[L']$, $\gen[\bar L']$ and $\gen[R']$,
see \appref{app:levelone}.

\section{Level-one generators}
\label{app:levelone}

In this appendix we will give explicit expressions for the coproducts
as well as single-field actions of the level-one Yangian generators
\[
\genyang[P]_{\dot\alpha \beta},\quad
\genyang[Q]_{a \beta}, \quad
\genyang[\bar Q]_{\dot\alpha}{}^b,\quad
\genyang[R]^a {}_b\quad
\text{and}\quad \genyang[B].
\]
These are the level-one generators which commute with the ordinary momentum $\gen[P]$
(up to gauge artefacts)
and hence their single-field action can be expected to have
no explicit dependence on the position $x$.

\paragraph{Level-one momentum.}

The level-one momentum $\genyang[P]$
has the following single field action:
\begin{align}
\label{eq:yang_sfa_app}
\genyang[P]_{\dot\alpha\beta}\appliedto \Phi^{cd} &:= 0
,\nln
\genyang[P]_{\dot\alpha\beta}\appliedto \Psi^c{}_\delta
&:=-\varepsilon_{\beta\delta} \acomm!{\Phi^{ce}}{\bar\Psi_{\dot\alpha e}}
,\nln
\genyang[P]_{\dot\alpha\beta}\appliedto \bar\Psi_{\dot\gamma d}
&:=-\varepsilon_{\dot\alpha \dot\gamma}\acomm!{\bar\Phi_{de}}{\Psi^e{}_\gamma}
,\nln
\genyang[P]_{\dot\alpha\beta}\appliedto A_{\dot\gamma\delta}
&:= \rfrac{i}{4}\varepsilon_{\dot\alpha \dot\gamma} \varepsilon_{\beta\delta}
\acomm{\Phi^{ef}}{\bar\Phi_{ef}}.
\end{align}
The coproduct reads:
\begin{align}
\label{eq:yang_pahtcopro_app}
\copro\genyang[P]_{\dot\alpha\beta}
&= \genyang[P]_{\dot\alpha\beta} \otimes 1
+  1\otimes \genyang[P]_{\dot\alpha\beta}
\nln &\qquad
+ \gen[P]_{\dot\alpha\gamma} \wedge\gen[L]^\gamma{}_\beta
+  \gen[P]_{\dot\gamma\beta} \wedge\gen[\bar L]^{\dot\gamma}{}_{\dot\alpha}
+  \gen[P]_{\dot\alpha\beta} \wedge \gen[D]
- \tfrac{i}{2} \gen[Q]_{c\beta} \wedge \gen[\bar Q]_{\dot\alpha}{}^c
\nln
&=
\genyang[P]_{\dot\alpha\beta} \otimes 1
+  1\otimes \genyang[P]_{\dot\alpha\beta}
- \gen[L]'^\gamma{}_\beta \wedge\gen[P]_{\dot\alpha\gamma}
- \gen[\bar L]'^{\dot\gamma}{}_{\dot\alpha} \wedge\gen[P]_{\dot\gamma\beta}
- \tfrac{i}{2} \gen[Q]_{c\beta} \wedge \gen[\bar Q]_{\dot\alpha}{}^c
.\end{align}
The expression on the second line uses the
operators defined in \appref{app:superconformal}
which superficially depend on the extra generator $\gen[B]$.
The benefit of this expression is that
in explicit calculations
it avoids the generation of several intermediate terms
which would be cancelled by other terms.

\paragraph{Level-one supersymmetry.}

The level-one supersymmetries $\genyang[Q]$ and $\genyang[\bar Q]$
have a non-trivial single-field action only on one of the fermionic fields:
\[
\genyang[Q]_{a \beta} \appliedto \Psi^c{}_\delta
= -\rfrac{1}{2} \delta^c_a \varepsilon_{\beta \delta} \acomm!{\Phi^{ef}}{\bar{\Phi}_{ef}},
\qquad
\genyang[\bar Q]_{\dot\alpha}{}^b \appliedto \bar{\Psi}_{\dot\gamma d}
= -\rfrac{1}{2} \delta^b_d \varepsilon_{\dot\alpha \dot\gamma} \acomm!{\Phi^{ef} }{\bar{\Phi}_{ef}}.
\]
The single-field action on all the remaining fields $\field$
of the theory is:
\[
\genyang[Q]_{a \beta} \appliedto \field = 0,
\qquad
\genyang[\bar Q]_{\dot\alpha}{}^b \appliedto \field = 0.
\]
The coproduct for $\genyang[Q]$ is now:
\begin{align}
\label{eq:yang_qhatcopro_app}
\copro \genyang[Q]_{a \beta}
&= \genyang[Q]_{a \beta} \otimes 1 + 1 \otimes \genyang[Q]_{a \beta}
\nln &\qquad
+ \gen[Q]_{a \gamma} \wedge \gen[L]^\gamma {}_\beta
+ \rfrac{1}{2}   \gen[Q]_{a \beta} \wedge \gen[D]
-i \gen[P]_{ \dot\gamma \beta} \wedge \gen[\bar S]_a {}^{\dot\gamma}
- \gen[Q]_{c \beta} \wedge \gen[R]^c{}_a
\nln
&= \genyang[Q]_{a \beta} \otimes 1 + 1 \otimes \genyang[Q]_{a \beta}
+ \gen[Q]_{a \gamma} \wedge \gen[L']^\gamma {}_\beta
-i \gen[P]_{ \dot\gamma \beta} \wedge \gen[\bar S]_a {}^{\dot\gamma}
- \gen[Q]_{c \beta} \wedge \gen[R']^c{}_a,
\end{align}
and the corresponding expression for $\genyang[\bar Q]$ reads:
\begin{align}
\copro \genyang[\bar{Q}]_{\dot\alpha}{}^b
&= \genyang[\bar{Q}]_{\dot\alpha}{}^b \otimes 1
+ 1 \otimes \genyang[\bar{Q}]_{\dot\alpha}{}^b
\nln &\qquad
+ \gen[\bar{Q}]_{\dot\gamma}{}^b \wedge \gen[\bar{L}]^{\dot\gamma}{}_{\dot\alpha}
+ \rfrac{1}{2}  \gen[\bar{Q}]_{\dot\alpha}{}^b \wedge \gen[D]
+ i  \gen[P]_{\dot\alpha \gamma} \wedge \gen[S]^{\gamma b}
+  \gen[\bar{Q}]_{\dot \alpha} {}^c \wedge \gen[R]^b {}_c
\nln
&= \genyang[\bar{Q}]_{\dot\alpha}{}^b \otimes 1
+ 1 \otimes \genyang[\bar{Q}]_{\dot\alpha}{}^b
+ \gen[\bar{Q}]_{\dot\gamma}{}^b \wedge \gen[\bar{L}']^{\dot\gamma}{}_{\dot\alpha}
+ i  \gen[P]_{\dot\alpha \gamma} \wedge \gen[S]^{\gamma b}
+  \gen[\bar{Q}]_{\dot \alpha} {}^c \wedge \gen[R']^b{}_c.
\end{align}

\paragraph{Level-one internal rotations.}

The generator $\genyang[R]^a{}_b$ has a trivial single-field action
\[
\genyang[R]^a{}_b \appliedto \field =0.
\]
Heuristically, this is easy to understand:
The generator $\genyang[R]$ carries zero mass dimension
so its action on a field $\field$
should have the same mass dimension as $\field$,
i.e.\ $1$ for bosonic fields or $\vfrac{3}{2}$ for fermionic fields.
Furthermore, the action should have no explicit position dependence
as argued above.
Together, this implies that the result can only be a single field.
However, the parity-inverting property of the level-one generators
requires to have at least two fields in the result.

The following coproduct thus defines this symmetry completely:
\begin{align}
\copro \genyang[R]^a {}_b
&= \genyang[R]^a{}_b \otimes 1 + 1 \otimes \genyang[R]^a{}_b
+ \gen[R]^a{}_c \wedge \gen[R]^c{}_b
\nln &\qquad
- \rfrac{1}{2} \gen[S]^{\gamma a} \wedge \gen[Q]_{b \gamma}
- \rfrac{1}{2} \gen[\bar{S}]_b {}^{\dot\gamma} \wedge \gen[\bar{Q}]_{\dot\gamma}{}^a
+ \rfrac{1}{8} \delta^a_b\gen[S]^{\gamma d} \wedge \gen[Q]_{d \gamma}
+ \rfrac{1}{8} \delta^a_b\gen[\bar{S}]_d{}^{\dot\gamma} \wedge \gen[\bar{Q}]_{\dot\gamma}{}^d.
\end{align}

\paragraph{Level-one bonus symmetry.}

For the same reasons as for $\genyang[R]$,
the level-one bonus symmetry $\genyang[B]$
has a trivial single-field action
\[
\genyang[B] \appliedto \field =0.
\]
The coproduct reads:
\[
\copro \genyang[B]
= \genyang[B] \otimes 1
+ 1 \otimes \genyang[B]
- \quarter\gen[\bar{S}]_b {}^{\dot\alpha} \wedge \gen[\bar{Q}]_{\dot\alpha}{}^b
- \quarter\gen[S]^{\alpha b} \wedge \gen[Q]_{b \alpha}.
\]
In fact we can now define a modified level-one internal rotation operator:
\[
\genyang[R]'^a {}_b := \genyang[R]^a {}_b + \half \delta^a_b\genyang[B],
\]
whose trace is $2\genyang[B]$.
This generator has a slightly simplified coproduct
as compared to $\genyang[R]$:
\begin{align}
\copro \genyang[R]'^a {}_b
&= \genyang[R]'^a{}_b \otimes 1 + 1 \otimes \genyang[R]'^a{}_b
+ \gen[R]^a{}_c \wedge \gen[R]^c{}_b
- \rfrac{1}{2} \gen[S]^{\gamma a} \wedge \gen[Q]_{b \gamma}
- \rfrac{1}{2} \gen[\bar{S}]_b {}^{\dot\gamma} \wedge \gen[\bar{Q}]_{\dot\gamma}{}^a.
\end{align}

\bibliographystyle{nb}
\bibliography{N4Yang}

\end{document}